\documentclass[journal]{IEEEtran}
\newcommand{\RNum}[1]{\uppercase\expandafter{\romannumeral#1\relax}}
\usepackage{diagbox}
\usepackage{graphicx}
\usepackage{url}
\usepackage{hyperref}
\usepackage{gensymb}
\usepackage{amsmath}
\usepackage{float}
\usepackage{booktabs}
\usepackage{epstopdf}
\usepackage[justification=centering]{caption}
\usepackage{cite}
\usepackage{amsfonts}
\usepackage{tabularx}
\usepackage{booktabs}
\usepackage{threeparttable}
\usepackage{upgreek}
\usepackage{subfigure}
\usepackage{color,soul}
\usepackage{mathtools}
\usepackage{multirow}
\usepackage[figuresright]{rotating}
\usepackage{adjustbox}
\usepackage{textcomp}
\usepackage{nomencl}
\usepackage{bbding}
\usepackage{hyperref}
\usepackage{lipsum}

\usepackage{colortbl}
\usepackage{arydshln}
\usepackage{multicol}

\makenomenclature

\soulregister\cite7
\soulregister\citep7
\soulregister\citet7
\soulregister\ref7
\soulregister\pageref7

\hypersetup{
	colorlinks=true,
	linkcolor=red,
	filecolor=blue,
	urlcolor=blue,
	citecolor=green,
}

\newtheorem{theorem}{Theorem}

\newtheorem{lemma}{Lemma}

\newtheorem{corollary}{Corollary}

\newtheorem{assumption}{Assumption}

\begin{document}
	%ÉÏÒ»¸ö°æ±¾ÊÇSecurityLEOSCS£¬ ±í¸ñ¶ÔÓ¦µÄCSVÒÔ¼°Ö®Ç°µÄ×öµÄ±¸×¢¶¼ÔÚ¸ÃÎÄ¼þÖÐ
	%ÕâÀï×îºóÓÃ²»ÓÃThreats	
	% ½ö¸üÐÂÁËË«À¸°æ±¾µÄhanzo Respon letter 0823 ×îºó°æ±¾
	\title{Low Earth Orbit Satellite Security and Reliability: Issues, Solutions, and the Road Ahead}
	
	\author{
		\normalsize
		%		Pingyue Yue,~\IEEEmembership{\normalsize Student Member,~IEEE},
		Pingyue Yue,~\IEEEmembership{\normalsize Student Member,~IEEE},
		Jianping An,~\IEEEmembership{\normalsize Senior Member,~IEEE},
		Jiankang Zhang,~\IEEEmembership{\normalsize Senior Member,~IEEE},
		\\Jia Ye,~\IEEEmembership{\normalsize Member,~IEEE},
		Gaofeng Pan,~\IEEEmembership{\normalsize Senior Member,~IEEE},
		Shuai Wang,~\IEEEmembership{\normalsize Member,~IEEE},
		\\Pei Xiao,~\IEEEmembership{\normalsize Senior Member,~IEEE},
		and Lajos Hanzo,~\IEEEmembership{\normalsize Life Fellow,~IEEE}
		
		\thanks{
			% \emph{\textbf{(Corresponding author)}}
			
			This work was supported  in part by the National Key Research and Development Program of China under Grant 2021YFC3320200 and 2022YFC3331102, and in part by the National Natural Science Foundation of China under Grant 62171031.
						
			Pingyue Yue is with the School of Information and Electronics, Beijing Institute of Technology, Beijing 100081, China (e-mails: ypy@bit.edu.cn).
			
			Jianping An \textbf{(Corresponding author)}, Gaofeng Pan, and Shuai Wang are with the School of Cyberspace Science and Technology, Beijing Institute of Technology, Beijing 100081, China (e-mails: an@bit.edu.cn; gfpan@bit.edu.cn; swang@bit.edu.cn).
			
			Jiankang Zhang is with the Department of Computing and Informatics, Bournemouth University, Bournemouth BH12 5BB, U.K. (e-mail: jzhang3@bournemouth.ac.uk).
			
			Jia Ye is with the school of Electrical Engineering, Chongqing University, Chongqing, 400044, China (yejiaft@163.com).
			
			Pei Xiao is with the 5GIC \& 6GIC, Institute for Communication Systems, University of Surrey,
			GU2 7XH, U.K. (e-mail: p.xiao@surrey.ac.uk).
			
			Lajos Hanzo is with the School of Electronics and Computer Science, University of Southampton, Southampton SO17 1BJ, U.K. (e-mail: lh@ecs.soton.ac.uk).
			
		}
	}
	\maketitle
	
	\begin{abstract}				
		Low Earth Orbit (LEO) satellites undergo a period of rapid development driven by ever-increasing user demands, reduced costs, and technological progress. Since there is a lack of literature on the security and reliability issues of LEO Satellite Communication Systems (SCSs), we aim to fill this knowledge gap. Specifically, we critically appraise the inherent characteristics of LEO SCSs and elaborate on their security and reliability requirements. In light of this, we further discuss their vulnerabilities, including potential security attacks launched against them and reliability risks, followed by outlining the associated lessons learned. Subsequently, we discuss the corresponding security and reliability enhancement solutions, unveil a range of trade-offs, and summarize the lessons gleaned. Furthermore, we shed light on several promising future research directions for enhancing the security and reliability of LEO SCSs, such as integrated sensing and communication, computer vision aided communications, as well as challenges brought about by mega-constellation and commercialization. Finally, we summarize the lessons inferred and crystallize the take-away messages in our design guidelines.		
	\end{abstract}
	
	\section{Introduction}
\begin{figure}[ht]
		\centering
		
		\includegraphics[width=1\columnwidth]{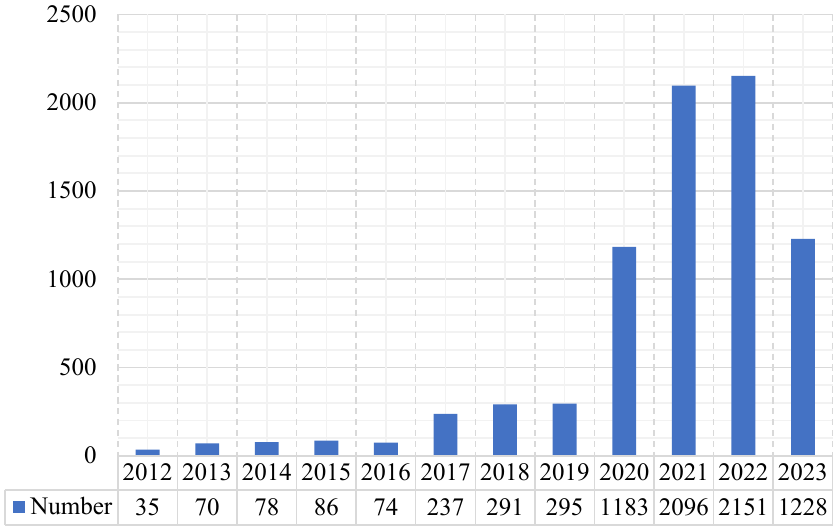}
		
		\caption{Number of LEO satellites launched from 2012 to the second quarter of 2023.}
		\label{satellitenum}
	\end{figure}

		Driven by the explosive proliferation of smart devices and the escalation of data traffic, the Sixth-generation (6G) \cite{LEOOFTS, THZCOMM2022, Netw2022} concept aims for building a large-dimensional and autonomous global network capable of supporting seamless coverage and ubiquitous services. 
		As evidenced by the literature \cite{Xiaohuyou}, it has been proposed that future wireless networks must be able to seamlessly interface with terrestrial and satellite networks. Compared to Medium Earth Orbit (MEO) and Geostationary Earth Orbit (GEO) satellites, Low Earth Orbit (LEO) satellites \cite{LEOSANWC2022, iOt2022, MATWC} are closer to the Earth. Hence, they are more suitable for supporting delay-sensitive communications worldwide \cite{GaoFengTMC2022}. Additionally, rocket recovery and multi-satellite launching technologies have substantially reduced the average launch cost and deployment time. From 2012 to the second quarter of 2023, about 7824 LEO satellites have been successfully launched, as shown in Fig.~\ref{satellitenum}. As a benefit, LEO Satellite Communication Systems (SCSs) have found a plethora of applications, including the Internet of Remote Things (IoRT), smart city, and emergency rescue \cite{XiaohanTWC2023, MatingWC2022, LEOSatBroad}. 
	
	%	From 2012 to the first quarter of 2023, about 7220 LEO satellites have been successfully launched, accounting for nearly 91.8\% of the total launch volume of all types of satellites, and LEO satellites are proliferating, as shown in Fig.~\ref{satellitenum}. 
	
\begin{figure*}[]
		\centering
		
		\includegraphics[width=2\columnwidth]{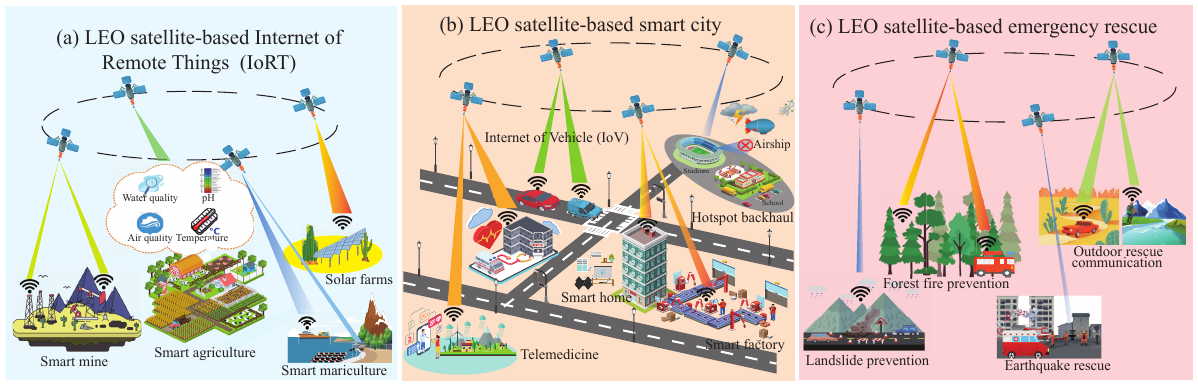}
		
		\caption{The application scenarios of LEO SCSs.}
		\label{application}
	\end{figure*}

	Fig.~\ref{application} illustrates the application scenarios of LEO satellites. In Fig.~\ref{application}(a), the LEO satellite-based IoRT concept is illustrated, where LEO satellites are deployed to support seamless wireless access to remote geographical areas \cite{MingtingTCOM2022}. Since they are closer to the earth, they have low propagation loss, which reduces the transmit power requirements of power-limited sensors. A large number of sensors deployed in mines, farms, mariculture farms, and solar power plants collect voltage, temperature, pH, and other status information and then separately upload them to LEO satellites. LEO satellites deliver these sensory data to remote operators for further analysis and processing.

	In Fig.~\ref{application}(b), the LEO satellite-based smart city scenario is illustrated, where they are employed in support of telemedicine, the Internet of Vehicles (IoV), smart factories, and homes for improving urban services, the city's sustainability, and the factory's production efficiency \cite{HoangTAES2023}. For instance, the IoV allows vehicles to communicate with the surrounding environment, such as neighboring cars and roadside infrastructure, supporting a wide range of `on-the-go' services such as road safety, congestion control, and location-dependent services. It is imperative to leverage LEO satellites to serve vehicles anywhere and anytime by exploiting their respective advantages in terms of low latency, and seamless coverage \cite{HanTVT2022}. Long-distance diagnosis, consultation, and treatment may be provided for the wounded and sick in case of emergencies. In addition, in massive connectivity scenarios at schools, sporting events, or rallies, it may be permitted to deploy airships to provide temporary access but it is difficult to establish stable wireless backhaul links due to the weather conditions, such as wind and rain. Therefore, LEO satellites constitute a promising solution for coverage extension and backhaul links since they tend to be more immune to weather conditions \cite{LEOTVT20221}. In Fig.~\ref{application}(c), an LEO satellite-based emergency rescue scenario is portrayed \cite{TII2023}. People in remote mountains or deserts may use their terminals to send distress signals or even may have access to real-time voice services via LEO satellites in an emergency. Disasters, such as landslides, forest fires, and earthquakes, may cause the loss of life and property, which motivate emergency responses based on LEO satellites to support enhanced situational awareness, automated decision-making, and a whole host of other prompt responses \cite{LEOemerge1}.

	Given the wide range of applications of LEO SCSs seen in Fig.~\ref{application}, their security is also of paramount importance. For example, the data collected by sensors in mines contains confidential information, including mineral types and reserves, which attracts potential commercial competitors to exploit their satellites for eavesdropping. Butun {\itshape et al}. \cite{SecurityIoT} revealed that the operational telemedicine systems lack the strong security services that prevent patient privacy from disclosure. In addition, given the ongoing deployment of dense LEO mega-constellations, the electromagnetic environment becomes more complex, which may jam or disrupt communications altogether.
	
	\section{Existing Literature and Contributions}
		In this section, we delve deeper into a discussion of existing literature and summarize their contributions and limitations, which has motivated our research. Subsequently, we detail our contributions. Finally, the organization of this paper is provided.
		
		\subsection{Existing Literature}
		
		\begin{table*}[ht]
			\centering
			\caption{Comparison with literature}
			%		\tiny
			\renewcommand\arraystretch{1.3}
			\begin{tabular}{|l|llllll|lllll|l|l|}
				\hline
				\hline
				& \multicolumn{6}{l|}{Security and reliability issues}                                                                  & \multicolumn{5}{l|}{Security and reliability}                                                                        &  &      \\ \cline{2-7}
				Ref.                                                 & \multicolumn{4}{l|}{Security attacks}                                                                                                                                                                                                 & \multicolumn{2}{l|}{Reliability risks}                                               & \multicolumn{5}{l|}{enchancement solutions}                                                                          & Future & Design     \\ \cline{2-12}
				& \multicolumn{1}{l|}{Eavesdropping} & \multicolumn{1}{l|}{Jamming} & \multicolumn{1}{l|}{\begin{tabular}[c]{@{}l@{}}Message\\ modification\end{tabular}} & \multicolumn{1}{l|}{\begin{tabular}[c]{@{}l@{}}User\\ privacy\end{tabular}} & \multicolumn{1}{l|}{CCI} & \begin{tabular}[c]{@{}l@{}}Physical\\ threats\end{tabular} & \multicolumn{1}{l|}{PLS} & \multicolumn{1}{l|}{CR} & \multicolumn{1}{l|}{AI} & \multicolumn{1}{l|}{QKD} & Blockchain & trends & guidelines \\ \hline
				\cite{PLSSIN}&  \multicolumn{1}{l|}{$\checkmark$}              & \multicolumn{1}{l|}{}        & \multicolumn{1}{l|}{}                                                               & \multicolumn{1}{l|}{}                                                       & \multicolumn{1}{l|}{}    &                                                           & \multicolumn{1}{l|}{$\checkmark$}    & \multicolumn{1}{l|}{}   & \multicolumn{1}{l|}{}   & \multicolumn{1}{l|}{}    &            &   $\checkmark$     &            \\ \hline
				
				\cite{SatNet}&  \multicolumn{1}{l|}{$\checkmark$}              & \multicolumn{1}{l|}{$\checkmark$}        & \multicolumn{1}{l|}{}                                                               & \multicolumn{1}{l|}{}                                                       & \multicolumn{1}{l|}{}    &                                                           & \multicolumn{1}{l|}{$\checkmark$}    & \multicolumn{1}{l|}{}   & \multicolumn{1}{l|}{}   & \multicolumn{1}{l|}{}    &            &   $\checkmark$     &            \\ \hline

				\cite{BlockchainSAGIN1}&  \multicolumn{1}{l|}{}              & \multicolumn{1}{l|}{}        & \multicolumn{1}{l|}{$\checkmark$}                                                               & \multicolumn{1}{l|}{$\checkmark$}                                                       & \multicolumn{1}{l|}{}    &                                                           & \multicolumn{1}{l|}{}    & \multicolumn{1}{l|}{}   & \multicolumn{1}{l|}{}   & \multicolumn{1}{l|}{}    &   $\checkmark$          &        &            \\ \hline
				
				\cite{SatNet1}&  \multicolumn{1}{l|}{$\checkmark$}              & \multicolumn{1}{l|}{}        & \multicolumn{1}{l|}{$\checkmark$}                                                               & \multicolumn{1}{l|}{}                                                       & \multicolumn{1}{l|}{}    &                                                           & \multicolumn{1}{l|}{}    & \multicolumn{1}{l|}{}   & \multicolumn{1}{l|}{$\checkmark$}   & \multicolumn{1}{l|}{}    &    $\checkmark$         &       &            \\ \hline
				
				\cite{Liujiajia2018}&  \multicolumn{1}{l|}{$\checkmark$}              & \multicolumn{1}{l|}{$\checkmark$}        & \multicolumn{1}{l|}{}                                                               & \multicolumn{1}{l|}{}                                                       & \multicolumn{1}{l|}{}    &                                                           & \multicolumn{1}{l|}{$\checkmark$}    & \multicolumn{1}{l|}{}   & \multicolumn{1}{l|}{}   & \multicolumn{1}{l|}{}    &            &       &            \\ \hline
				
				\cite{PLS5GSat}&  \multicolumn{1}{l|}{$\checkmark$}              & \multicolumn{1}{l|}{}        & \multicolumn{1}{l|}{}                                                               & \multicolumn{1}{l|}{}                                                       & \multicolumn{1}{l|}{}    &                                                           & \multicolumn{1}{l|}{$\checkmark$}    & \multicolumn{1}{l|}{}   & \multicolumn{1}{l|}{}   & \multicolumn{1}{l|}{}    &            &   $\checkmark$    &            \\ \hline
				
				\cite{ORBCOMM}&  \multicolumn{1}{l|}{$\checkmark$}              & \multicolumn{1}{l|}{}        & \multicolumn{1}{l|}{}                                                               & \multicolumn{1}{l|}{}                                                       & \multicolumn{1}{l|}{$\checkmark$}    &                                                           & \multicolumn{1}{l|}{$\checkmark$}    & \multicolumn{1}{l|}{$\checkmark$}   & \multicolumn{1}{l|}{}   & \multicolumn{1}{l|}{}    &            &   $\checkmark$    &            \\ \hline
				
				\cite{PLS6G}&  \multicolumn{1}{l|}{$\checkmark$}              & \multicolumn{1}{l|}{}        & \multicolumn{1}{l|}{}                                                               & \multicolumn{1}{l|}{}                                                       & \multicolumn{1}{l|}{}    &                                                           & \multicolumn{1}{l|}{$\checkmark$}    & \multicolumn{1}{l|}{}   & \multicolumn{1}{l|}{}   & \multicolumn{1}{l|}{}    &            &   $\checkmark$    &            \\ \hline			
				
				\cite{COMST6G1}&  \multicolumn{1}{l|}{$\checkmark$}              & \multicolumn{1}{l|}{$\checkmark$}        & \multicolumn{1}{l|}{$\checkmark$}                                                               & \multicolumn{1}{l|}{$\checkmark$}                                                       & \multicolumn{1}{l|}{}    &                                                           & \multicolumn{1}{l|}{}    & \multicolumn{1}{l|}{}   & \multicolumn{1}{l|}{$\checkmark$}   & \multicolumn{1}{l|}{$\checkmark$}    &   $\checkmark$         &   $\checkmark$    &            \\ \hline	
				
				\cite{COMST6G20214}&  \multicolumn{1}{l|}{$\checkmark$}              & \multicolumn{1}{l|}{$\checkmark$}        & \multicolumn{1}{l|}{$\checkmark$}                                                               &  \multicolumn{1}{l|}{$\checkmark$}                                                       & \multicolumn{1}{l|}{}    &                   $\checkmark$                                        & \multicolumn{1}{l|}{}    & \multicolumn{1}{l|}{$\checkmark$}   & \multicolumn{1}{l|}{$\checkmark$}   & \multicolumn{1}{l|}{$\checkmark$}    &          &   $\checkmark$    &            \\ \hline
				
				\cite{LEO6GACCESS} &  \multicolumn{1}{l|}{}              & \multicolumn{1}{l|}{}        & \multicolumn{1}{l|}{}                                                               &  \multicolumn{1}{l|}{}                                                       & \multicolumn{1}{l|}{$\checkmark$}    &                                                           & \multicolumn{1}{l|}{}    & \multicolumn{1}{l|}{$\checkmark$}   & \multicolumn{1}{l|}{}   & \multicolumn{1}{l|}{}    &          &       &            \\ \hline		
				
				\cite{BlockchainSAGIN}&  \multicolumn{1}{l|}{}              & \multicolumn{1}{l|}{}        & \multicolumn{1}{l|}{$\checkmark$}                                                               &  \multicolumn{1}{l|}{$\checkmark$}                                                       & \multicolumn{1}{l|}{}    &                                                           & \multicolumn{1}{l|}{}    & \multicolumn{1}{l|}{}   & \multicolumn{1}{l|}{$\checkmark$}   & \multicolumn{1}{l|}{}    &   $\checkmark$       &       &            \\ \hline
				
				\cite{MinruiXu}&  \multicolumn{1}{l|}{$\checkmark$}              & \multicolumn{1}{l|}{}        & \multicolumn{1}{l|}{$\checkmark$}                                                               &  \multicolumn{1}{l|}{}                                                       & \multicolumn{1}{l|}{}    &                                                          & \multicolumn{1}{l|}{}    & \multicolumn{1}{l|}{}   & \multicolumn{1}{l|}{$\checkmark$}   & \multicolumn{1}{l|}{$\checkmark$}    &          &     &            \\ \hline
				
				\cite{ShengmingComst}&  \multicolumn{1}{l|}{}              & \multicolumn{1}{l|}{}        & \multicolumn{1}{l|}{$\checkmark$}                                                               &  \multicolumn{1}{l|}{}                                                       & \multicolumn{1}{l|}{$\checkmark$ }    &                                                          & \multicolumn{1}{l|}{}    & \multicolumn{1}{l|}{$\checkmark$}   & \multicolumn{1}{l|}{$\checkmark$}   & \multicolumn{1}{l|}{}    &          &   $\checkmark$    &            \\ \hline
				
				\cite{ReinformComst}&  \multicolumn{1}{l|}{$\checkmark$}              & \multicolumn{1}{l|}{}        & \multicolumn{1}{l|}{$\checkmark$}                                                               &  \multicolumn{1}{l|}{$\checkmark$}                                                       & \multicolumn{1}{l|}{}    &                                                          & \multicolumn{1}{l|}{}    & \multicolumn{1}{l|}{$\checkmark$}   & \multicolumn{1}{l|}{}   & \multicolumn{1}{l|}{}    &          &   $\checkmark$    &            \\ \hline
				
				\cite{SatAir}&  \multicolumn{1}{l|}{}              & \multicolumn{1}{l|}{$\checkmark$}        & \multicolumn{1}{l|}{$\checkmark$}                                                               &  \multicolumn{1}{l|}{}                                                       & \multicolumn{1}{l|}{}    &                                                          & \multicolumn{1}{l|}{}    & \multicolumn{1}{l|}{}   & \multicolumn{1}{l|}{$\checkmark$}   & \multicolumn{1}{l|}{}    &          &       &            \\ \hline
				
				\cite{SatSpectNet}&  \multicolumn{1}{l|}{}              & \multicolumn{1}{l|}{}        & \multicolumn{1}{l|}{$\checkmark$}                                                               &  \multicolumn{1}{l|}{$\checkmark$}                                                       & \multicolumn{1}{l|}{}    &                                                           & \multicolumn{1}{l|}{}    & \multicolumn{1}{l|}{}   & \multicolumn{1}{l|}{$\checkmark$}   & \multicolumn{1}{l|}{$\checkmark$}    &          &       &            \\ \hline
				
				\cite{SatIoTCOMST}&  \multicolumn{1}{l|}{$\checkmark$}              & \multicolumn{1}{l|}{$\checkmark$}        & \multicolumn{1}{l|}{}                                                               &  \multicolumn{1}{l|}{}                                                       & \multicolumn{1}{l|}{}    &                                                          & \multicolumn{1}{l|}{}    & \multicolumn{1}{l|}{$\checkmark$}   & \multicolumn{1}{l|}{}   & \multicolumn{1}{l|}{}    &          &       &            \\ \hline	
				
				\cite{COMST6G2}&  \multicolumn{1}{l|}{}              & \multicolumn{1}{l|}{$\checkmark$}        & \multicolumn{1}{l|}{$\checkmark$}                                                               &  \multicolumn{1}{l|}{$\checkmark$}                                                       & \multicolumn{1}{l|}{}    &                                                           & \multicolumn{1}{l|}{}    & \multicolumn{1}{l|}{}   & \multicolumn{1}{l|}{$\checkmark$}   & \multicolumn{1}{l|}{$\checkmark$}    &          &   $\checkmark$    &            \\ \hline
				
				\cite{LEODisturb}&  \multicolumn{1}{l|}{$\checkmark$}              & \multicolumn{1}{l|}{$\checkmark$}        & \multicolumn{1}{l|}{$\checkmark$}                                                               &  \multicolumn{1}{l|}{$\checkmark$}                                                       & \multicolumn{1}{l|}{$\checkmark$}    &                                                         & \multicolumn{1}{l|}{}    & \multicolumn{1}{l|}{$\checkmark$}   & \multicolumn{1}{l|}{$\checkmark$}   & \multicolumn{1}{l|}{$\checkmark$}    &          &   $\checkmark$    &            \\ \hline			
				
				\cite{Cyber2020}&  \multicolumn{1}{l|}{$\checkmark$}              & \multicolumn{1}{l|}{$\checkmark$}        & \multicolumn{1}{l|}{$\checkmark$}                                                               &  \multicolumn{1}{l|}{$\checkmark$}                                                       & \multicolumn{1}{l|}{}    &                                                          & \multicolumn{1}{l|}{$\checkmark$}    & \multicolumn{1}{l|}{}   & \multicolumn{1}{l|}{}   & \multicolumn{1}{l|}{$\checkmark$}    &          &   $\checkmark$    &            \\ \hline	
				
				\cite{Satarxiv}&  \multicolumn{1}{l|}{$\checkmark$}              & \multicolumn{1}{l|}{$\checkmark$}        & \multicolumn{1}{l|}{$\checkmark$}                                                               &  \multicolumn{1}{l|}{$\checkmark$}                                                       & \multicolumn{1}{l|}{}    &                                                           & \multicolumn{1}{l|}{$\checkmark$}    & \multicolumn{1}{l|}{$\checkmark$}   & \multicolumn{1}{l|}{$\checkmark$}   & \multicolumn{1}{l|}{$\checkmark$}    &          &   $\checkmark$    &            \\ \hline
				
				\cite{Marko2022}&  \multicolumn{1}{l|}{}              & \multicolumn{1}{l|}{}        & \multicolumn{1}{l|}{}                                                               &  \multicolumn{1}{l|}{}                                                       & \multicolumn{1}{l|}{$\checkmark$ }    &                   $\checkmark$                                        & \multicolumn{1}{l|}{}    & \multicolumn{1}{l|}{$\checkmark$}   & \multicolumn{1}{l|}{$\checkmark$}   & \multicolumn{1}{l|}{$\checkmark$}    &          &   $\checkmark$    &            \\ \hline								
				
				\begin{tabular}[c]{@{}l@{}}This\\ paper\end{tabular} & \multicolumn{1}{l|}{\CheckmarkBold}              & \multicolumn{1}{l|}{\CheckmarkBold}        & \multicolumn{1}{l|}{\CheckmarkBold}                                                               & \multicolumn{1}{l|}{\CheckmarkBold}                                                       & \multicolumn{1}{l|}{\CheckmarkBold}    &        \CheckmarkBold                                                   & \multicolumn{1}{l|}{\CheckmarkBold}    & \multicolumn{1}{l|}{\CheckmarkBold}   & \multicolumn{1}{l|}{\CheckmarkBold}   & \multicolumn{1}{l|}{\CheckmarkBold}    &  \CheckmarkBold          &    \CheckmarkBold    &    \CheckmarkBold        \\ \hline
			\end{tabular}
		\end{table*}
		
		%In recent years, a range of short magazine papers \cite{SINCM, SatAir, SatNet1, PLS5GSat, SatNet, SatSpectNet, BlockchainSAGIN1,MinruiXu} and survey papers \cite{JK2016, YAN2019, ORBCOMM, PLSSIN, Cyber2020, PLS6G, SatIoTCOMST, COMST6G1, Satarxiv, COMST6G20214, LEODisturb, LEO6GACCESS, COMST6G2, BlockchainSAGIN,Marko2022} 
		
		In recent years, a range of important papers have been conceived on the security of LEO SCSs. Although these papers have played a certain role in how to safeguard LEO SCSs, they also have shortcomings. Firstly, some of them focused on the security of Space Information Networks (SIN) relying on LEO SCSs. More specifically, Li {\itshape et al.} \cite{PLSSIN} considered the security performance as their pivotal target, focused on eavesdropping security attacks, and presented a security design of the SCSs from the perspective of Physical Layer Security (PLS). Han {\itshape et al.} \cite{SatNet} critically appraised a secure architecture to safeguard the SIN, where relays relying on hopped beams were deployed for mitigating both the jamming attacks of the uplink and the eavesdropping attacks of the downlink. In fact, the SIN has also suffered security attacks, including user privacy and message modification, which are not covered in this paper. To address this, Bao {\itshape et al.} \cite{BlockchainSAGIN1} presented blockchain techniques for dealing with user privacy and message modification. But blockchain alone cannot solve physical layer attacks like jamming. Therefore, the protection of SIN cannot depend on one technique alone, but on the collaboration of several techniques. 
		
		Secondly, substantial efforts were dedicated to integrated networks containing LEO SCSs, e.g., Space-air-ground Integrated Networks (SAGIN), and satellite-terrestrial networks. Li {\itshape et al.} \cite{SatNet1} conceived the integration of Artificial Intelligence (AI) and blockchain for improving data security (e.g., eavesdropping and malicious message modification) in 6G. Similarly, this paper still lacked a discussion on malicious jamming and corresponding measures. Liu {\itshape et al.} \cite{Liujiajia2018} discussed the core issues of cross-layer design, resource management, and allocation in SAGINs, with a brief reference to PLS techniques to address eavesdropping. Lin {\itshape et al.} \cite{PLS5GSat} surveyed the current activities and system architecture of converged 5G and satellite networks. A novel metric, termed as an effective and achievable rate relied on the PLS technique, was conceived for quantifying the trade-off between reliability and security. However, security is mainly focused on eavesdropping. Wang {\itshape et al.} \cite{ORBCOMM} highlighted the convergence of satellite and terrestrial networks, where the former was deemed to be more vulnerable to security violation risks and eavesdropping threats. Multiple Input Multiple Output (MIMO) antenna-aided PLS techniques were discussed as their solutions. Another security concern addressed in this paper is reliability degradation due to spectrum scarcity. As a remedy, the Cognitive Radio (CR) technique is adopted for dealing with this issue. But other security attacks, such as jamming and message modification, were not mentioned in this paper. Lorenzo {\itshape et al.} \cite{PLS6G} focused their attention on PLS based on Artificial Noise(AN) as well as Reflective Intelligent Surfaces (RISs) to address eavesdropping in 6G. Nguyen {\itshape et al.} \cite{COMST6G1} focused their research on the emerging security risks, such as learning-empowered attacks and massive data breaches, caused by the plethora of devices and a suite of novel technologies emerging as part of the recent 6G. Security and privacy issues were discussed in the context of the physical, connection, and service layers. The assessments of the prospective techniques, such as PLS, Quantum Key Distribution (QKD), and distributed ledgers, were also outlined. Guo {\itshape et al.} \cite{COMST6G20214} surveyed the security threats in SAGINs and divided them into four research areas, i.e. operation threats, network threats, and data threats. Furthermore, a variety of attack methodologies and their corresponding solutions were discussed. This paper discusses security issues comprehensively, especially the first-time discussion of physical threats, such as earthquakes and floods, that affect reliability. However, these threats are targeted at terrestrial networks in SAGINs and do not involve LEO satellites. Xie {\itshape et al.} \cite{LEO6GACCESS} focused on the future development of key technologies and challenges for LEO mega-constellations for 6G global coverage. This paper introduced interference coordination techniques, such as CR, to mitigate interference between LEO satellites and terrestrial networks as well as GEO satellites. Wang {\itshape et al.} \cite{BlockchainSAGIN} presented a comprehensive survey of the family of blockchain solutions designed for SAGINs. This paper classified security attacks into data-related, identity-related, service-related, and so on, but neglected signal-related security attacks, such as eavesdropping and jamming, in wireless environments. Xu {\itshape et al.} \cite{MinruiXu} focused their attention on QKD solutions for the sake of providing ultimate security for the space, aerial, and ground nodes of the emerging SAGIN systems. Zhou {\itshape et al.}\cite{ShengmingComst} presented a comprehensive survey of aerospace-integrated network innovation for 6G and discussed AI measures to deal with security attacks. In addition, Lu {\itshape et al.} \cite{ReinformComst} discussed the reinforcement learning-based cross-layer security and privacy protection methods conceived for enhancing physical layer security and user privacy in 6G. By integrating the physical layer, media access control layer, and network layer, the paper optimized the network security decision-making mechanism using reinforcement learning, while maintaining the user experience and performance. However, Lu {\itshape et al.} only provided an overview of reinforcement learning-based security solutions for UAV communications in the Non-terrestrial Networks (NTNs) of 6G.
		
		Thirdly, some authors have discussed security issues in specific satellite-based applications. Liu {\itshape et al.} \cite{SatAir} focused on the security of satellite-based Automatic Dependent Surveillance-Broadcast (ADS-B). This paper mainly discussed the employment of machine learning for dealing with malicious injection and modification attacks. Hao {\itshape et al.} \cite{SatSpectNet} considered the security and privacy issues encountered in satellite-based radio spectrum monitoring and outlined the compelling benefit of blockchain in security and privacy protection dispensing with centralized authorization. Centenaro {\itshape et al.} \cite{SatIoTCOMST} surveyed the satellite-based IoT and suggested the employment of optical Inter-satellite Links (ISLs) for mitigating jamming and eavesdropping. However, this paper lacked a discussion on the protection of data security and user privacy. Vaezi {\itshape et al.} \cite{COMST6G2} studied the most prevalent attacks targeted at the satellite-based Internet of Things (IoT), and categorized them into physical attacks, software attacks, and network attacks based on their entry point. Then a host of Deep Learning (DL) and federated learning techniques were proposed as their corresponding solutions. This paper primarily focused on the security attacks on cellular IoT, but there was relatively little content about satellite IoT. Hraishawi {\itshape et al.} \cite{LEODisturb} discussed the deployment challenges of LEO SCSs, including their coexistence with GEO SCSs and terrestrial communication systems. Both PLS techniques and QKD schemes were also introduced as the means of mitigating the associated security threats, such as eavesdropping and jamming. With the wider development of LEO SCSs, some concomitant security issues have come along. Manulis {\itshape et al.} \cite{Cyber2020} analyzed a whole host of past satellite security threats and discussed their motivations and characteristics. Moreover, they also discussed the emerging security risks posed by advanced technologies, such as Commercial Off The Shelf (COTS) components, Software Defined Radios (SDRs), and cloud computing. Tedeschi {\itshape et al.} \cite{Satarxiv} surveyed the security of SCSs, with an emphasis on PLS and cryptography. More specifically, anti-jamming strategies and anti-spoofing schemes were discussed in PLS, while authentication, key agreement, and key distribution based on emerging quantum domain techniques were also studied. However, these two papers lacked a discussion on the reliability of the continuous deployment of LEO satellites, such as the Co-channel Interference (CCI) between LEO satellites and terrestrial networks as well as GEO satellites due to the spectrum scarcity.  In addition, Marko {\itshape et al.} \cite{Marko2022} emphasized the importance of space safety for sustainable satellites and discussed hot issues of space traffic management, debris detection, and spectrum sharing of SCSs. However, this paper lacked a discussion on eavesdropping and malicious jamming as well as corresponding solutions.

		Again, Table~\uppercase\expandafter{\romannumeral1} boldly and explicitly contrasts this survey against the existing magazine and survey papers, indicating that a wider survey is provided by this literature by critically appraising as many as 23 citations \cite{PLSSIN,SatNet,BlockchainSAGIN1,SatNet1,Liujiajia2018,PLS5GSat,ORBCOMM,PLS6G,COMST6G1,COMST6G20214,LEO6GACCESS,BlockchainSAGIN,MinruiXu,ShengmingComst,ReinformComst,SatAir,SatSpectNet,SatIoTCOMST,COMST6G2,LEODisturb,Cyber2020,Satarxiv,Marko2022}. Furthermore, based on these we formulated explicit lessons to prevent pitfalls and design guidelines, which are unique for this paper, along with detailed discussions on reliability enhancement solutions.

		\subsection{Motivations}
		
		Although there are some early papers on the security and reliability of LEO SCSs, in the light of recent advances, it is timely to critically appraise them.

		\subsubsection{Insufficient Research of Existing Papers}
		As seen in Table I, the existing papers have incomplete coverage of security and reliability issues. In particular, while some papers have recognized the impact of physical threats, such as space debris \cite{SatAir}, on reliability, the existing research on this topic is limited.
		
		Only with a deeper assessment and understanding of the existing security and reliability issues, such as the degree of damage, reversibility, awareness, collateral damage, etc., can we design potent solutions. Hence we are inspired to fill this knowledge gap in the open literature.

		\subsubsection{ Lack of Consideration for the Inherent Characteristics of LEO SCSs}
		
		Lots of existing papers discussed LEO satellites as an important component of the future 6G or SAGINs, but only a few of them focus on the analysis of the inherent characteristics of LEO SCSs, such as being sandwiched between GEO satellites and terrestrial communication systems, their high mobility, the large number of LEO satellites, limited onboard resources, and so on. These characteristics are the key to considering the security and reliability requirements, issues, and solutions. 
		
		Given the crowded orbits and the ongoing deployments of dense LEO mega-constellations, their orbits are becoming increasingly overcrowded, which undoubtedly increases the probability of collisions. Frequent launch activities also generate space debris in LEOs, which hence threatens the safe operation of LEO satellites. In addition, the space environment is harsh. Many satellites have failed before accomplishing their missions, partly because cosmic radiation may impair the electronic devices on the satellite.
		
		The Doppler shift caused by high mobility seriously deteriorates the performance \cite{DopplerLEORL}. Additionally, LEO satellites are sandwiched between MEO satellites and terrestrial communication systems. Hence the CCI due to spectrum sharing among these systems and its corresponding solutions have to be investigated in detail.

		\subsubsection{Lack of Design Guidelines of Secure and Reliable LEO SCSs}
		
		Existing papers show that protecting the secure and reliable operation of LEO SCSs does not rely on a specific solution alone but on the cooperation of multiple solutions. Moreover, effective security and reliability enhancement solutions may be derived by carefully characterizing the relationship between their confidentiality, integrity, and latency. For example, due to size, memory, and power constraints, LEO satellites either have no operating system at all or can only run stripped-down versions of a sophisticated operating system. They are unsuitable for complex encryption algorithms. By contrast, usually complex encryption algorithms are used for authentication in the ground segment, given the abundant power supply and computing resources. Hence, several trade-offs must be struck in the design of secure LEO SCSs.
		
		However, the existing literature does not integrate these advanced solutions together to safeguard LEO SCSs. Hence in this literature, design guidelines for secure and reliable LEO SCSs. Indeed, design guidelines are distilled from the characteristics of LEO SCSs, security and reliability requirements, issues, and solutions, outlining the lessons learned and the various trade-offs.

		\subsection{Contributions}
		Against this backdrop, the main contributions of this survey are summarized as follows:
		
		\begin{itemize}	
			\item[$\bullet$] We discuss the inherent characteristics of LEO SCSs and outline their unique security and reliability challenges. Based on this, we also summarize their security and reliability requirements (Sec. \uppercase\expandafter{\romannumeral3}, \uppercase\expandafter{\romannumeral4}).
			
			\item[$\bullet$] Relying on recent research results and the unique security and reliability challenges encountered by LEO SCSs, we review their security attacks and discuss several reliability risks, such as Single Event Upsets (SEUs) and collisions with debris. Furthermore, the characteristics and impacts of these issues are analyzed and summarized in Table \uppercase\expandafter{\romannumeral6} at a glance. We also summarize several lessons learned from these issues (Sec. \uppercase\expandafter{\romannumeral5}).
			
			\item[$\bullet$] As a remedy, we review a rich suite of solutions and classify them into security and reliability enhancement solutions. Moreover, we further divide the family of security enhancement solutions into active and passive security enhancement solutions from the perspective of prevention, detection, and mitigation. Moreover, we discuss several trade-offs to be observed by the solutions and summarize the lessons learned from these solutions (Sec. \uppercase\expandafter{\romannumeral6}).
			
			\item[$\bullet$] Our discussions concerning the lessons learned from the analysis of solutions and gleaned from the remaining technical challenges inspire several promising future research directions, such as the employment of integrated sensing and communication, Computer Vision (CV)-aided secure communications, as well as the unique challenges imposed by mega-constellations and commercialization (Sec. \uppercase\expandafter{\romannumeral7}).	
			
			\item[$\bullet$] Again, the analysis of the inherent characteristics, security and reliability requirements, and issues as well as solutions, allows us to outline the lessons learned, leading to our design guidelines for secure and reliable LEO SCSs (Sec. \uppercase\expandafter{\romannumeral8}).		
			
		\end{itemize}

	\subsection{Paper Organization}
\begin{figure}[ht]
		\centering
		
		\includegraphics[width=0.9\columnwidth]{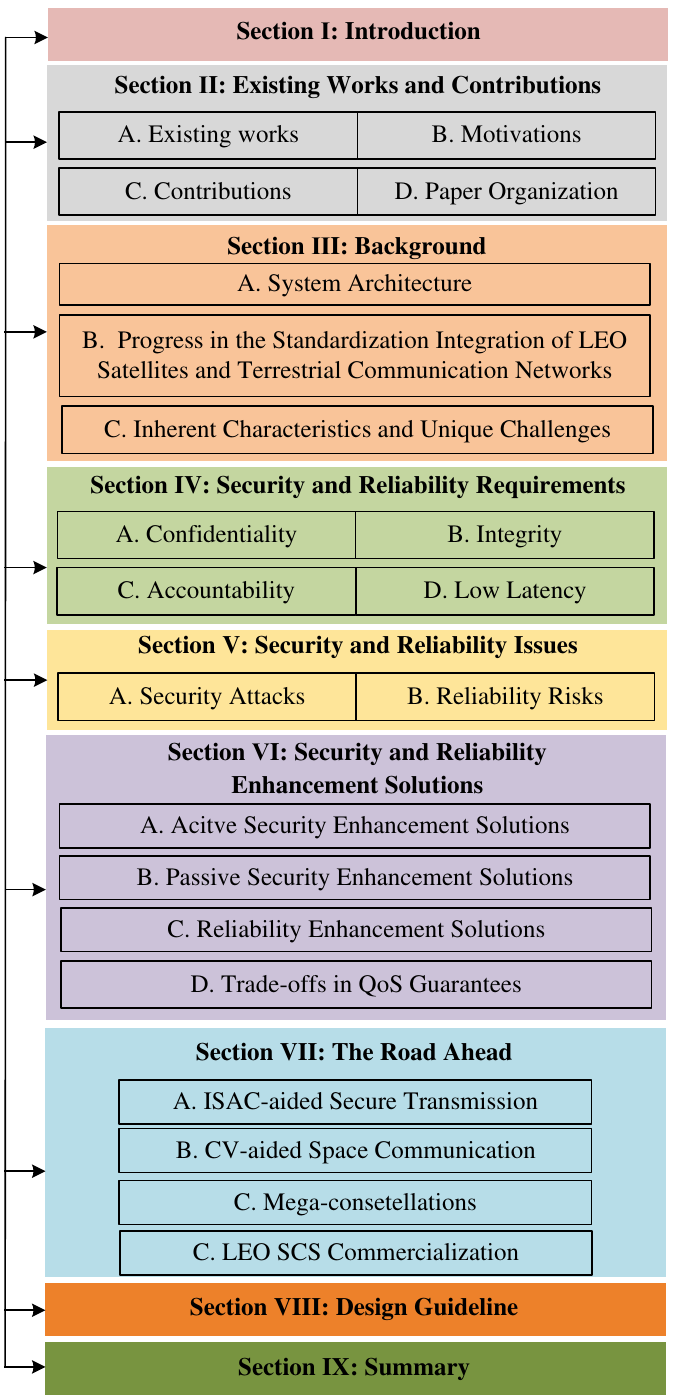}
		
		\caption{Organization of this paper.}
		\label{Organization}
	\end{figure}

	The organization of this paper is illustrated in Fig.~\ref{Organization}. Section \uppercase\expandafter{\romannumeral3} presents the background of LEO SCSs. Section \uppercase\expandafter{\romannumeral4} introduces security and reliability requirements in LEO SCSs. In Section \uppercase\expandafter{\romannumeral5}, the security and reliability issues encountered by LEO SCSs are categorized. Section \uppercase\expandafter{\romannumeral6} describes solutions for safeguarding LEO SCSs. In Section \uppercase\expandafter{\romannumeral7}, some open problems and research ideas concerning LEO SCSs are provided. Section \uppercase\expandafter{\romannumeral8} provides design guidelines for LEO SCSs. Finally, our concluding remarks for LEO SCSs are provided in Section \uppercase\expandafter{\romannumeral9}. The acronyms used in this paper can be found in Table \uppercase\expandafter{\romannumeral12} for convenience.
	
		\section{Background}
		
		In this section, we first outline the system architecture of LEO SCSs. Then we summarize the current developments concerning LEO constellations and the standard progress of Non-terrestrial Networks (NTNs) represented by LEO satellites. Finally, the inherent characteristics and unique challenges of LEO SCSs are introduced.	
		
		\subsection{System Architecture}
		
\begin{figure}[]
			\centering
			
			\includegraphics[width=1\columnwidth]{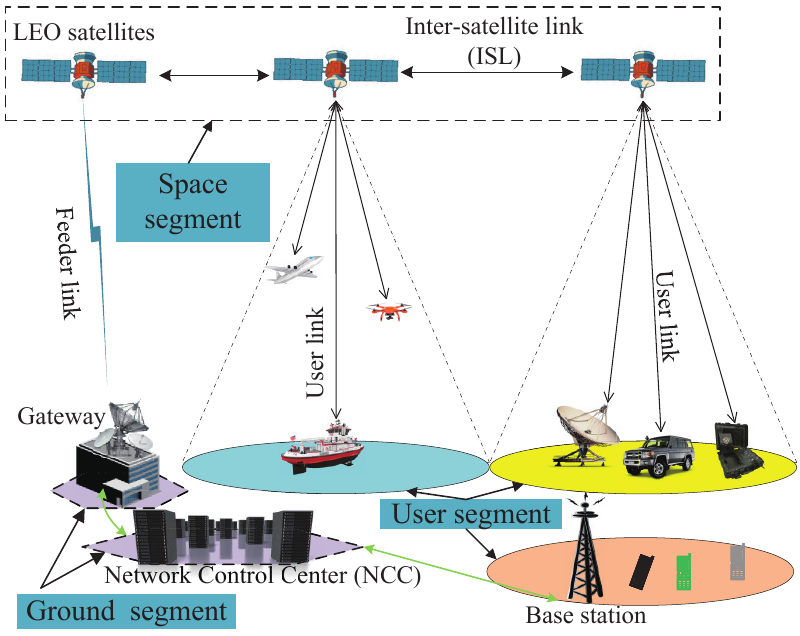}
			
			\caption{The system architecture of LEO SCSs.}
			\label{systemarchitecture}
		\end{figure}

		As shown in Fig.~\ref{systemarchitecture}, the system architecture of LEO SCSs is divided into the components of space segment, ground segment, and user segment. The space segment consists of LEO satellites and ISLs, where the LEO satellites are connected by ISLs. However, not all LEO SCSs have ISLs, a counter-example is OneWeb \cite{MegaCM}. The ground segment is composed of the gateway and Network Control Center (NCC). The gateway sets up the feeder links for tracking LEO satellites, while the NCC is the center of operation, management, and control for the entire LEO SCS. If there are no ISLs in the space segment, then we have to build enough gateways for ensuring that each LEO satellite is indeed visible. These gateways are connected by optical fiber to jointly ensure the reliable operation of all satellites. Additionally, the NCC is also responsible for the interaction of LEO SCSs with other systems, such as terrestrial mobile communication systems and Wireless Local Area Networks (WLAN). Finally, the user segment includes a large number of terminals. These terminals access LEO satellites via the user link.
		
		\subsection{Progress in the Standardization Integration of LEO Satellites and Terrestrial Communication Networks}
		
\begin{figure}[ht]
			\centering
			
			\includegraphics[width=1\columnwidth]{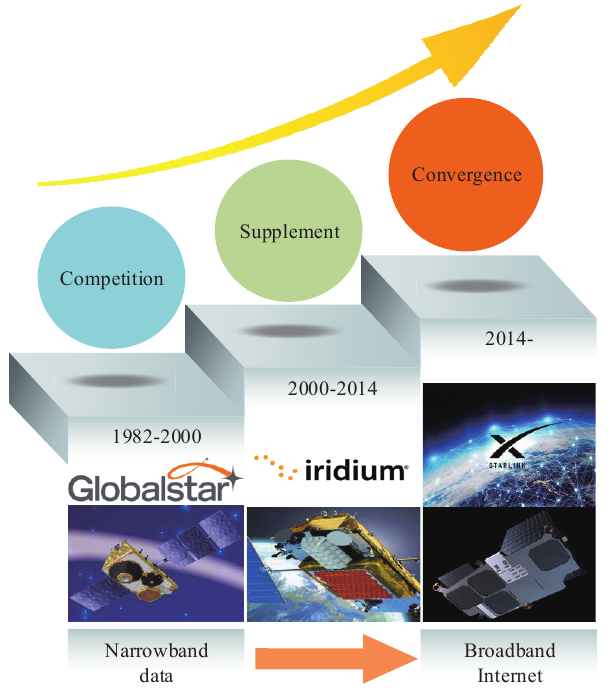}
			\caption{Development process of LEO SCSs.}
			\label{Development}
		\end{figure}

		LEO satellites were first launched over 50 years ago. The concept of LEO SCSs can be traced back to the 1990s, when the Iridium \cite{IridiumIntro}, Globalstar \cite{Globalstar1998}, and Orbcomm \cite{ORBCOMM} were designed to provide low-latency voice and data service. However, some of them ended up becoming bankrupt due to the high cost, immature technology, and modest communication capabilities \cite{LEObankruptcy}. But thanks to the development of advanced materials, sophisticated technology, and scale of economy, a new LEO SCS age has dawned. In recent years, as a benefit of the ever-increasing demands \cite{Oneweb}, reduced costs \cite{Onewebfactory}, and technological progress, LEO mega-constellations, such as OneWeb, Starlink, and Lightspeed, are making a renewed effort to provide services for `the other 3 Billion' who do not as yet have access to the Internet. At the time of writing, they tend to evolve towards a converged system, as shown in Fig.~\ref{Development}.
		
		Meanwhile, 5G services have been made available to consumers having smartphones, but there is also a significant demand among network operators to offer 5G services to massive Machine Type Communication (mMTC) devices, especially in remote areas \cite{MMTC2021}. As a result, both the research focus and industrial push are shifting towards NTNs represented by LEO satellites.
		
		The 3rd Generation Partnership Project (3GPP) is a primary international body responsible for defining technical specifications for mobile wireless networks. Enabling the 5G system to support NTNs requires a holistic and comprehensive design that spans numerous areas, primarily including Radio Access Networks (RAN) and Service Aspects (SA).
		
		\subsubsection{Radio Access Networks}
		
		The work on NTNs started in 2017 with a Study Item in Release-15 in 3GPP RAN \cite{T38811} that focused on deployment scenarios and channel models for NTNs.
		
		After completing this study, 3GPP followed up with the Release-16, which aimed for providing solutions for adapting 5G New Radio (NR) to support NTNs. The main objective of this study is to identify a set of features, including the architecture, higher-layer protocols, and physical layer components, that enable NTNs within the 5G system while minimizing the impact on the existing 5G system. The outcome of this study is summarized in \cite{T38821}.
		
		The Release-17 Work Item focused on addressing the issues left from the Release-16 study, such as the architecture, frequency synchronization, and Hybrid Automatic Repeat Request (HARQ). Specifically, two possible architectural options are discussed: transparent mode and onboard processing (regenerative) mode. Furthermore, UEs with GPS capabilities can employ their position and the NTNs’ ephemeris for predicting and compensating for the Doppler frequency shift \cite{T23737}.
		
		\subsubsection{Service Aspects}
		
		The 3GPP SA workgroup started to discuss the use cases of satellite-based NTNs as part of the Study Item on satellite access in 5G based on Release-15. This study identified three main categories of use cases and service requirements, including serving as an access point for User Equipment (UE) and a backhaul link, guaranteeing coverage for IoT devices, and supporting mission-critical access in disaster situations \cite{T22822}.
		
		Based on this, the 3GPP SA workgroup further identified the main network functions, how these functions were linked to each other, and the information they exchanged \cite{T237371}. 
		
		Subsequently, the scope of this work was to identify key issues and provide efficient solutions, such as mobility management, delay in satellite, and Quality of Service (QoS), which are captured in TR 28.808 \cite{T28808}.
		
		Additionally, the 3GPP workgroups approved the study on 5G core network architecture \cite{T28808} and the operation of IoT NTNs \cite{T24821}.
		
		The deployment of LEO mega-constellations and work at 3GPP provide a potential pathway for closer integration of terrestrial networks and NTNs. At the time of writing, many researchers and commercial companies are tirelessly striving for integrating these two types of networks.

		\subsection{Inherent Characteristics and Unique Challenges}
		
		Although the development of LEO SCSs is in full swing, they tend to suffer from numerous unprecedented challenges due to their inherent characteristics. Table II summarizes the characteristics of LEO, MEO, and GEO satellites. As seen in Table II, the inherent characteristics of LEO satellites are as follows, and the consequent challenges faced by LEO SCSs can also be clarified.
		
		\begin{table*}[ht]
			\begin{center}
				\renewcommand\arraystretch{1.3}
				\caption{Comparison of the main characteristics between GEO, MEO, and LEO satellites}
				
				\begin{tabular}{|m{2.4cm}<{\raggedright}|m{3.0cm}<{\raggedleft}|m{3.0cm}<{\raggedleft}|m{5cm}<{\raggedleft}|}
					\hline
					\hline
					Satellite feature          & GEO satellites     & MEO satellites           & LEO satellites            \\ \hline
					
					Orbital altitude                   & 35786 km          & 2000-20000 km           & 500-2000 km              \\ \hline
					
					Orbital period             & 24 hours          & 2 to 8 hours            & 10 to 50 minutes         \\ \hline
					
					Path loss           & High              & High                    & Least                    \\ \hline
					Propagation latency          & High              & High                    & Low                      \\ \hline
					Coverage                   & Largest           & Large                   & Small                    \\ \hline
					Satellite life             & 10-15 years       & 10-15 years             & From a few years up to 10-15 years   \\ \hline
					Satellite required         & At least 3        & At least 6              & Depends on the design                  \\ \hline	
					Deployment time            & \begin{tabular}[c]{@{}l@{}}Depending on the\\ deployment strategy\end{tabular} & \begin{tabular}[c]{@{}l@{}}Depending on the\\ deployment strategy\end{tabular} & \begin{tabular}[c]{@{}l@{}}Depending on the number of satellites\\ per launch and orbit parameters\end{tabular} \\ \hline
					
				\end{tabular}
			\end{center}
		\end{table*} 
		
		\textbf{The Specific Orbit of LEO Satellites Degrades Both the Security and Reliability:} 
		LEO satellites are sandwiched between terrestrial communication systems and MEO satellites, which are convenient for attackers. For example, compared to MEO and GEO satellites, ground attackers can achieve the same jamming effect at low jamming power, since LEO satellites are closer to the Earth. By contrast, MEO and GEO satellites can also act as jammers to launch malicious jamming to contaminate the downlink signals of LEO satellites. To make things worse, since the MEO or GEO satellites have a larger coverage area than LEO satellites, a single MEO or GEO satellite may affect many LEO satellites at the same time.
		
		The spectrum crunch imposed by the scarcity of radio resources results in inevitable spectrum sharing between SCSs and terrestrial communication systems. Moreover, the specific location of LEO satellites may also lead to CCI with both GEO SCSs and terrestrial communication systems. Typically, a large number of LEO satellites are sandwiched between the terrestrial communication systems and MEO SCSs. Severe CCI may arise whenever LEO satellites pass through the Line of Sight (LoS) path of a GEO satellite in spectrum sharing scenarios \cite{QianChen}.
		
		\textbf{The High Mobility of LEO Satellites Degrades Both the Security and Reliability:}
		Both the high mobility and limited coverage area of each LEO satellite result in limited time spent above the horizon, hence the ground segment of LEO SCSs should be responsible for the mobility management of terminals. Moreover, in order to prevent malicious intrusion, the ground segment is usually allowed to admit users following authentication \cite{AuthenBRL}. Therefore, both the mobility management and authentication of a massive number of terminals impose severe challenges on the ground segment. Additionally, if there are no ISLs in the space segment, many gateways have to be constructed to support the reliable operation of all satellites, which are prone to hacking attacks.
		
		On a different note, the dramatic increase in the number of LEO satellites will undoubtedly increase the probability of satellite collisions. Moreover, the proliferation of launch activities has caused a surge in LEO space debris, which fly at high speeds and impose severe challenges on the reliable operation of LEO satellites.
		
		\textbf{The Large Number of LEO Satellites or Gateways Degrades the Security:}
		LEO SCSs have to rely on a large number of gateways (transparent without ISLs, e.g., OneWeb \cite{ShengmingComst}) or multiple satellites (on-board processing with ISLs, e.g., Iridium \cite{Iridium}) to achieve global coverage. A large number of LEO satellites or gateways creates numerous opportunities, respectively.
		
		\textbf{The Limited Resources of LEO SCSs Degrades Both the Security and Reliability:}
		Both the LEO satellites and the terminals in LEO SCSs have limited power, storage capacity, and computation capability. On the one hand, in order to reduce both the satellite manufacturing and the launching cost, the weight of a typical Starlink satellite is only 227 kg, while the weight of a OneWeb satellite is less than 150 kg \cite{SatrlinkOneweb}. These satellites have to be equipped with small batteries as well as limited fuel and computing equipment.
		
		Compared to the terminals deployed in urban areas, terminals operating in remote areas lack stable power supplies. These terminals have to rely either on the battery carried or on solar panels \cite{PowerLimitedref1}. Hence, the limited computation and storage capacity of the LEO satellites makes them unsuitable for signal processing with high complexity.
		
		Additionally, the power-limited terminals result in low transmit power, which in turn leads to a low Signal-to-Noise Ratio (SNR) for the signal arriving from LEO satellites, posing serious challenges for reliable reception. Specifically, in order to improve the detection performance of weak signals under the condition of low SNR, the receiver has to increase the coherent integration time, indicating that more data needs to be stored and processed \cite{Acqnum}.
		
		\textbf{The Production of Low-cost Satellites Degrades Both the Security and Reliability:}
		The deployment of a large number of satellites in LEO mega-constellations has stimulated the transformation of their production and testing models. Hence, a large number of low-specification components used for LEO satellites are supplied by civilian manufacturers both for cost savings and for reducing the production cycle duration by relying on COTS components. For instance, OneWeb is known to be a pioneer in the mass production of satellites, whose satellite factory in Florida is expected to produce as many as two satellites per day \cite{Onewebfactory}. However, loopholes in production methods and inadequate testing may lead to potential defects in satellites.
		
		In order to support more diverse scenarios and iterative updates of functions, LEO satellites adopt a large number of Field Programmable Gate Arrays (FPGAs), which exhibit flexibility and programmability. However, these FPGAs are also susceptible to the impact of cosmic radiation \cite{SRAMFPGATMR}, which can affect the reliability of the programs and algorithms running onboard the satellites.
		
		\textbf{Quality of Service (QoS) Guarantee When Designing Security and Reliability Solutions:}
		A certain target QoS must be guaranteed by satellite communication service providers for the satellite links. Satellite providers can use these indicators to measure and optimize the performance of their systems, ensuring that users can consistently enjoy a high-quality service experience. 
		
		In contrast to terrestrial communication systems, there are different types of services due to the wide coverage of each satellite. With the emergence of compelling services based on LEO SCSs, as seen in Fig.~\ref{application}, they have to offer differentiated
		service capabilities. For example, smart mining only requires low-rate data transmission, but a large number of connections, while tele-medicine requires highly reliable, low-latency data transmission to support tele-consultations and remote patient monitoring. 
		
		Therefore, when designing solutions to improve security and reliability, it is also necessary to provide QoS guarantees for differentiated services.	
		
		\begin{figure*}[]
			\centering
			
			\includegraphics[width=1.5\columnwidth]{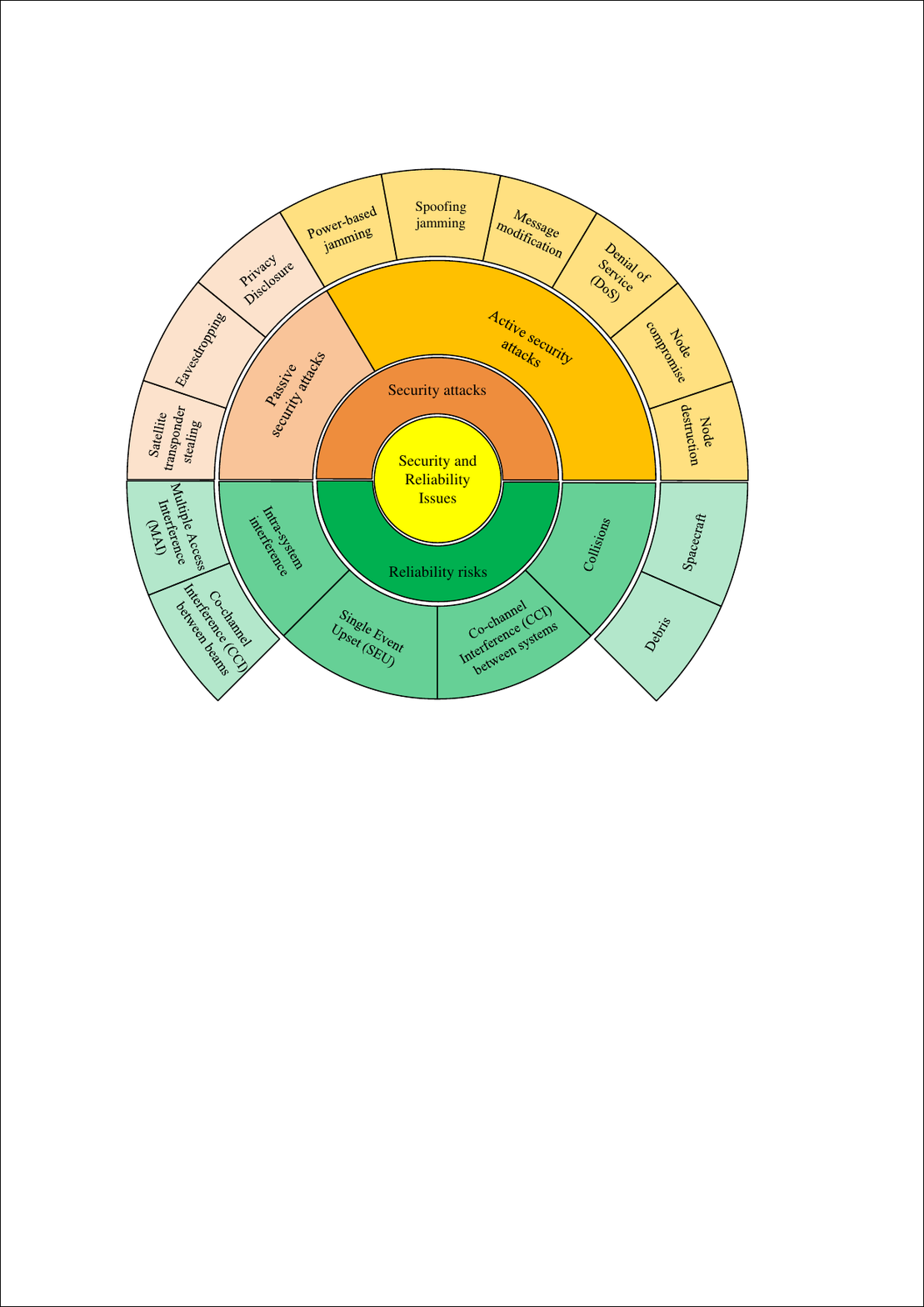}
			
			\caption{Classification of security and reliability issues.}
			\label{vulnerabilitiespic}
		\end{figure*}

		\section{Security and Reliability Requirements}
		
		Again, the LEO SCSs suffer from both security attacks and reliability risks. The security and reliability requirements of LEO SCSs are specified for the sake of preventing both these attacks and risks, exemplified by eavesdropping, jamming, SEUs, collisions, and so on. For example, maintaining the specific target integrity say in terms of BER constitutes a pivotal security requirement, which refers to reliable reception even in the face of malicious jamming~\cite{IntegrityRL}. Philosophically, secure and reliable LEO SCSs should satisfy confidentiality, integrity, availability, and accountability~\cite{Vulnerabilities1}, which will be discussed in deeper technical detail in this section.

		\subsection{Confidentiality}
		
		Confidentiality implies that the transmitted data or information is not disclosed to unauthorized users or groups. However, due to the broadcast nature of the wireless medium, it is vulnerable to eavesdropping, which may cause potential confidentiality violations \cite{CondiTAES2022}. In general, the PLS \cite{GaofengPanPLS} philosophy has been conceived for satellite-to-Earth links, which exploits the random physical characteristics of wireless channels to protect confidentiality. To make things worse, it is necessary to deploy numerous LEO satellites for achieving seamless global coverage. The large number of satellites provides convenience for attackers. Similarly, LEO SCSs without ISLs, such as Oneweb \cite{MegaCM}, should rely on a large number of gateways for seamless global coverage, which are also tempting for attackers. The above two statements impose more serious challenges to the confidentiality of LEO SCSs.
		
		\subsection{Integrity}
		
		Integrity characterizes the accuracy and completeness of confidential information, which must be safeguarded during its transmission. For example, a powerful adversary may jam the wireless user's link by contaminating it with high-power white noise across the entire frequency band. Compared to GEO satellites, LEO satellites are closer to the ground, which makes it easier to contaminate legitimate signals at a lower jamming power. As a solution, the Direct Sequence Spread Spectrum (DSSS) technique may be adopted to counteract it by exploiting its inherent jamming mitigation capability\cite{DSSSIterative}. 
		
		Furthermore, even in the absence of jamming, the Doppler frequency shift due to the high mobility of LEO satellites also affects the integrity of LEO SCSs.

		\subsection{Accountability}
		
		Each country or institution has the responsibility to use the space sustainably. As LEO mega-constellations are proliferating, the debris of transportation tools as well as operational or retired satellites make the low Earth orbit increasingly crowded, thus requiring debris removal measures, for reducing the above-mentioned deleterious impact on space. Additionally, considering the scarce spectral resources, LEO SCSs and GEO SCSs have to rely on spectrum sharing according to the International Telecommunication Union (ITU) regulations \cite{ITU1}. In this context, LEO SCSs shall not impose excessive interference on GEO SCSs.
		
		\subsection{Low Latency}
		
		Having low latency and high security is also a desirable feature of 6G networks. The end-to-end latency is given by the sum of the propagation delay, the processing latency, and the queuing delay. An LEO satellite at 600 km orbit altitude has, for example, a 4 Milliseconds (ms) uplink/downlink turn-around propagation delay at the speed of light, which is perceptually unobjectionable for voice calls. However, the G.729 speech code would add 10 ms \cite{G729RL} processing delay at both the encoder and decoder, which may escalate further owing to the channel coding and queuing delays. Hence it is imperative to conceive low-latency security solutions for supporting secure delay-sensitive services in tele-medicine and emergency rescue.

		\textbf{Remarks:}
		The QoS terminology includes numerous metrics, such as latency, traffic rate, Bit Error Rate (BER), voice and video quality, etc. 
		
		Although all the above-mentioned security and reliability requirements are of paramount importance, it is challenging to simultaneously satisfy all of them. Indeed, typically trade-offs must be struck. For example, to improve the confidentiality in the uplink without any powerful secrecy coding, the conventional approach is to reduce the terminal's transmit power. However, the system's integrity will also be reduced simultaneously. Conversely, by increasing the transmit power to improve the desired link's integrity, the probability of eavesdropping will also be inevitably increased \cite{YulongTradeoff}. Suffice it to say that further research is required for the multi-component Pareto-optimization of the system to determine all optimal operating points of the non-dominated set of solutions \cite{Jingjingwang}. None of the metrics in these optimal solutions can be improved without degrading at least one or several of the others.

	\section{Security and Reliability Issues}

	LEO SCSs support more and more civilian and military applications, thus it is of paramount importance to eliminate their issues. In this section, we focus our discussions on the issues of LEO SCSs, which are shown in Fig.~\ref{vulnerabilitiespic} at a glance. In addition to the security attacks by potential adversaries that the existing magazine and survey papers focus on \cite{Satarxiv, COMST6G20214, Confidentiality, SecurityIoT, Vulnerabilities1, Vulnerabilities2}, there is a whole host of other issues, which are not due to the presence of attackers as exemplified by SEUs, collisions, and so on. To this end, the issues of LEO SCSs can be classified into security attacks and reliability risks. These two categories also have their respective subclasses. Furthermore, based on this classification, we further analyze and infer the characteristics of these issues, such as attribution, reversibility, awareness, and collateral damage. We will continue by highlighting several lessons learned from these issues and use them as a springboard for conceiving potent security and reliability enhancements.

	\subsection{Security Attacks}
	
	LEO SCSs provide a powerful platform for military applications, which are hence prime targets for hostile attacks. Their ground segments are responsible for all interactions with other terrestrial communication systems, and these facilities create opportunities for attackers.
	
	Considering the activity of the attack, security attacks can be further classified into passive and active security attacks, and both of them are detailed below.
	
	\subsubsection{Passive Security Attacks}
	
	The most crucial thing in passive security attacks is that the victim does not get informed about the attack. Passive security attacks may cause the loss of confidentiality. Two types of passive security attacks are eavesdropping and satellite transponder stealing.
	
	\textbf{Eavesdropping:} The open nature of wireless propagation makes legitimate transmissions vulnerable to the interception and interpretation of signal or message. Eavesdropping attacks do not require high technical capabilities, only individuals or commercial competitors can deploy a number of drones to obtain an opportunity to overhear the user link due to frequent access caused by the high mobility of LEO satellites. Furthermore, the large number of LEO satellites also provide convenience for eavesdroppers. Additionally, eavesdroppers will analyze and extract useful information to create future attacks. DSSS and PLS techniques are separately used for mitigating the eavesdropping \cite{DSSSPLS20221, DSSSPLS20222, DSSSPLS20223}.
	
	Furthermore, perturbing the normal behavior or stealing secret information may also occur during the design and during runtime due to hardware issues. To proceed one step further, given the explosive proliferation of LEO satellites, many manufacturers would prefer using COTS components to increase their production rate at a reduced cost. However, some COTS components may also open the door for attackers. For example, the authors of \cite{RISCV} discussed the security threats that arise from the adoption of the well-known Reduced Instruction Set Computer V microprocessor operating on board of satellites. They demonstrated how hardware trojan horses and microarchitectural side-channel attacks might compromise the overall system's operation by stealing confidential information.
	
\begin{figure}[ht]
		\centering
		
		\includegraphics[width=1\columnwidth]{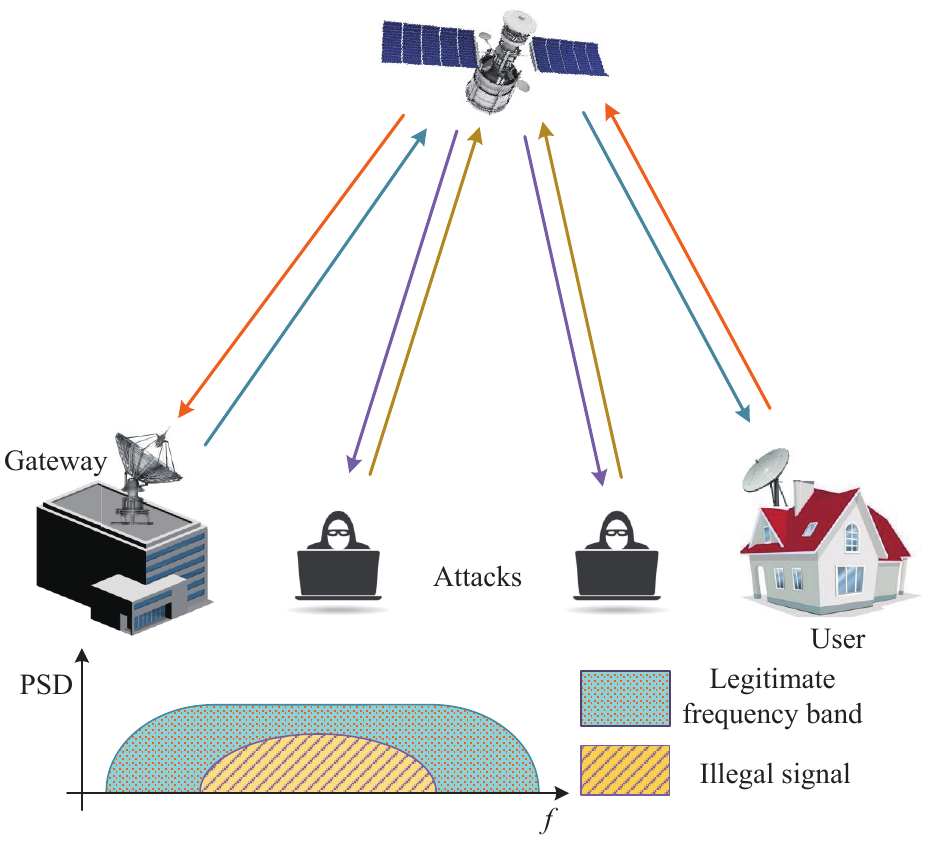}
		
		\caption{Illustration of transponder stealing.}
		\label{Stealing}
	\end{figure}

	\textbf{Satellite Transponder Stealing:} Satellites with the merits of wide coverage and free from natural disasters have attracted widespread attention. However, their high cost and advanced technology make it possible for only a few countries or institutions in the world to produce. In this case, Some criminals without satellite production capabilities exploit existing satellites to quietly complete their own transmissions. Existing satellites mainly include on-board processing, and transparent forwarding \cite{Broadband}. Between them, transparent forwarding satellites are more likely to be exploited by criminals seen in Fig.~\ref{Stealing}, because they do not perform any signal processing \cite{terrestrialNOMA1}. Hence, it is not possible to determine whether the received data is from a legitimate user. When attackers send their illegal signals, the satellite will still forward the signals \cite{pavur2019secrets}.
	
	In order to use the satellite transponder secretly, attackers need to conduct some research to obtain the specific parameters of the satellite transponder, such as operating frequency, satellite orbit information, etc. In addition, the DSSS techniques at a low Power Spectral Density (PSD) are adopted to bury themselves under the legitimate frequency band, as shown in Fig.~\ref{Stealing}. Regular replacement of satellite operating parameters, such as operating frequency, may prevent this type of attack.
	
	\textbf{Privacy Disclosure:} As mentioned in \cite{DataSharing,LEOOFDMSIoT,Multidomain2023}, compared to GEO satellites, LEO satellites have the merits of low path loss and propagation delay. Hence they are capable of supporting access to the IoRT in upcoming 6G communication systems, as shown in Fig.~\ref{application}(a). However, the sharing and process of LEO satellite-based IoRT data may lead to the invasion of user privacy. Mass data of the registered User Equipments (UEs) has to rely on a large number of gateways (transparent without ISLs, e.g., OneWeb \cite{ShengmingComst}) or multiple satellites (on-board processing with ISLs, e.g., Iridium \cite{Iridium}) to exchange worldwide. However, some of the data may contain sensitive information. For example, to reduce the delay of handover \cite{HandoverComst} among LEO satellites for achieving high-precision continuous user tracking, the location of UEs has to be shared among gateways or satellites \cite{SAGINRL}. Usually, the location of each UE can only be accessed by authorized individuals (e.g., designated service providers or insurance agents). Once this data is shared, it is beyond the owner's control and may be illegally accessed by unauthorized parties, leading to a high risk of data misuse. Moreover, once an attacker obtains access to user location information, this may trigger subsequently irreversible attacks, even potential physical destruction.

	\subsubsection{Active Security Attacks}
	
	For active security attacks, malicious acts are performed to disrupt or even damage the system operation. Hence the victim gets informed about these attacks. Active security attacks are dangerous to integrity as well as availability. The most common forms of active security attacks contain power-based jamming, spoofing jamming, message modification, DoS, node compromise, and node destruction. Detection-based methods are adopted for minimizing the impacts of issues and speeding remediation \cite{AttackDetection1, AttackDetection2, AttackDetection3}.
	
	\textbf{Power-based Jamming:} A simple strategy to disrupt the legitimate signal reception by releasing the power-based jamming upon wireless user link of LEO SCSs \cite{JammingCOMST}. Most of the on-orbit satellites adopt the so-called bent-pipe\footnote{Many satellites send back to Earth what goes into the satellite with only amplification and a shift from uplink to downlink frequency, like a bent pipe. A bent-pipe satellite does not demodulate and decode the signal.} transponder without digital signal processing, so it is easy to encounter signal power-based jamming attacks. Attackers may easily perturb the satellite's operation by transmitting high-power jamming signals\cite{Caojamming}. If the jamming power is too high, it may at worst `fry' the receiver front end of the satellite. There are many types of jamming signals and classification methods. Zou {\itshape et al}. \cite{Yulong} classified jamming based on the grade of difficulty generating them and compared the different types of jamming schemes in terms of their energy efficiency, how disruptive their interference is, their complexity, and the prior knowledge.
	
\begin{figure*}[ht]
		\centering
		
		\includegraphics[width=1.5\columnwidth]{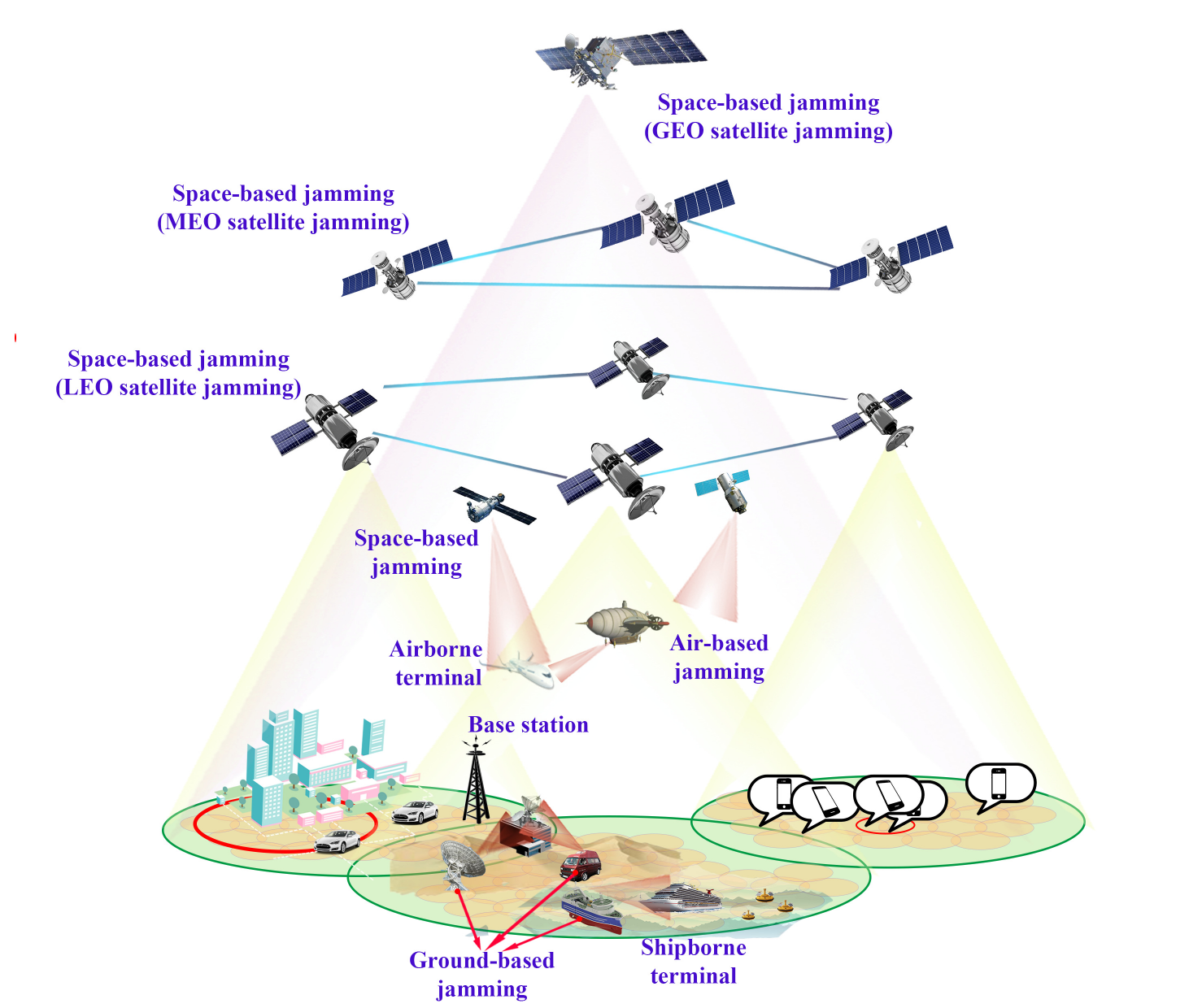}
		
		\caption{Sources of power-based and spoofing jamming contaminating LEO SCSs.}
		\label{interference}
	\end{figure*}

	Due to the long open wireless link between LEO satellites and the Earth, the adversary may contaminate it by jamming at different locations, which can be divided into the types illustrated in Fig.~\ref{interference}.
	The space-based jamming is mainly released by spacecraft. This type of jamming has an extensive range over which it may disrupt the downlink transmission, but it has limited jamming time and power owing to having limited time above the horizon.
	
	The adversary may generate air-based electronic jamming from aircraft or airships. As electronic-jamming aircraft and airships have more flexibility than their space-based counterparts, they are suitable for releasing burst-type jamming. Compared with space-based jamming, the power of air-based jamming is typically higher. Because airships are generally located between ground users and LEO satellites, they can interfere with the desired communication during both uplink and downlink transmissions.
	
	The power of ground-based jamming is typically high, and the jamming style is diverse because ground-based jamming is maliciously released by large-scale fixed, vehicle-mounted, or ship-borne jamming stations having abundant resources and power. Ground-based jamming mainly affects the uplink transmissions. There are many types of ground-based jamming, but distance is not a dominant factor. Ground-based jamming is usually of a blocking nature, which directly blocks the satellite transponder. These three types of power-based jamming are compared in Table \uppercase\expandafter{\romannumeral3}. As a remedy, both temporal domains adaptive filtering \cite{LMS2} and transform domain adaptive filtering \cite{TransformSNR} are efficient methods for jamming mitigation.
	
	\begin{table}[]
		\renewcommand\arraystretch{1.5}
		\begin{center}
			\caption{Comparison of power-based jamming}
			\begin{tabular}{|m{1.6cm}<{\raggedright}|m{1.8cm}<{\raggedleft}|m{1.8cm}<{\raggedleft}|m{1.8cm}<{\raggedleft}|}
				\hline
				\hline
				Jamming types & Space-based jamming & Air-based jamming & Ground-based jamming	\\ \hline
				
				Jamming power & Low & Medium & High	\\ \hline	
				
				Jamming time & Burst & Burst & Continuous 	\\ \hline	
				
				Resources & Limited & Limited & Rich 	\\ \hline	
				
				Mobility & Poor & Strong & Poor 	\\ \hline	
				
				Sphere of action & Large & Medium & Small	\\ \hline	
				
				Scenarios & Downlink & Downlink Uplink & Uplink	\\ \hline										
			\end{tabular}	
		\end{center}
	\end{table}	
	
	\textbf{Spoofing Jamming:} Spoofing jamming is a form of more insidious electronic attack where the attacker tricks a receiver into believing in the genuine nature of a malicious signal produced by the attacker. Compared with power-based jamming, spoofing jamming is more technical. The attacker must fully understand the signal characteristics, including physical layer waveform, frame structure, etc., to forge its equivalent and confuse legitimate receivers. For example, spoofing jamming often occurs in the civilian GPS. It is easy for the adversary to release spoof GPS signals to provide false information because the format of the civilian GPS signal is known \cite{SpoofingUAV}. Similar to GPS, there is usually a dedicated downlink pilot channel for broadcasting channel status, user management information, call information, etc., as exemplified by Iridium \cite{Iridium}. Attackers can imitate the dedicated pilot channel to broadcast false information to legitimate users, causing network paralysis. Fortunately, there are some standard methods to alleviate spoofing jamming, such as energy detection, multiple antennas \cite{Spoofing2022}, and authentication. However, energy detection and multiple antennas increase the terminal complexity. Hence, the most effective approach is to apply authentication for LEO SCSs. The authors of \cite{UAVasist} proposed an Unmanned Aerial Vehicle (UAV)-assisted authentication method to tackle spoofing jamming.

	\textbf{Message Modification:} Message modification means that a hacker intercepts messages and changes their contents, which contains message change, message insertion, and message deletion. Message modification is more likely to occur in the ground segment, where the hacker illegally obtains the data operation permission and modifies the message. Subsequently, these modified messages may result in some wrong decisions. To combat the message modification attack, existing SCSs typically consider the employment of Intrusion Detection Systems (IDS) and encryption algorithms \cite{IDS1}.
	
	\textbf{Denial of Service:} A DoS attack is that a hacker means to shut down a device or network, making it inaccessible to its intended users. A DoS attack tends to occur in the ground segment and the space segment of LEO SCSs. There are many methods for carrying out DoS attacks. The most common method of attack occurs when a hacker floods a network server with traffic. In this type of DoS attack, the hacker sends requests to the target satellite all the time. The target satellite is busy responding to these illegal requests, resulting in authorized users being ignored. On the other hand, the ground segment is responsible for the authentication of legitimate users. The hacker may forge a legitimate user in the ground segment to request authentication. As the junk requests are processed constantly, the ground segment is overwhelmed.
	
	Attackers also exploit issues or device weaknesses to orchestrate a synchronized DoS attack to a single target, which is co-called Distributed DoS (DDoS). The IoT botnet in which malware source code was leaked in early 2015 is a typical paradigm of DDoS attacks \cite{IoTbonet} in terrestrial IoT networks. IoT botnets, created as hackers, infect numerous IoT devices and recruit them to launch large-scale DDoS attacks. Furthermore, with the continued proliferation of LEO  mega-constellations, the large number of LEO satellites also become the potential target of DDoS attacks \cite{IoSatDDoS}. These attacks are difficult to detect and mitigate because they use hit-and-run tactics that originate from numerous IoT vectors distributed around the world \cite{DDOS1}.
	
	\textbf{Node Compromise:} A legitimate node may be attacked by an attacker under the control of malicious algorithms, programs, or software, potentially threatening the entire network \cite{satellitecomponet}. These compromised nodes may lead to some collateral damage. For example, these compromised nodes can deliberately leak confidential information to attackers. These compromised nodes may also trick other legitimate nodes into compromised nodes \cite{Sensorsecurity}. Moreover, an attacker may exploit a compromised node and pretend to be a legitimate user and device again to usurp system resources \cite{Masquerade}. It is challenging to detect compromised nodes because the behaviors between these compromised nodes and legitimate nodes are hard to distinguish. Using code patches is a common method of mitigating the probabilities of these events \cite{YulongzouPLS}.
	
	Moreover, with limited memory and processing capacity, many satellites do not even have complicated encryption algorithms to protect themselves. To this end, some hackers may hijack a satellite as a compromised node by taking over its feeder link. For example, A group of hackers once controlled a satellite by its feeder link and further tried to change its orbit. Hackers also used the hijacked satellite for extortion \cite{SkyNet}. Even worse, hackers could control satellites to achieve self-destruction by malicious commands, or they can use special tools to trick satellites and ultimately use them to attack other satellites or space assets.
	
	\textbf{Node Destruction:} Both the space segments as well as ground segments and terminals are subject to the risk of being destroyed. In the satellite-IoT applications supported by LEO SCSs, the power-limited terminals, such as oceanic buoys, operating without advanced security protection algorithms may become captured by an adversary \cite{IoTsecurity1}. Additionally, LEO satellites are potential targets for anti-satellite weapons, such as missiles and high-power laser beams.
	
	{\it Lessons Learned:} 
		Table \uppercase\expandafter{\romannumeral4} summarizes the important differences between active and passive security attacks. As mentioned in \cite{PLSSIN}, passive security attacks, such as eavesdropping, typically aim for stealing confidential information, such as passwords and messages. By contrast, active security attacks \cite{Yulong, YulongzouPLS}, including message modification, DoS, node compromise, and so on, may be carried out based on the results of passive security attacks. Attackers often exploit the confidential information stolen during passive eavesdropping attacks for performing active security attacks. Moreover, active security attacks may cause severe collateral damage when hostile nodes pretend to be legitimate ones and occupy valuable resources \cite{Masquerade}.

	\begin{table*}[ht]
		\centering
		\renewcommand\arraystretch{1.3}
		%		\begin{center}
			\caption{The differences between the active and passive security attacks}	
			\begin{tabular}{|m{3cm}<{\raggedright}|m{6cm}<{\raggedleft}|m{6cm}<{\raggedleft}|}
				\hline
				\hline
				Characteristics & Passive security attacks & Active security attacks \\ \hline
				Awareness & Not be aware & Aware \\ \hline
				Against on & Confidentiality & Integrity as well as availability \\ \hline
				Impact on system & There is no any harm to system & System is damaged, its degree of damage depends on the type of active attacks  \\ \hline
				Countermeasure & Prevention and mitigation &  Detection and mitigation\\ \hline		
				Technical capacity & Simple to implement & Requires sophisticated technical capacities \\ \hline		
				Degree of difficulty to deal with & Easy to mitigate compared with active attacks & Tough to restrict \\ \hline		
			\end{tabular}
			%		\end{center}
	\end{table*}
	
	\subsection{Reliability Risks}
	
	Apart from security attacks, the harsh working environment, crowded orbits, and spectrum crunch result in reliability risks, which may threaten the normal operation of LEO SCSs. These threats include intra-system interference, CCI between systems, SEUs, and collisions, which will be described in detail.
	
	\subsubsection{Intra-system Interference} Intra-system interference contains MAI \cite{MAIInterferencecite1}, and CCI between beams \cite{InterferenceComMag}, which are separately caused by physical waveform selection and the scarce spectrum.
	
	\textbf{MAI:} Spread Spectrum (SS) techniques are eminently suitable for LEO SCSs in military applications, which are immune to most types of interference to a certain extent. However, it is difficult to avoid the near-far effect caused by MAI. Power control and multi-user detection are common methods of mitigating these near-far effects \cite{Globalstar1998}. Additionally, the careful choice of SS codes may mitigate the near-far effects. Orthogonal complementary codes have been chosen to substantially mitigate MAI \cite{OCC1, OCC2, OCC3}. However, these orthogonal codes are sensitive to frequency shifts, which must be mitigated by future research.
	
\begin{figure}[ht]
		\centering
		
		\includegraphics[width=0.9\columnwidth]{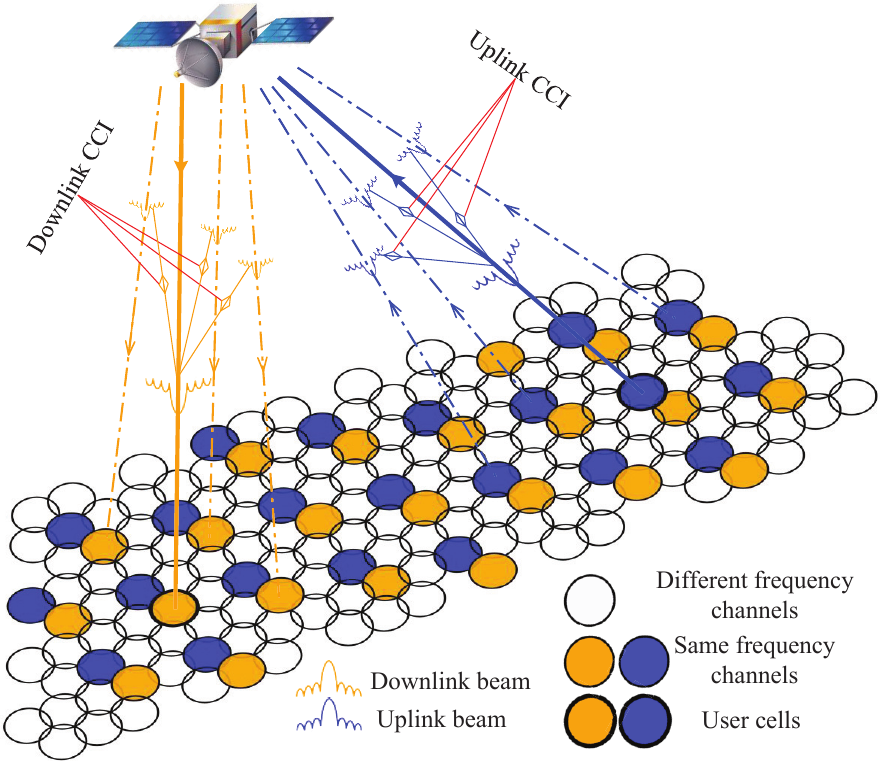}
		
		\caption{Depiction of the satellite uplink and downlink CCI.}
		\label{cochannelinterference}
	\end{figure}

	\textbf{CCI between Beams:} Multi-beam satellites reuse the available frequencies within their coverage to increase capacity. However, frequency reuse among beams may cause CCI in the overlapping areas when some beams rely on the same frequency \cite{InterferenceComMag}, especially in adjacent beams using the same frequency. The angular side-lobes of the beam radiation patterns create interference leakage, seen in Fig.~\ref{cochannelinterference}. The interference level is typically quantified in terms of the Carrier to Interference Ratio (CIR) \cite{CCIBeam}. Clearly, the interference limits the attainable capacity. To improve the capacity, Transmit Precoding (TPC) techniques relying on transmitter-side channel state information can be applied to mitigate the interference. A potent scheme based on hybrid wide-spot beams was designed to alleviate this source of interference in~\cite{Broadband}. The main philosophy of this scheme is that the space-borne payload generates several fixed wide beams for providing wide-range coverage to increase the frequency reuse distance. On this basis, the space-borne payload also adopts some high-gain spot beams for enhancing the capacity in tele-traffic hot spots.
	
	{\it Lessons Learned:}
		Compared to terrestrial communication systems, LEO satellites have a wider coverage area. Some public areas have higher access requirements, such as airports or railway stations. By contrast, some regions, such as deserts, oceans, etc., have a low access demand. Hence it is suboptimal to have fixed frequency reuse within each satellite. For example, although the frequency reuse pattern may be appropriately adjusted to satisfy the high access demands in hotspot areas, the associated CCI between beams also affects the regions having low access requirements. By contrast, the frequency pattern may also be adjusted for reducing CCI between beams, but then it cannot satisfy the access requirements of hotspot areas. To make things worse, LEO satellites fly over many regions, hence the resultant time-varying and unbalanced services pose a more serious challenge for the control of frequency reuse.
	
	%与地面移动通信系统相比，低轨卫星的覆盖范围较大，每颗星覆盖区内存在业务不平衡特性，即有些地区需要更高接入需求较高比如体育馆，车站等；而有些地区接入需求较低比如沙漠，海洋等。这导致每颗星内以固定的频率复用并不合适，举个例子，虽然提高频率复用满足热点地区较高的接入需求，但是随之而来的CCI也影响着对接入较低需求地区。相反，降低频率复用有利于降低CCI但是无法满足热点地区的接入需求。更严重的是，LEO satellites fly over many regions. 时变的且不平衡的业务负载对频率复用的确定提出了更加严峻的挑战。
	
	\subsubsection{CCI between Systems}
	
\begin{figure}[ht]
		\centering
		
		\includegraphics[width=0.8\columnwidth]{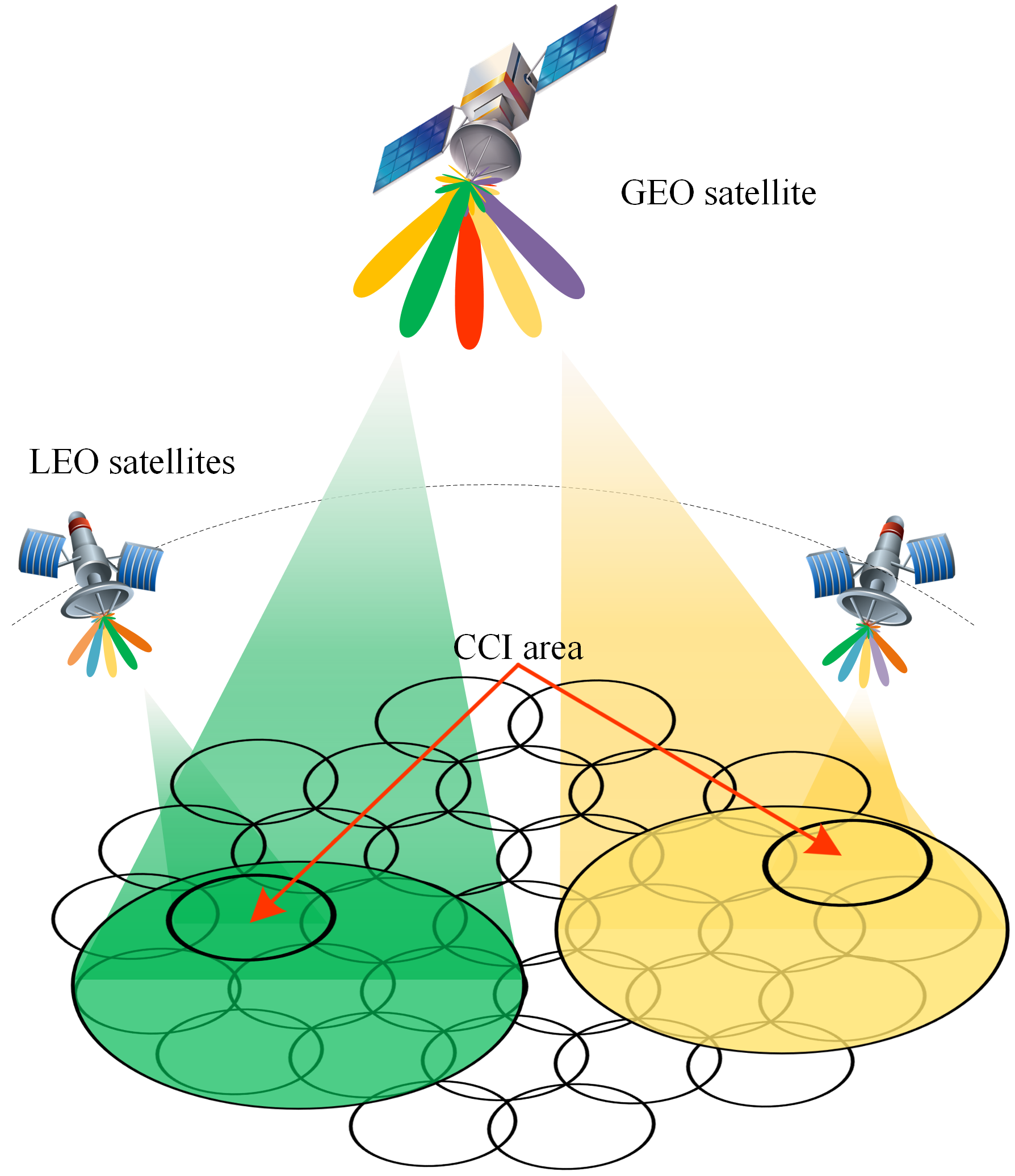}
		
		\caption{CCI between LEO and GEO SCSs.}
		\label{Onewebpitch}
	\end{figure}

	CCI between systems is essentially spectrum sharing between LEO SCSs and other systems, such as GEO SCSs and terrestrial mobile communication systems. An increasing number of LEO satellites has been deployed over the last few years, but the available radio spectrum remains limited. So LEO SCSs require high spectral efficiency to address the spectrum scarcity problem. Furthermore, GEO SCSs have to coexist within the same spectrum to achieve this objective. Consequently, the high-level CCI between LEO and GEO SCSs is unavoidable. When LEO satellites \cite{NoNGEOYe} approach the equator, they tend to inflict increased interference upon GEO satellites operating within the same frequency band, as shown in Fig.~\ref{Onewebpitch}. According to current ITU regulations \cite{ITU1}, it is mandated to consider the spectrum sharing between GEO and LEO SCSs. LEO SCSs shall not impose unacceptable interference on GEO SCSs. In other words, GEO SCSs are regarded as the Primary User (PU), while LEO SCSs are regarded as the Secondary User (SU). Thus interference coordination is imperative for mitigating the interference.
	
	{\it Lessons Learned:} 		
		The impact of CCI encountered by LEO SCSs is set to become more serious in the future, since the next-generation networks will provide ubiquitous connectivity through the convergence of terrestrial systems, LEO SCSs, and GEO SCSs \cite{Shanzhichen6G}. However, the coexistence of LEO SCSs and GEO SCSs has to be carefully planned. Adding terrestrial systems to the mix makes an already complicated picture more complex. 
		
		Since the GEO SCSs \cite{PowercontrolMinJia} and the terrestrial mobile communication systems \cite{powercontrol2016} have higher priority access to the existing spectrum, LEO SCSs having lower priority have to do their best to mitigate the CCI. The evolution of 6G systems stimulates the explosive proliferation of LEO satellites \cite{MassiveaccessLEO}, which undoubtedly increases the probability of CCI between LEO SCSs. Therefore, it will continue to attract wide attention.	

	\subsubsection{Single Event Upsets}
	
	The particles existing in cosmic radiation generate a large number of electrons and holes in the incident path by ionization. Electronic devices like FPGAs collect these charges, which may cause transient faults. If the charge exceeds the maximum level that the device can withstand without SEUs, the logic state of the circuit will be inverted. However, the circuit can be restored to its original working state by rewriting or resetting. Hence, SEUs constitute reversible soft errors \cite{SEUindicate}.
	
	The probability of SEUs is related to the orbit altitude and orbit inclination\footnote{Orbital inclination measures the tilt of an object's orbit around a celestial body. It is expressed as the angle between a reference plane, and the orbit plane or axis of direction of the orbiting object \cite{Inclination}.}. The authors of \cite{SEUFuqiang} investigated their effects on SEUs, and the results showed that at altitudes below 2000 km, the higher the orbit altitude, the higher the probability of SEUs occurrence. On the other hand, the closer the orbit inclination is to 90$^{\circ}$, the higher the probability of SEUs occurrence.
	
	The nature of SEUs is hardware-dependent. Compared to FPGAs, Application Specific Integrated Circuits (ASICs) exhibit better resistance to SEUs, but they lack flexibility. Therefore, FPGAs are widely used in LEO satellites as a benefit of their high performance and flexibility. To ensure the reliable operation of FPGAs in-orbit, it is necessary to employ SEUs prevention measures, such as Triple Module Redundancy (TMR) technique and periodical refreshing.
	
	\subsubsection{Collisions}
\begin{figure}[ht]
		\centering
		
		\includegraphics[width=1.05\columnwidth]{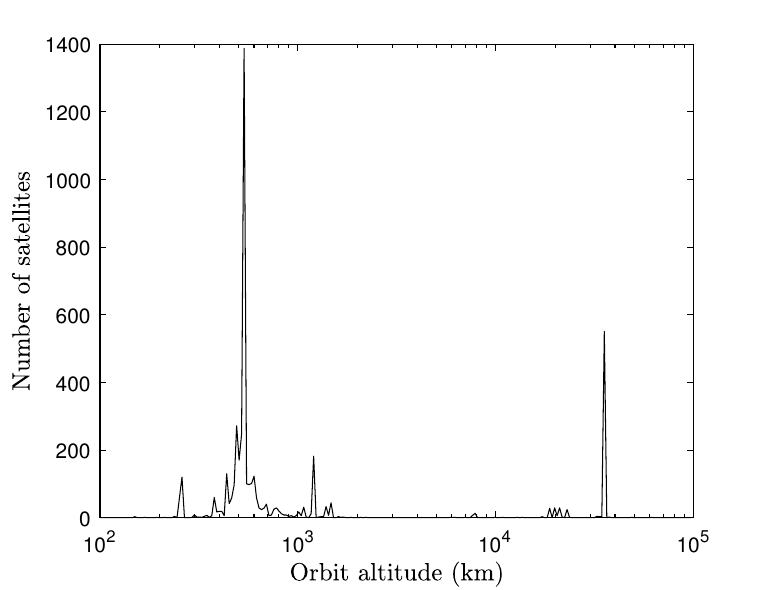}
		
		\caption{Launch of satellites in different orbits during Jan. 2012 and Dec. 2022\cite{UCS}.}
		\label{Debrispic}
	\end{figure}

	In recent years, the launch activities have been increasing for LEO, MEO, and GEO satellites. The different orbit regions are unevenly populated. It is seen from Fig.~\ref{Debrispic} that the LEO orbits between 800 and 1400 km constitute the most crowded space fuelled by the miniaturization of satellites and the deployment of mega-constellations. Crowded space in LEOs increases the risk of collisions, threatening the regular operation of satellites or spacecraft. Even worse, LEO satellites or spacecraft move around the planet at about 7 km/s, and their relative speed may be 10 km/s or higher. At this speed, even a tiny piece of debris presents a serious hazard for satellites or spacecraft. Hence, it is clear that LEO has to be treated with a special interest. Collisions occur not only between spacecraft but also between spacecraft and space debris. Table \uppercase\expandafter{\romannumeral5} summarizes the publicly reported collision avoidance and collision accidents in LEOs in the past 20 years.
	
	\begin{table*}[]
		\centering
		\renewcommand\arraystretch{1.3}
		\begin{center}
			\caption{Publicly reported collision avoidances and collision accidents in LEOs}
			\begin{tabular}{m{0.5cm}<{\centering}|m{14cm}<{\raggedright}}
				\hline
				\hline
				2007& Orbital debris completely penetrated one of the radiator panels of the shuttle Endeavour \cite{Endeavour}.\\
				2009& The active Iridium 33 and the derelict Russian military Kosmos 2251 collided above Siberia \cite{collision}. \\
				2013& Russian Satellite called BLITS crashed with the derelict Chinese Fengyun 1C satellite \cite{BLITS}.\\
				2013& Ecuador's NEE-01 Pegaso collided with Argentina's CubeBug-1 \cite{Ecuador}.\\
				2013& A tiny Ecuadoran satellite collided in space with the remains of a Soviet rocket \cite{Ecuador1}.\\
				2015& A millimeter-sized debris hit a solar panel on the ESA Sentinel-1A satellite \cite{Sentinel1A}.\\
				2016& A piece of space debris chipped one of the International Space Station's huge windows \cite{windows}.\\
				2019& Aeolus satellite belong to ESA performed a maneuver to avoid a potential approach to the Starlink 44 \cite{Aeolusstarlink}.\\
				2021& Yunhai-1 02 collided with the debris from the Zenit-2 rocket body launched Tselina-2 in 1996 \cite{Ecuador1}.\\
				2021& China Space Station has successfully conducted two evasive maneuvers to avoid potential collisions with Starlink separately in July and October \cite{ChinaSpace}.\\
				2021& The Arirang-1 satellite raised its orbit to avoid collision with debris \cite{Arirang}.\\
				2021& The Canadarm2 robot arm on the International Space Station was struck by space debris \cite{ISSArm}.\\
				\hline
			\end{tabular}
		\end{center}
	\end{table*}
	
	\begin{table*}[ht]
		\begin{center}
			
			\caption{Analysis, classification, and comparison of issues}
			\renewcommand\arraystretch{1.3}
			\begin{tabular}{|l|r|r|r|r|r|r|r|}
				\hline
				\hline
				Issue                  & Damaged                         & Damaged                      & Security\&reliability                        & \multirow{2}{*}{Reversibility} & \multirow{2}{*}{Apparency}  & Intended                    & Collateral            \\
				types                           & location                        & type                         & requirements                     &                                &                             & solutions             & damage                \\ \hline
				\multirow{2}{*}{Eavesdropping} & \multirow{2}{*}{Wireless link}  & \multirow{2}{*}{Security} & \multirow{2}{*}{Confidentiality} & \multirow{2}{*}{Reversible}    & \multirow{2}{*}{Inapparent} & Prevention                  & Could create          \\
				&                                 &                              &                                  &                                &                             & Mitigation                  & active attacks        \\ \hline
				Satellite transponder          & \multirow{2}{*}{Wireless link}  & \multirow{2}{*}{Security} & \multirow{2}{*}{Confidentiality} & \multirow{2}{*}{Reversible}    & \multirow{2}{*}{Inapparent} & \multirow{2}{*}{Prevention} & \multirow{2}{*}{None} \\
				stealing                       &                                 &                              &                                  &                                &                             &                             &                       \\ \hline
				Power-based                    & \multirow{2}{*}{Wireless link}  & \multirow{2}{*}{Security} & \multirow{2}{*}{Integrity}       & Depending on                   & \multirow{2}{*}{Apparent}   & \multirow{2}{*}{Mitigation} & Could leave target    \\
				jamming                        &                                 &                              &                                  & the attackers                  &                             &                             & disabled              \\ \hline
				Spoofing                       & User segment                    & \multirow{2}{*}{Security} & \multirow{2}{*}{Availability}    & \multirow{2}{*}{Reversible}    & \multirow{2}{*}{Apparent}   & Detection                   & Could leave target    \\
				jamming                        & Space segment                   &                              &                                  &                                &                             & Mitigation                  & disabled              \\ \hline
				Message                        & \multirow{2}{*}{Ground segment} & \multirow{2}{*}{Security} & \multirow{2}{*}{Integrity}       & \multirow{2}{*}{Reversible}    & \multirow{2}{*}{Apparent}   & Detection                   & Could lead to         \\
				modification                   &                                 &                              &                                  &                                &                             & Mitigation                  & wrong decisions       \\ \hline
				\multirow{2}{*}{DoS}           & Ground segment                  & \multirow{2}{*}{Security} & \multirow{2}{*}{Availability}    & \multirow{2}{*}{Reversible}    & \multirow{2}{*}{Apparent}   & Detection                   & Could leave target    \\
				& Space segment                   &                              &                                  &                                &                             & Mitigation                  & disabled              \\ \hline
				Node                           & User segment                    & \multirow{2}{*}{Security} & \multirow{2}{*}{Availability}    & \multirow{2}{*}{Irreversible}  & \multirow{2}{*}{Apparent}   & Detection                   & Could leave target    \\
				compromise                     & Space segment                   &                              &                                  &                                &                             & Mitigation                  & disabled              \\ \hline
				Node                           & User segment                    & \multirow{2}{*}{Security}    & \multirow{2}{*}{Availability}    & \multirow{2}{*}{Irreversible}  & \multirow{2}{*}{Apparent}   & Detection                   & Could generate        \\
				destruction                    & Space segment                   &                              &                                  &                                &                             & Mitigation                  & more space debris     \\ \hline
				\multirow{2}{*}{MAI}           & Ground segment                  & \multirow{2}{*}{Security} & Integrity                        & \multirow{2}{*}{Reversible}    & \multirow{2}{*}{Apparent}   & \multirow{2}{*}{Mitigation} & \multirow{2}{*}{None} \\
				& Space segment                   &                              & Availability                     &                                &                             &                             &                       \\ \hline
				CCI between                    & Ground segment                  & \multirow{2}{*}{Security} & Integrity                        & \multirow{2}{*}{Reversible}    & \multirow{2}{*}{Apparent}   & \multirow{2}{*}{Mitigation} & \multirow{2}{*}{None} \\
				beams                          & Space segment                   &                              & Availability                     &                                &                             &                             &                       \\ \hline
				\multirow{2}{*}{SEU}           & \multirow{2}{*}{Space segment}  & \multirow{2}{*}{Reliability}    & \multirow{2}{*}{Integrity}       & \multirow{2}{*}{Reversible}    & \multirow{2}{*}{Apparent}   & Prevention                  & Could lead to         \\
				&                                 &                              &                                  &                                &                             & Mitigation                  & wrong decisions       \\ \hline
				CCI between                    & Ground segment                  & \multirow{2}{*}{Security} & Integrity                        & \multirow{2}{*}{Reversible}    & \multirow{2}{*}{Apparent}   & \multirow{2}{*}{Mitigation} & \multirow{2}{*}{None} \\
				systems                        & Space segment                   &                              & Availability                     &                                &                             &                             &                       \\ \hline
				Collisions with                & \multirow{2}{*}{Space segment}  & \multirow{2}{*}{Reliability}    & \multirow{2}{*}{Availability}    & \multirow{2}{*}{Irreversible}  & \multirow{2}{*}{Apparent}   & Detection                   & Could generate        \\
				spacecraft                     &                                 &                              &                                  &                                &                             & Mitigation                  & more space debris     \\ \hline
				Collisions with                & \multirow{2}{*}{Space segment}  & Reliability                     & \multirow{2}{*}{Availability}    & \multirow{2}{*}{Irreversible}  & \multirow{2}{*}{Apparent}   & Detection                   & Could generate        \\
				space debris                   &                                 &                              &                                  &                                &                             & Mitigation                  & more space debris     \\ \hline
			\end{tabular}
		\end{center}
	\end{table*}
	
	\textbf{Collisions with Spacecraft:} Given so many spacecraft belonging to different agencies entering LEOs, it is difficult to manage them collaboratively. Even worse, the orbits are constantly changing under the action of non-sphericity of the earth, ocean tides, and atmospheric damping, which results in the spacecraft deviating from their pre-set orbits. As a matter of fact, in 2009, the Iridium 33 satellite collided with the scrapped Russian Cosmos over Siberia, producing at least thousands of debris \cite{collision}. This space debris was fixed only a few months later, distributed between 500 km and 1300 km. As a remedy, collision avoidance control has to be carried out to reduce the risk of collisions with LEO satellites. On Sep. 2, 2019, the European Space Agency (ESA) made an emergency steering of the Aeolus satellite, successfully avoiding a space `car accident' with Starlink-44 \cite{Aeolusstarlink}. As reported by United Nation Office for Outer Space Affairs, the China Space Station has successfully conducted two evasive maneuvers to avoid potential collisions with the Starlink-1095 satellite on Jul. 1, 2021, and the Starlink-2305 satellite on Oct. 21, 2021, respectively \cite{ChinaSpace}.
	
	\textbf{Collisions with Debris:} Again, such frequent deployment activities have also led to a surge in space debris. Most orbit debris is human-generated objects, such as pieces of spacecraft, tiny flecks of paint from a spacecraft, parts of rockets, and decayed satellites. According to the ESA, there are approximately 1036500 debris objects larger than 1 cm estimated by statistical models to be in orbit \cite{ESASpacedebris}. There are close to 6000 tons of materials in LEOs. Most `space debris' moves fast, reaching speeds of 18000 miles per hour, almost seven times that of bullets. They expose LEO satellites to the Kessler phenomenon\footnote{The Kessler phenomenon, proposed by National Aeronautics and Space Administration (NASA) scientist Donald J. Kessler in 1978, is a chain reaction in which the resulting space debris would destroy other satellites and so on, with the result that LEO would become unusable \cite{DRMOLA201829}.}. Specifically, the density of space debris in LEO is high enough to cause cascade collisions, which adversely affects space exploration. With the advent of standardized production, the satellite development cycle and constellation deployment cycle have been substantially shortened, but there are also satellite failures, potentially requiring replacements during the deployment. Hence, Kessler's hypothesis is becoming a reality.
	
	As a matter of fact, collisions with debris at LEO orbits have already occurred \cite{Hubble, Endeavour, Sentinel1A, ISSArm}. Explicitly, ESA has showcased the solar cells retrieved from the Hubble Space Telescope, which have been damaged by various collisions with space debris. In 2007, orbit debris completely penetrated one of the radiator panels of the shuttle Endeavour. On Aug. 23, 2015, ESA engineers discovered that a solar panel on the Sentinel-1A satellite was hit by a piece of millimeter-sized debris, according to space-borne cameras. Fortunately, this satellite still remained capable of operating normally. A piece of space debris struck the International Space Station's Canadarm2 robot arm, which was spotted on May 12, 2021.
	
	Because of these incidents, it is routine for operators of satellites in dense orbits to spend time tracking the collision risk. When the probability of collision exceeds a specific limit, debris avoidance maneuvers have to be planned. For example,
	Indian Space Research Organisation (ISRO) reported that they monitored 7600 satellite collision threats in 2021 and avoided 60 since 2015 \cite{ISRO}. Moreover, the International Space Station has carried out as many as 29 debris avoidance maneuvers since 1999 \cite{Debrisavoidance}. However, due to its excessive fuel consumption, the technical solutions in \cite{Debrisavoidance} are not suitable for low-cost LEO satellites with limited energy. Debris tracking \cite{Debristracking1, Debristracking2, Debristracking3}, space probe \cite{Debrisdetection1, Debrisdetection2}, debris removal \cite{Debrismitigation} are separately efficient methods for detecting and preventing collisions.

	{\it Lessons Learned:} Table \uppercase\expandafter{\romannumeral6} summarizes, classifies, and compares the issues encountered by LEO SCSs in terms of their types, damaged locations, security and reliability requirements, and so on. In summary, the critical lessons learned from the in-depth review of the issues are as follows.
		
		The authors of \cite{Vulnerabilities2} discussed the specific characteristics of attackers, while the authors of \cite{Satarxiv, COMST6G20214, Vulnerabilities1} classified the related issues. Indeed, the identity and technical capabilities of attackers determine the type of security attacks, and different types of security attacks result in different levels of damage. Since eavesdropping attacks inflict no harm upon the entire system, only those malicious individuals, who know the target satellite's operating frequency and orbit information, can have the opportunity to steal confidential information, for example, by launching eavesdropping drones \cite{DroneRL}. Suffice it to say that irreversible damage may be inflicted upon satellites by anti-satellite weapons owned by a national army, for example, because individuals normally do not have the capability of manufacturing weapons.
		
		Compared to the ground segment and the user segment, the security of the space segment is more critical. As detailed in \cite{JK2016}, the power, storage, and computing capability of LEO satellites is severely limited, rendering the specific class of security algorithms having high complexity and storage requirements inapplicable. Moreover, it is inconvenient to modify a satellite for incorporating security enhancements from an operational perspective. Additionally, owing to their harsh environment, LEO satellites tend to suffer from the risk of both SEUs as well as collisions, and the consequences of collisions are irreversible. To make things worse, debris generated by collisions could potentially cause further collisions \cite{Research}.
		
		Proficient orbit selection is extremely critical. Specifically, in order to reduce the risks of SEUs and collisions, orbits having only a few satellites on their adjacent orbits should be preferentially picked, as shown in Fig. 10. The authors of \cite{SEUFuqiang} provided the evidence that the higher the orbit altitude, the higher the probability of SEUs occurrence in LEOs. Therefore, it is difficult to find a beneficial orbit altitude that guarantees both low collision and low SEUs probability. Having an orbit altitude chosen for reducing the probability of collisions has a higher priority than that reducing the probability of SEUs. Given that the damages caused by collisions are irreversible \cite{Sentinel1A}, it is difficult to conceive solutions to repair the damaged satellites. In addition, the closer the orbit inclination is to 90$^{\circ}$, the more seamless the global coverage becomes \cite{pratt2019satellite}, but the probability of SEUs is also increased.

	\begin{figure*}[]
		\centering
		
		\includegraphics[width=1.4\columnwidth]{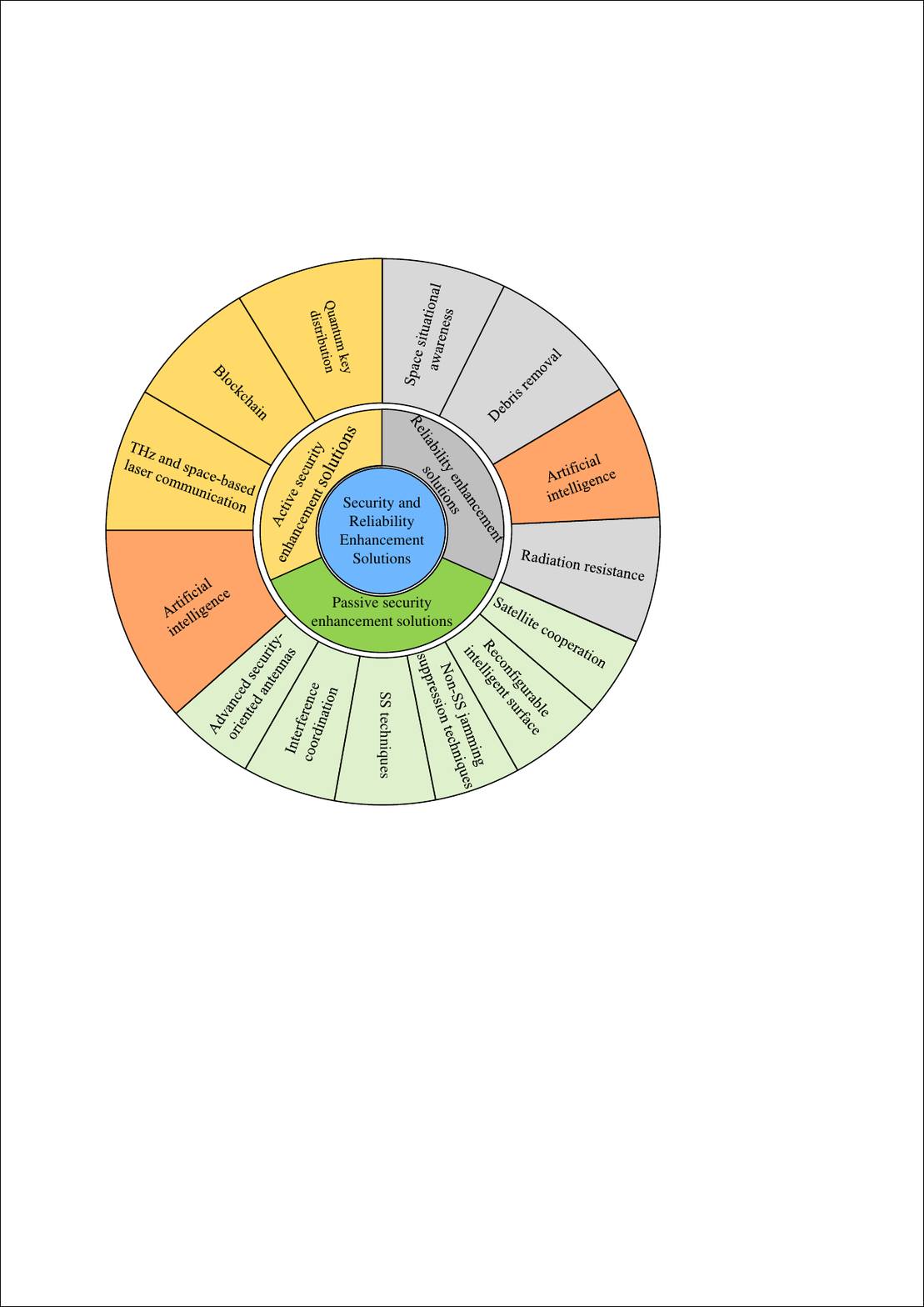}	
		\caption{Classification of solutions dealing with security and reliability in LEO SCSs.}
		\label{Keytechniquespic}
	\end{figure*}	
	
	\section{Security and Reliability Enhancement Solutions}
	
	In this section, \textit{prevention}, \textit{detection}, and \textit{mitigation} constitute essential principles closely linked to both security and reliability enhancement solutions \cite{SecurityIoT}.
	
	\begin{itemize}
		\item[$\bullet$]\textbf{Prevention}: Prevention focuses on protecting LEO SCSs from issues before they are exposed to LEO SCSs. The employment of Terahertz (THz) and laser techniques is capable of coping with CCI by avoiding frequency-reuse in the immediate vicinity \cite{THzJSAC}. Moreover, deploying firewalls and antivirus software and applying patches for the issues identified can dramatically reduce the probability of successful attacks. Additionally, prevention is also vitally critical concerning 'fatal' issues, such as collisions, because the resultant damage is clearly irreversible. Although prevention is desired to avoid potential security incidents, it is not always feasible.
		
		\item[$\bullet$]\textbf{Detection}: Prevention usually aims for improving its existing defense mechanism. However, once an attacker succeeds in circumventing the existing preventative solutions, this must be detected to minimize its impact. Usually, IDS is adopted for detecting the attacks and for mitigating the damage inflicted \cite{MLIDS}.
		
		\item[$\bullet$]\textbf{Mitigation}: Mitigation refers to the specific solutions put in place to help prevent issues as well as limit the extent of damage when security attacks do happen. Again, SS techniques are still popular due to their immunity to jamming and eavesdropping \cite{DSSSTVT2021}.
	\end{itemize}
	
	In light of this, we further classify solutions into active and passive solutions. Active solutions include the functions of prevention and detection, thus making LEO SCSs more proactive in the face of issues. By contrast, passive solutions must directly face these issues and reduce or eliminate their impact as far as possible. To this end, a series of security and reliability enhancement solutions are presented in Fig.~\ref{Keytechniquespic}. In addition, several trade-offs and the critical lessons learned from these solutions are also summarized.

	\subsection{Active Security Enhancement Solutions}
	Active security enhancement solutions, including QKD, blockchain, THz, space-based laser communications, and AI, aim for preventing or actively detecting impending deleterious issues. Among them, QKD constitutes a symmetric secret key negotiation protocol capable of maintaining information-theoretic security, and it has evolved from academic research to off-the-shelf commercialization~\cite{QKDCOMST}. Furthermore, blockchain is capable of satisfying the security requirement of decentralization, making LEO SCSs more robust. As a further advance, the progress of THz and laser-based communications is conducive to dealing with the CCI caused by the spectrum crunch. Finally, intelligent data-driven model-based AI-aided solutions are suitable for traffic prediction, telemetry-based data mining, and anomaly detection.

	\begin{table*}[]
		\begin{center}
			%	\tiny
			\scriptsize
			\renewcommand\arraystretch{1.4}
			\caption{Major achievements in the implementation of QKD}
			\begin{threeparttable}
				\begin{tabular}{m{0.5cm}<{\centering}|m{14cm}<{\raggedright}}
					\hline
					\hline
					
					2017 & 1200 km satellite-to-ground QKD at 1.1 kbit/s  \cite{QKDR6}        \\
					
					2017 & 1000 km satellite-to-ground entanglement-based QKD at 3.5 bit/s   \cite{QKDR7}        \\
					
					2018 & 7600 km apart ground gateways with satellite relay QKD and encryption demonstration at key volume 100 kB \cite{QKDR8} \\
					
					2019 & Continuous-variable QKD over 100 km fiber link at 0.14 kbit/s based on a photonic integrated quantum system  \cite{QKD201912}        \\
					2020 & Point-to-point discrete-variable QKD over 509 km fiber link at 0.1 bit/s \cite{DVQKD20202}                                      \\
					2020 & First wavelength division multiplexing of 194 continuous-variable QKD at 172.6 Mbit/s over 25 km \cite{CVQKD20204}                \\
					2020 & Point-to-point continuous-variable QKD over 13 km fiber link at 0.88 Mbit/s \cite{CVQKD20205}                                       \\
					2020 & Point-to-point continuous-variable QKD over 202.81 km fiber link at 6.214 bit/s \cite{CVQKD20206}                                   \\
					2020 & Discrete-variable QKD over 1200 km free space optical link at 31 bit/s using Micius \cite{QKDR6}                           \\
					2020 & Continuous-variable QKD over 180 km fiber link at 31 bit/s based on a photonic integrated quantum system \cite{QKDR7}          \\ 
					
					2020 & 1120 km apart ground stations entanglement-based QKD  at 0.12 kbit/s  \cite{QKDR9} \\ 
					
					2021 & 4600 km apart ground stations entanglement-based QKD  at 47.8 kbit/s  \cite{QKDR10}  \\ 
					
					\hline

				\end{tabular}
			\end{threeparttable}
		\end{center}
	\end{table*}
	
	\subsubsection{Quantum Key Distribution}					
	The conceptually simplest encryption method relies on generating a pseudo-random secret key and then taking the modulo-two function of the key and the information to be encrypted, which is termed as plain text. Naturally, the key has to be as long as the data sequence is transmitted, which implies imposing an overhead of 100\%.
	
	Then the resultant so-called ciphertext may be transmitted from the source to the destination over a public channel. Given the knowledge of the secret key, the receiver can recover the original plaintext using the secret key. Since the key must remain confidential for the communications of the two parties, it must be shared between them over a secure channel.
	
	The family of legacy cryptography schemes was conceived under the assumption that it would require an excessive amount of time, even upon using the most powerful computers by the eavesdropper to infer the key. However, given the threat of powerful quantum computers, it is no longer safe to rely on the above-mentioned antiquated assumption.
	
	Similarly, simple principles may be used in QKD systems for the encryption/decryption process, but the negotiation of the secret key relies on a quantum channel as well as on an insecure public channel plus an authenticated public channel. The family of satellite-based QKD systems was richly characterized in \cite{HanzoQKD1}, along with diverse satellite channels using detailed examples. 

	The transmission medium of QKD-based key negotiation typically relies on optical fibers and free space. Optical fiber has a low loss of about $0.3$ dB/km and a high stability, hence it is more suitable for transmitting quantum signals. In recent years, numerous theoretical and experimental QKD designs have been proposed for improving the achievable secret-key rate vs distance trade-off \cite{QKD201912,DVQKD20202,CVQKD20204,CVQKD20205,CVQKD20206}.
		
		There has also been substantial progress in QKD experiments relying on free-space optical links, culminating in the launch of the world's first quantum satellite-based QKD experiment in 2016, as reported in \cite{QKDR6,QKDR7,QKDR8,QKDR9,QKDR10}. The significant milestones achieved in the implementation of QKD systems are chronologically arranged in Table \uppercase\expandafter{\romannumeral7}.
		
		Additionally, many other countries or organizations, such as the ESA and Canada, are also aiming for providing their satellite-based QKD services. 
		
		However, as the terminology suggests, QKD remains a key negotiation and distribution protocol, where the secret key is used by classical systems. By contrast, Quantum Secure Direct Communications (QSDC) \cite{HanzoQKD2} is a fully-fledged quantum communication protocol, in which confidential messages are transmitted directly over a quantum channel without requiring a secret key. It has hence enjoyed a rapid evolution, as documented in \cite{HanzoDeng, HanzoDeng2, Hanzoyan2004scheme, Hanzowang2005quantum, Hanzozhou2020measurement, Hanzohuang2018implementation, HanzoQKDTCOM1, HanzoQKDTCOM2}.
		
		{\it Lessons Learned:} 	
		QKD has already found numerous commercial applications \cite{QKDRL}, such as finance and healthcare. But it still has numerous open challenges. Specifically, the operational QKD networks only tend to provide point-to-point key distribution or short-distance network services by relying on optical switches and routers. Explicitly, their distance is limited, since the quantum-domain signal must not be amplified. Otherwise, it collapses back into the classical domain. Continued focus in this area is required to facilitate large-scale deployments. Moreover, in the FSO-based QKD and QSDC scenarios \cite{HanzoQKD1}, the clouds may affect QKD and QSDC transmission owing to dispersion imposed by atmospheric eddies. Specifically, the water molecules in the cloud layer cause scattering and absorption of optical signals, whereas the attenuation of optical signals by the cloud causes signal strength reduction. Furthermore, free-space QKD and QSDC usually require the precise alignment of telescopes. When there are clouds, the transmission of visible light and infrared light is limited, hence the telescope cannot be accurately aligned. This affects the stability of the QKD and QSDC systems. Even worse, the high mobility of LEO satellites exacerbates the above challenges.

	\subsubsection{Blockchain}
	
	Given the increasing number of LEO satellites and users supported by LEO SCSs, managing their security becomes a new challenge. As a remedy, the blockchain technique becomes a promising solution for the secure decentralized management of LEO SCSs. Briefly, the blockchain technique \cite{Blockchainintro}  is a structure that stores transactional records in several databases, known as the `block', in a peer-to-peer network constituted by connected nodes, known as the `chain'. Typically, this storage is referred to as a digital ledger.
	
	The blockchain technique satisfies several of the above-mentioned security and reliability requirements, namely confidentiality, accountability, and decentralization. By relying on an encrypted database, users must have the correct key to read information from this database or write to it. Moreover, once the information is updated, all the related information is updated together as a block and appended to the previous version, thereby creating an immutable tamper-proof record. The premise is that the majority of participants check and verify this information. Otherwise, the information cannot be updated on the blockchain.
	
	Additionally, decentralization is another compelling feature of the blockchain technique. If a failure occurs on one or several nodes of a blockchain network, the other nodes still retain their data, and the network continues to function. Hence, a blockchain is often referred to as a distributed ledger because the information resides on multiple devices in a peer-to-peer network, where each device replicates and holds an identical copy of the ledger and updates it independently.
	
	Given these benefits, many researchers have harnessed the blockchain technique for dealing with security attacks. Han {\itshape et al.} \cite{blockchainspoofing} exploited the blockchain technique to share and verify location information in a UAV network to detect spoofing jamming. By contrast, the blockchain technique is adopted in \cite{ChenNode, BlockchainPUF} for protecting information from modification in resource-constrained IoT devices. Briefly, Chen {\itshape et al.} \cite{ChenNode} conceived a stochastic blockchain scheme for protecting the integrity of IoT data. A fraction of the nodes were randomly selected for broadcasting their IoT data, which led to uncertainty for the attacker. As a further development, Yuan {\itshape et al.} \cite{BlockchainPUF} exploited the characteristics of the Physical Unclonable Functions (PUF) as part of the key agreement without storing sensitive keys in their lightweight broadcast authentication protocol in the blockchain.
	
	As a benefit of its distributed ledgers and consensus operations, the blockchain technique is immune to both the DoS and the DDoS attacks \cite{BlockchainSAGIN}. For example, Georgios {\itshape et al.} \cite{BlockchainDDOS} employed lightweight agents for exchanging outbound traffic information governed by blockchain to identify possible victims of DDoS attacks, which ensured the integrity of both the procedure and information exchanged.
	
	For instance, to detect a compromised node in the DoS attack scenario, Kumar {\itshape et al.} \cite{Blockchainfilter2019} proposed a blockchain-based deterministic en-route report filtering scheme, which is capable of dropping false reports. As a further benefit, their scheme did not require any critical exchange between sensor nodes for data endorsement or authentication, thus reducing both the associated key storage overhead and communication overhead.

	{{\it Lessons Learned}:} As evidenced by the literature, LEO SCSs empowered by the blockchain exhibit high robustness and fault tolerance even in the context of high-mobility LEO satellites. Since blockchain-based security services usually require substantial storage space and computing power, their employment in resource-constrained terminals and LEO satellites requires radical innovation.
		
		Again, LEO SCSs have to share the user's sensitive information among multiple gateways or LEO satellites. In this case, the users' sensitive information may become exposed, hence potential security risks and data misuse issues may occur without user awareness. Therefore, during the blockchain implementation process, the protection of user privacy also has to be carefully considered.
	
	\subsubsection{THz and Space-based Laser Communication} 
\begin{figure*}[]
		\centering
		
		\includegraphics[width=1.3\columnwidth]{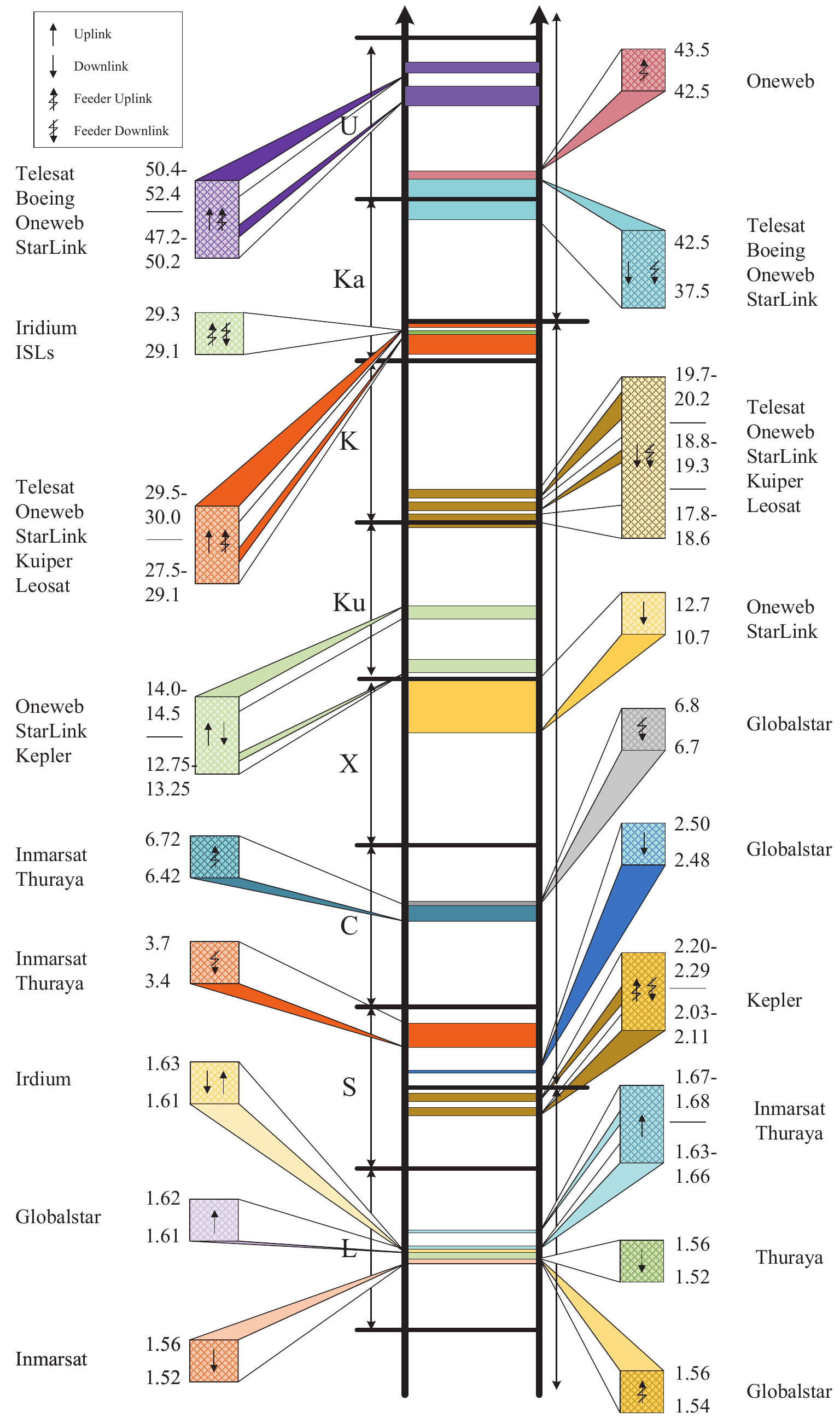}
		
		\caption{Frequency allocations of several commercial constellations between 1 and 60 GHz.}
		\label{Frequency}
	\end{figure*}

The frequency allocations of several commercial LEO satellite constellations are shown in Fig.~\ref{Frequency}. Observe that many LEO satellites operate in the decimeter wave and centimeter wave bands such as Iridium and Globalstar. At the time of writing, the Millimeter Wave (MmWave) band is attracting research attention as a benefit of its rich spectral resources \cite{HanzommWave}. Many LEO satellite manufacturers such as Boeing, Starlink, and OneWeb sought permission to launch satellites operating in the 50.2-52.4 Gigahertz (GHz) bands \cite{Oneweb1, StarlinkLEO, Boeing524}. However, these frequency resources are becoming congested. A potential solution is to increase the operating frequency to the THz or even optical bands. Thanks to the development of device and communication technology, these emerging bands are gradually entering commercialization \cite{Haichuan1, Haichuan3}.
	
	The THz band has a vast amount of available bandwidth, which has to be further explored. Radio frequencies above 100 GHz are largely untapped for specific applications by the ITU. Hence they might become available for SCSs. In the presence of water vapor molecules and other propagation effects, the THz band suffers a limitation in transmission distance, which is not suitable for the satellite-Earth link \cite{Terahertz3SE}. Hence, the employment of THz communications for ISLs \cite{THZRL, TerahertzISL}, which operates above the Earth's atmosphere, could be an attractive alternative. According to \cite{terahertztransceivers}, THz transmitters and receivers could be designed to circumvent the disadvantages of microwave bands. Although the attenuation of the THz band is high, this may potentially be compensated by large-scale antennas used for Beamforming (BF) on a space-borne payload. The beam width of the large-scale antennas in the THz band is narrower than that of common microwave ISLs, which enhances their ability to resist eavesdropping.
	
	However, observe that the longest communication distance was 21 km at 140 GHz \cite{140G5G12KM}, which is insufficient for ISLs. Therefore, a large antenna array and high-power devices operating in the THz band should be developed to overcome the extremely high propagation loss and power limitations of the space-borne transceivers in harsh operating environments.
	
	The laser band is far above the electromagnetic spectrum. Thus it has a strong anti-interference capability. Laser communications cannot be detected by spectrum analyzers or RF meters since the laser beam is highly directional, which makes it a strong candidate for ISLs and cross-layer links \cite{Laserhighsecurity}. Additionally, laser offers several advantages over microwave communications in terms of size, weight, and power dissipation compared to the MmWave band under the same data rate conditions \cite{toyoshima2005trends, Laserhighsecurity2}.
	
	Many research institutions across the world have conducted numerous experiments, which are summarized in Table \uppercase\expandafter{\romannumeral8} at a glance. Additionally, Starlink tested `space lasers' between two satellites, relaying hundreds of Gbytes of data in Sep. 2020. At the time of writing, Starlink is engaged in rolling out further laser cross-links amongst their satellites to minimize the number of ground facilities and to extend the coverage to remote areas \cite{Starlinklaser1, Starlinklaser2}.

	\begin{table*}[]
		\centering
		%	\tiny
		\scriptsize
		%	\footnotesize
		%	\small
		%	\normalsize
		%	\large
		%	\Large
		%	\LARGE
		%	\huge
		%	\Huge
		\renewcommand\arraystretch{1.3}
		\caption{The evolution of space-based laser communications}
		\begin{tabular}{m{1cm}<{\centering}m{1.4cm}<{\centering}m{1.4cm}<{\centering}m{1.6cm}<{\centering}m{2.0cm}<{\centering}m{2.5cm}<{\centering}m{1.9cm}<{\centering}m{1cm}<{\centering}}
			\hline
			\hline
			Year                  & Project                           & Type      & Country/Region          & Data rate (Mbps)        & Modulation              & Distance (km) & Ref.                              \\ \hline
			2001                  & SILEX                             & GEO-LEO   & Europe                  & 50                      & IMDD                    & 45000         & \cite{SPOT4}                      \\
			2006                  & OICETS                            & LEO-OGS   & Japan                   & 50                      & IMDD                    & 610           & \cite{OICETS}                     \\
			\multirow{2}{*}{2010} & \multirow{2}{*}{TerraSAR} & LEO-OGS   & \multirow{2}{*}{Europe} & \multirow{2}{*}{5625}   & \multirow{2}{*}{BPSK}   & 500-1000      & \multirow{2}{*}{\cite{TESAT}}     \\
			&                                   & LEO-LEO   &                         &                         &                         & 1000-5000     &                                   \\
			2011                  & BTLS                              & LEO-OGS   & Russia                  & 125                     & IMDD                    & 400           & \cite{RussiaLaser}                \\
			2013                  & LLCD                              & Lunar-OGS & US                      & 622                     & PPM                     & 400000        & \cite{NASALLCD}                   \\
			2013                  & Alphasat                          & GEO-LEO   & Europe                  & 1800                    & BPSK                    & 45000         & \cite{Alphasat}                   \\
			2014                  & OPALS                             & LEO-OGS   & US                      & 50                      & IMDD                    & 400           & \cite{OPALS}                      \\
			2014                  & SOTA                              & LEO-OGS   & Japan                   & 10                      & OOK/IMDD                & 642           & \cite{SOTA}                       \\
			2016                  & MICIUS                            & LEO-OGS   & China                   & 5120                    & DPSK                    & 1500          & \cite{MICIUSChina}                \\
			2016                  & OCSD                              & LEO-OGS   & US                      & 200                     & IMDD                    & 450           & \cite{janson2016nasa}             \\
			2017                  & VSOTA                             & LEO-OGS   & Japan                   & 10                      & ------                  & 1000          & \cite{ takenaka2017satellite }    \\
			2017                  & SJ-13                             & GEO-OGS   & China                   & 4800                    & IMDD                    & 36000         & \cite{ SJ13 }                     \\
			2020                  & EDRS-C                            & GEO-LEO   & Europe                  & 1800                    & BPSK                    & 45000         & \cite{ EDRSC }                    \\
			2020                  & SJ-20                             & GEO-OGS   & China                   & 10000                   & OOK/BPSK/QPSK           & 36000         & \cite{Shijian20}                  \\ \hline
			\multirow{2}{*}{2023} & \multirow{2}{*}{CubeSOTA}         & GEO-LEO   & \multirow{2}{*}{Japan}  & \multirow{2}{*}{10000}  & \multirow{2}{*}{DPSK}   & 39693         & \multirow{2}{*}{\cite{Japan2021}} \\
			&                                   & LEO-OGS   &                         &                         &                         & 1103          &                                   \\
			2025                  & EDRS-D                            & GEO-GEO   & Europe                  & 3600-10000              & BPSK                    & 80000         & \cite{SJ13,EDRSD}                 \\
			\multirow{2}{*}{2025} & \multirow{2}{*}{ScyLight}         & GEO-LEO   & \multirow{2}{*}{Europe} & \multirow{2}{*}{100000} & \multirow{2}{*}{------} & ------        & \multirow{2}{*}{\cite{ScyLight}}  \\
			&                                   & LEO-OGS   &                         &                         &                         & 80000         &                                   \\ \hline
		\end{tabular}
	\end{table*}

	{\it Lessons Learned:} 		
		Since the THz and laser beam are highly directional, they have a limited coverage area. This imposes serious challenges for signal alignment (acquisition and tracking) in the context of high-mobility LEO satellites. In the existing satellite-Earth communication experiments, the experiments in Table VIII were conducted between satellites and an Optical Ground Station (OGS). The OGS exploits its own position and the position of the satellites to assist with signal alignment. However, the Doppler shift still has to be mitigated with the aid of sophisticated processing algorithms to eliminate their adverse effects.
		
		Furthermore, signal alignment is more challenging for ISLs in LEO SCSs. The high velocity and the jitter of the space-borne payload {\cite{Jitterlaser} make the accurate alignment of the beam a challenge. Furthermore, even higher Doppler frequency shifts may be observed for the space-borne laser terminals in the `reverse seam' \cite{Iridium}, where the adjacent satellites move in opposite directions. The authors of \cite{yang2009doppler} analyzed the Doppler frequency shift of LEO SCSs relying on laser links. Inadequate Doppler frequency shift compensation results in a loss of frequency synchronization at the receiver, ultimately resulting in data loss.

		\subsubsection{Aritificial Intelligence}

		In recent years, the success achieved by AI in terrestrial wireless communication systems has also pervaded the family of LEO SCSs. Among the AI techniques, Machine Learning (ML) has matured and emerged as a valuable tool capable of learning empirical models and tracking changes in data patterns during space missions \cite{li2022machine}. This aspect of ML proves particularly advantageous in the context of security solutions since it enables the analysis of various types of data from different perspectives.
		
		\textbf{Tele-traffic Data}:
		Accurate processing of tele-traffic in LEO SCSs holds paramount significance. Predicted traffic patterns play a critical role in optimizing routing paths and pre-scheduling networking resources, hence mitigating the CCI and minimizing both transmission outages and the probability of congestion. Moreover, early detection of abnormal traffic generated by malicious attackers can proactively prevent congestion from occurring. The authors of \cite{na2018distributed, bie2019combined} adopted the extreme learning machine (ELM), a neural network having a simple structure and high computation speed in accurately characterizing the traffic load of LEO satellites. As a further development, an accurate long-short-term memory (LSTM) prediction model based on a deep recurrent neural network was proposed in \cite{han2019prediction,jia2022adaptive}. LSTM constitutes an ideal ML method for handling multi-variate time-series data since it is capable of modeling the complex interactions of a system. However, the LSTM prediction model imposes a high computational burden. To circumvent this, Li {\it et al.} developed a Gated Recurrent Unit (GRU) based neural traffic prediction algorithm having a reduced gate structure \cite{li2021research}, whose training efficiency and accuracy can be substantially improved by cooperating with the powerful techniques of transfer learning and online training.
		
		\textbf{Housekeeping Data}:
		The housekeeping data collected from satellites consists of numerous measurements and readings that reflect the status of the satellite and its surrounding environment. By analyzing the abnormalties within the housekeeping data, potential failures and imminent alerts can be inferred, allowing the satellite to intelligently make proactive decisions for mitigating the risk of failure. In this context, Fuertes {\it et al.} developed a Support Vector Machine (SVM)-based anomaly detection algorithm \cite{fuertes2016improving}, which is capable of recognizing anomalies with high detection sensitivity, but comes at the price of a high false alarm rate. The properties of housekeeping data are summarized in \cite{yairi2017data}, including their high dimensionality, multi-modality, and heterogeneity, which lays the foundations for ML-based detection relying on probabilistic clustering. LSTM still represents a significant leap forward in efficiently processing historical data for future prediction and anomaly detection \cite{wang2022deep,hundman2018detecting,tariq2019detecting,zeng2022satellite}. In \cite{wang2022deep}, the LSTM method is combined with the Gaussian model of the training errors for the sake of detecting anomalies. Notably, this solution mitigates the probability of false alarms resulting from misconstrued anomalies, unknown incidents, and sparse samples using the Deviation Divide Mean over Neighbors method. Additionally, the authors of \cite{zeng2022satellite} propose exploiting the causality of time series to construct a causal network, which exhibits plausible interpretability, robustness, and adaptability. Comprehensive comparisons among different ML techniques, including the Recurrent Neural Network (RNN), LSTM, and GRU, used for the prediction of the LEO satellite data are conducted in \cite{ibrahim2018machine}. The evaluation of prediction accuracy using battery temperature, power bus voltage, and load current data shows that LSTM achieves the highest accuracy, while GRU exhibits the shortest running time. More recently, the temporal convolution network also raised much attention for time series prediction with favorable parallel processing ability and temporal characterization \cite{wang2021anomaly}, which shows superior operational efficiency compared to LSTM. The predicted telemetry data based on the aforementioned prediction techniques can then be further analyzed to detect any potential future failure. A common approach is to compare the prediction error to a predefined or automatically adjusted threshold to detect which data is anomalous \cite{hundman2018detecting,tariq2019detecting,wang2021anomaly}. 
		
		\textbf{Attack Data}: 
		Furthermore, the interconnected nature of LEO satellite networks results in vulnerabilities to cyber threats and malicious satellites because of the poor network security of inter-satellite networks. It was thus important to detect cyber-attacks from satellite networks while preserving data privacy. More recently, the distributed LSTM technique was utilized in \cite{moustafa2022dfsat} for identifying cyber-attacks in each smart satellite network, such as reconnaissance, fuzzes, and denial of service attacks. The results were then further processed by a federated learning architecture to form a more private and secure intrusion detection system. 
		
		\textbf{Power System Data}:
		Power system data plays a critical role in maintaining the safe and stable operation of a specific mission. However, the structure of the power system is complex, where faults may occur in cables, solar arrays, batteries, power distribution switches, power controllers, and so on. Quantifying the correlation between different fault types and the causality between faults and the corresponding fault predictions is challenging. Fortunately, a large amount of power system data can be gathered, which allows ML methods to build an accurate fault classification model for characterizing the relationship between abnormal data and fault association \cite{article}. In particular, the LSTM method discussed above can also be utilized for predicting the parameters and for performing anomaly detection in the satellite’s power system \cite{dong2019deep,cheng2021research}. In \cite{7819878}, the authors specifically focus on detecting faults in solar arrays, which have the highest failure rate among all components in orbit. Similar to \cite{fuertes2016improving}, an SVM-based regression is utilized in \cite{coulter2021online} for detecting potential threats in a generic spacecraft power system. This system exhibits excellent learning and prediction capabilities.
		
		\textbf{Spectrum Data}: 
		ML tools are crucial for processing spectrum data in satellite communication systems to detect anomalies and interference. Similarly to other time-dependent telemetry data, the spectrum data can be utilized in an LSTM-based prediction model for intelligently managing future signal spectrum as well as for detecting anomalies and interference \cite{gunn2018anomaly,henarejos2019deep,pellaco2019spectrum}. Furthermore, the features of interference can be leveraged for interference classification to identify and avoid interfering sources. For example, in \cite{henarejos2019deep}, the authors utilize an intelligent LSTM interference classifier that harnesses four features, including the magnitude and phase of the temporal and frequency domain signals. The ML classifier model developed in \cite{qin2022interference} consists of a backbone network, neck network, and head network, which characterize interference based on its type, bandwidth, intensity, and frequency. These features are classified into six interference patterns: single-frequency interference, frequency-hopping interference, single-frequency sweeping interference, round-trip frequency sweeping interference, low interference, and other interference. The experiments conducted demonstrated nearly 100\% accuracy in spectrum detection, enabling accurate satellite interference management.
		
		In summary, the most popular ML techniques for conducting prediction, anomaly detection, and classification on satellite data are neural networks. Again, a representative technique is LSTM, which is a kind of neural network associated with recurrent connections. These connections allow the network to retain valuable information in internal memory. GRU is similar to LSTM in terms of having recurrent connections and memory, but with a simpler structure and less information-flows within the network compared to LSTM. Both LSTM and GRU constitute specialized variants of the traditional RNN, where each neuron receives its input from the previous time step and passes the output to the next time step. On the other hand, ELM, as adopted in \cite{na2018distributed,bie2019combined}, is a feed-forward neural network without any recurrent connections. In other words, ELM does not have explicit memory to store past information, but it is known for its fast training. Fig.~\ref{LSTM} illustrates the entire process of the ML-based security solutions conceived for satellite data reviewed in this paper.
		
		\begin{figure}[ht]
			\centering
			
			\includegraphics[width=1\columnwidth]{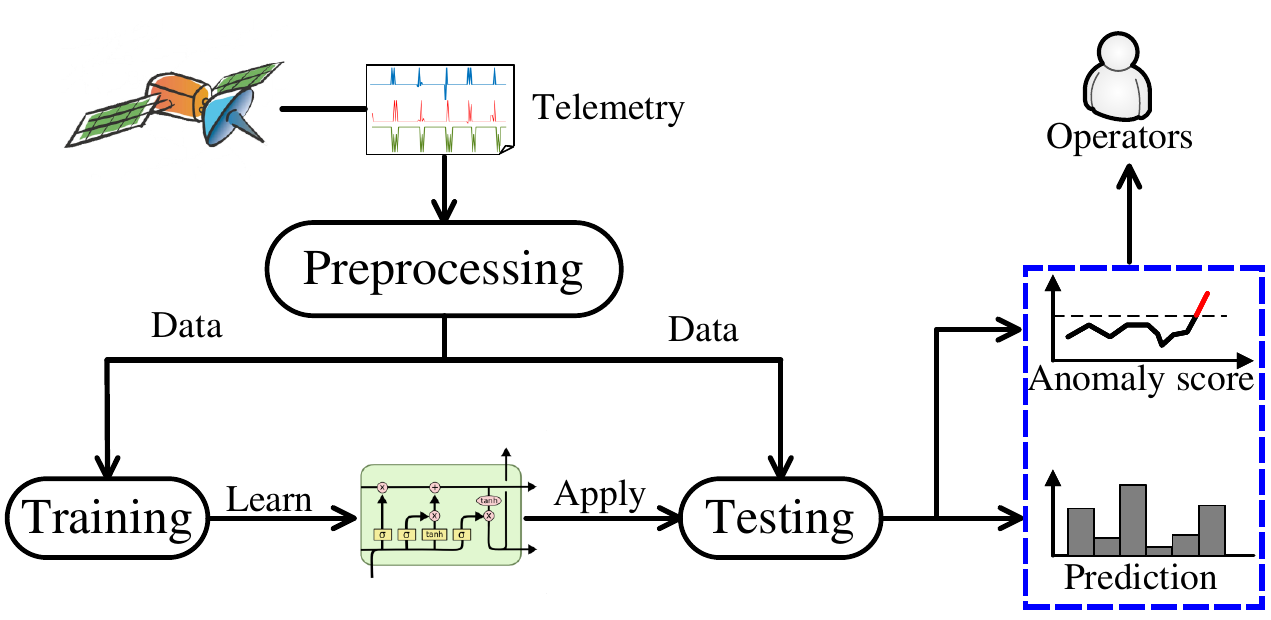}
			
			\caption{The entire process of the ML-based security solutions for satellite data.}
			\label{LSTM}
		\end{figure}
		
		When it comes to ML techniques, one of the key bottlenecks in implementation is the availability of datasets. However, most of the existing literature does not make the associated datasets gleaned from real satellites publicly available \cite{li2021research,yairi2017data,wang2022deep,hundman2018detecting,tariq2019detecting,ibrahim2018machine,article,pellaco2019spectrum}. Hence some authors resorted to simulated artificial data generated from software \cite{bie2019combined,han2019prediction,dong2019deep,qin2022interference}. Since accessing real satellite data is challenging, some of the literature utilizes data generated on the ground as a substitute. For example, traffic \cite{na2015research,zhu2020lstm}, housekeeping \cite{CCAD,e17085868}, and attack \cite{AccessIIOT,UNSWNB15} data on the ground were used for representing satellite data in the above papers. Additionally, the authors of \cite{na2018distributed} mentoioned an avaliable internet network traffic dataset :http://ita.ee.lbl.gov/index.html, which provides a collection of data related to network traffic, including packet-level information, network flow statistics, and other measurements. Although the data traces available on the website were last updated in April 2018, they still serve as valuable resources for studying network dynamics, usage characteristics, and growth patterns.}

	\begin{figure*}[ht]
		\centering
		
		\subfigure[]{
			\label{PA1}
			\includegraphics[width=0.3\textwidth]{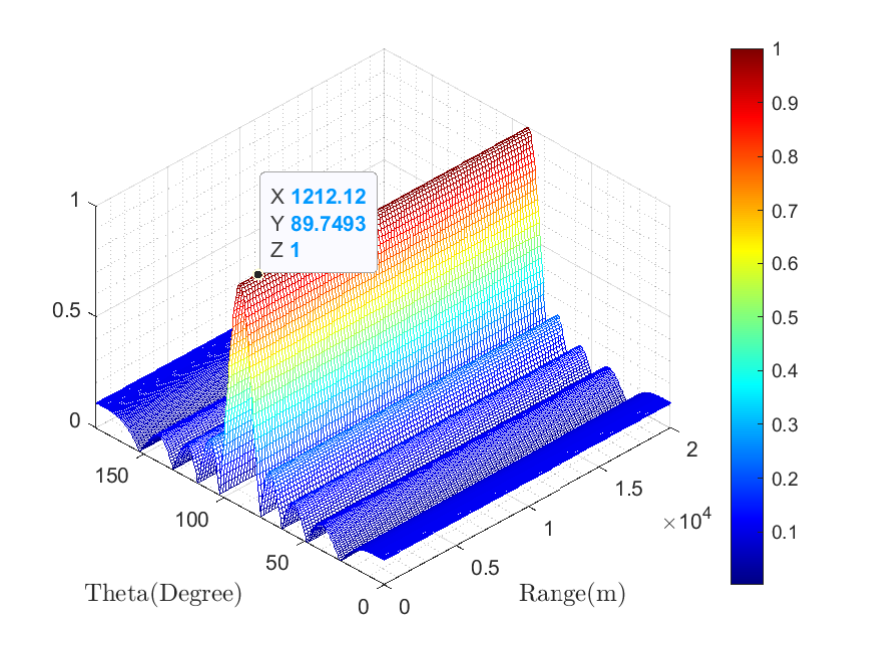}}	
		\subfigure[]{
			\label{LFDA1}
			\includegraphics[width=0.3\textwidth]{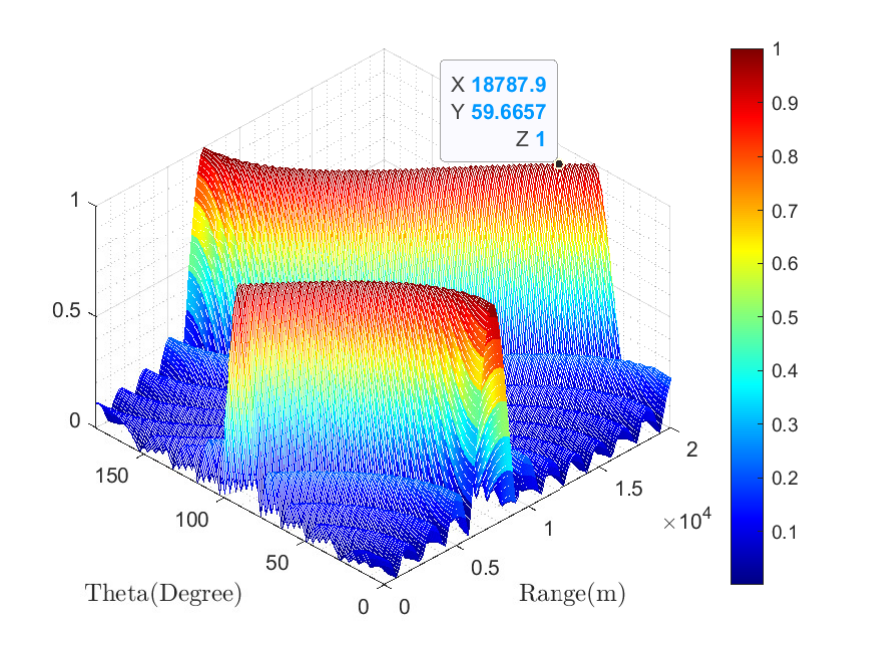}}		
		\subfigure[]{
			\label{RFDA1}
			\includegraphics[width=0.3\textwidth]{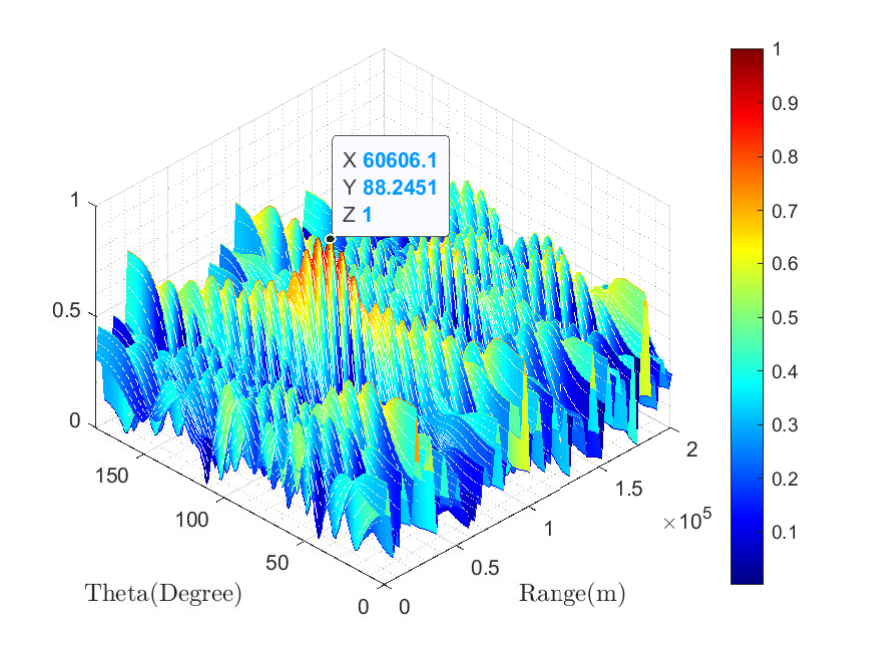}}
		
		\subfigure[]{
			\label{PA2}
			\includegraphics[width=0.3\textwidth]{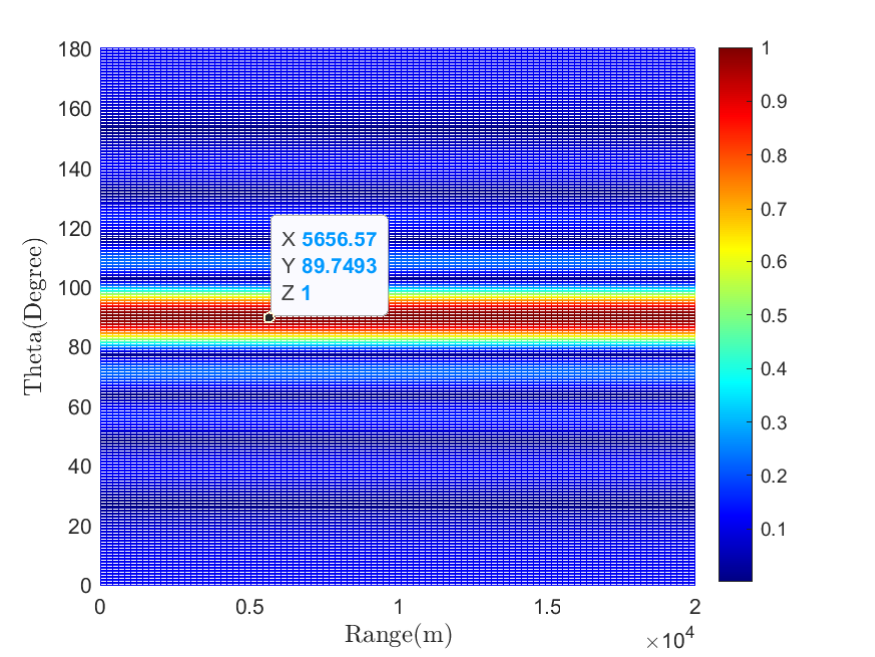}}	
		\subfigure[]{
			\label{LFDA2}
			\includegraphics[width=0.3\textwidth]{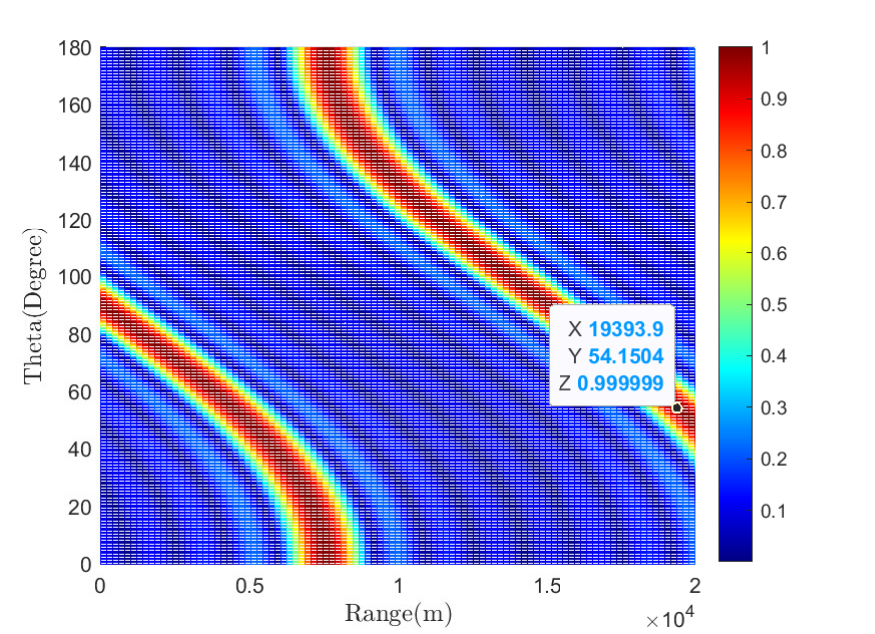}}	
		\subfigure[]{
			\label{RFDA2}
			\includegraphics[width=0.3\textwidth]{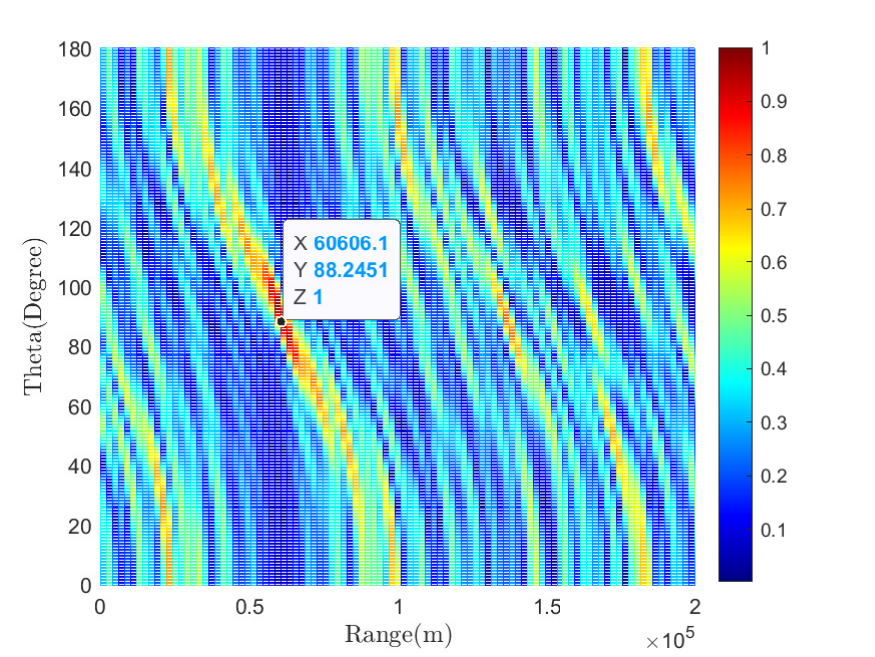}}
		
		\caption{The beam patterns of PA, LFDA, and logarithmic FDA. (a), (b), and (c) separately present the 3D beam pattern when $t=0$. While (d), (e), and (f) show the projection of their beam pattern on the distance-direction plane when $t=0$. The simulation parameters are as follows. The operating frequency $f_{c}$ and the antenna interval $d$ are given by 10 GHz and 0.15 m, respectively. The number of the array is 9. The $\Delta f$ of LFDA and logarithmic FDA is 20 kHz.}
		\label{beampattern}
	\end{figure*}
	
	\subsection{Passive Security Enhancement Solutions}
	Passive security enhancement solutions tend to rely on advanced security-oriented antennas, interference coordination, SS techniques, Non-SS jamming suppression techniques, reconfigurable intelligent surfaces, satellite cooperation, and AI tools, which are adopted for mitigating eavesdropping, CCI between systems, and malicious power-based jamming. These solutions are discussed in the following.
	
	\subsubsection{Advanced Security-oriented Antennas}
	Advanced security-oriented antennas combining PLS and multiple-antenna-aided techniques can effectively mitigate eavesdropping and power-based jamming, which are detailed below.
	
	\textbf{Eavesdropping Mitigation:} There are recent studies on advanced security-oriented antennas for secure transmissions since they are capable of reinforcing the radiation pattern in the direction of the desired receiver while suppressing the pattern in most of the other directions. However, an eavesdropper equipped with a sensitive receiver may still be capable of intercepting the communication link via a side lobe. To tackle this problem, side-lobe randomization\cite{ANTWC} may be used for alleviating side-lobe information leakage. The advanced security-oriented antennas employ BF, and Artificial Noise (AN) \cite{ANHanzo} in the downlink to transmit AN in the direction of eavesdroppers for actively suppressing eavesdropping \cite{ANTCOM2016, ANTVT2020, ANHanzo}.
	
	However, eavesdroppers may be able to penetrate the main-lobe direction anywhere between the LEO satellite and the Earth. It may frequently occur in LEO SCSs, because LEO satellites are always orbiting overhead, inevitably making eavesdroppers fall within the main-lobe direction. In this scenario, the Phased Array (PA) no longer works, as its beams are only angle-dependent. The Frequency Diverse Array (FDA)\cite{FDAattennaXIdian} can be employed to address this problem.
	
	The authors of \cite{FDAattenna} introduced a Linear Frequency Diverse Array (LFDA) that can generate a beam pattern depending on both the angle and the distance by linearly shifting the carrier frequencies across different antennas. However, the length and direction of the beam pattern generated are coupled. Hence it may still be possible for the eavesdropper to intercept the message of the legitimate user at certain positions. To tackle this problem, several kinds of non-Linear frequency offset schemes, including logarithmic offset, exponential offset, and random offset, are proposed for decoupling distance and direction of the beam pattern \cite{RFDA1, RFDA2, ShuFeng}. Fig.~\ref{beampattern} shows the beam pattern of PA, LFDA, and logarithmic FDA. Explicitly, recall from Fig.~\ref{beampattern} that the peak position of the PA is independent of distance. While the peak position of the LFDA and the logarithmic FDA is related to both the direction and the distance. Between them, the peak positions of LFDA are distributed as an `S' shape due to the coupling of distance and direction. Its beam is periodic, and the period is determined by the frequency offset $\Delta f$. But the logarithmic FDA can form a spot beam owing to its enabling distance and angle decoupling. Both of their detailed derivations are contained in \cite{FDAattenna, RFDA1}. Although the FDA is capable of providing additional security in the distance dimension, its beam pattern is time-variant, which limits its field of application \cite{wenqin2018}.
	
	\textbf{Jamming Mitigation:}
	The advanced security-oriented antennas also allow the beam pattern to be adjusted in response to power-based jamming conditions. Explicitly, the beam pattern can be adjusted in azimuth to minimize the jamming impinging from the left or right of an antenna or in elevation \cite{Antennas, Antennasspacetime, AntennasspacetimeTSP}.
	
	\begin{figure*}[]
		\centering
		
		\includegraphics[width=1.6\columnwidth]{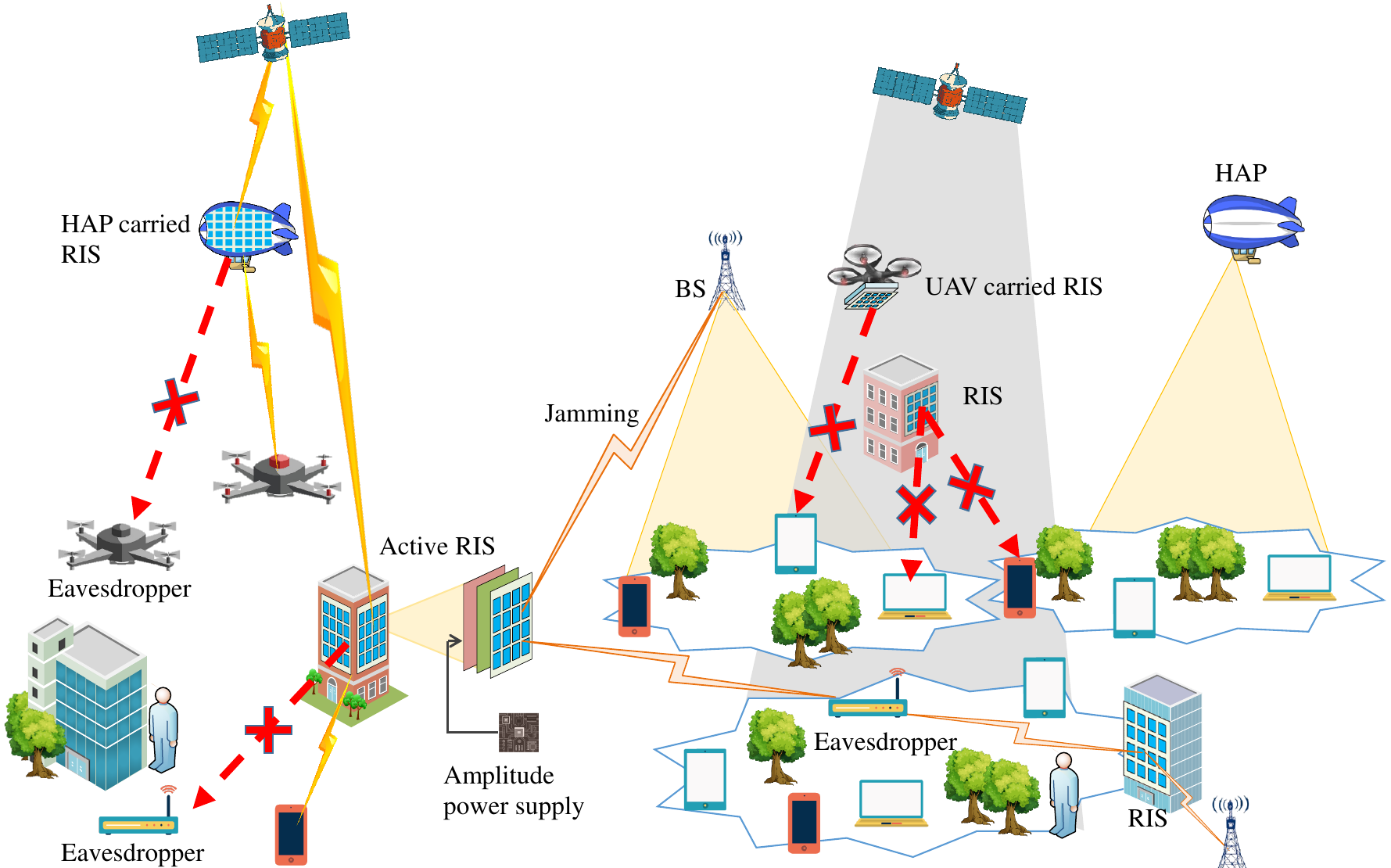}
		
		\caption{Application scenarios for RIS-enabled passive security enhancement solutions.}
		\label{RISpassive}
	\end{figure*}
	
	\subsubsection{Reconfigurable Intelligent Surfaces} The above techniques mainly rely on specifically designing the signals to prevent eavesdropping and mitigate interference. At the time of writing, the innovative technology of RISs has generated excitement in the wireless community, which is capable of beneficially ameliorating the wireless communication environment. Specifically, a RIS is capable of manipulating the phase and even the amplitude of advanced reflecting elements. This property allows the system to mitigate the blockage of the LoS component in satellite communication systems. Additionally, RISs can also mitigate security and interference problems.

	\textbf{Security Safeguards:}
	The preliminary contributions in the field of RIS-assisted secure satellite communications appeared in \cite{xu2021intelligent}, where the authors used a RIS to reflect the terrestrial interference signals to the eavesdroppers on the ground, who aimed for overhearing the satellite downlink transmission. The transmit beamformer weights of the terrestrial BS and the reflection coefficients of RIS are thus jointly designed to ensure that the interference generated can be tolerated by the satellite user while guaranteeing reliable satellite communication. As a benefit, the proposed RIS-assisted cooperative jamming strategy achieves lower SINR at the eavesdroppers than the conventional one operating without a RIS. This allows the RIS to enhance security. The authors also showed that the RIS having reflecting elements imposes increased jamming power on the eavesdroppers, thereby improving security. In addition to safeguarding the conventional satellite downlink,  the authors of \cite{huangdeep} proposed deploying a RIS in a full-duplex relaying aided satellite communication system.
	
	However, a terrestrial RIS cannot get close to the eavesdroppers and to objects flying in the air, thus it has eroded gains. Owing to the lightweight and conformal geometry of the RIS, a HAP carrying a RIS is proposed in \cite{yuan2022secure} for securing the communication link between an LEO satellite and a UAV receiver in the presence of a UAV eavesdropper. Even without the CSI knowledge of the eavesdropper, the legitimate user can still have a secure system by increasing the number of reflecting elements of the RIS in a hostile environment by simply maximizing the received signal power. 
	
	Additionally, the authors conceived a RIS optimization strategy to maintain a higher level of security, when either the statistical or the perfect CSI is known. The importance of a suitable RIS design is highlighted by characterizing a system at serious risk, namely when the RIS coefficients are random. The authors of \cite{yuan2022secure} also reveal the impact of the phase quantization at the RIS on the secrecy performance. Explicitly, they demonstrated that 3 bits are sufficient to avoid substantial secrecy degradation.
	
	As we mentioned before, the amplitude of reflected elements can be adjusted. However, this kind of RIS does not work in a passive manner, since it consumes additional power to bring about additional amplitude gain. It is therefore termed as an active RIS. Another difference with respect to the passive RIS is that the active one inevitably introduces thermal noise, which is also amplified along with the incident signal. The investigation of an active RIS in terms of securing satellite communication was conducted in \cite{ge2022active}, where a cooperative jamming strategy similar to that in \cite{xu2021intelligent} is adopted to allow the terrestrial network to secure the legitimate satellite downlink transmission under the assistance of an active RIS. However, the impact of an active RIS on the secrecy performance is not fully revealed in \cite{ge2022active}. The authors of \cite{wang2022secure} showed that a well-designed active RIS outperforms its passive counterpart when it comes to the secrecy energy efficiency of the conventional satellite downlink overheard by terrestrial eavesdroppers. However, it is still an open question, whether the security is enhanced by an active RIS. Moreover, the authors of \cite{wang2022secure} considered a GEO satellite without considering the unique mobility-induced propagation properties of LEO satellites. But again, the benefits of the active RIS over the passive one in securing LEO satellite systems, as well as the impact of the amplification power budget, are still awaiting further investigation. 				
	
	\textbf{Interference Mitigation:} %,
	Similar to protecting legitimate signals from eavesdroppers, RISs can also be utilized for mitigating both intra-system interference and CCI. The authors of \cite{cao2022robust} reveal the benefit of RIS in terms of improving the sum rate of LEO satellite systems, which is directly related to the SINR at the receivers. The superiority of the RIS in spectrum-sharing-based integrated terrestrial-LEO satellite networks was investigated by Dong \textit{et al.} in \cite{dong2021weighted,dong2021towards, dong2022intelligent}. The improvement of the received SINR becomes more pronounced upon increasing the number of RISs, the number of reflecting elements, and the phase shift resolution. The benefits of RIS were also observed in \cite{xu2021robust} where instead of a terrestrial network, a HAP-aided scenario was considered, where the SINR was the constraint rather than the optimization objective in the design of the RIS. This paper emphasized that both the channel estimation error and the multipath effect should be carefully addressed when designing the reflecting elements. The interference reduction capability of RIS was investigated in \cite{liu2022irs} for a UAV-mounted RIS (U-RIS). The U-RIS is shown to have the ability to enhance signal transmission within the terrestrial network, while mitigating the interference generated by the uplink signals transmitted from the ground stations to the satellite, thereby improving the SINR at the intended terrestrial users.
	
	{\it Lessons Learned}: 		
		The literature reviewed above has been summarized in Fig. \ref{RISpassive}. Observe that there is a paucity of literature on the security of LEO SCSs relying on RISs. Moreover, most of the existing literature where the RIS acts as the security safeguard or operates as the interference canceller either does not specify the type of the satellite at all or only targets GEO satellites. Hence, they ignore the effect of the limited above-the-horizon communication period, the frequent handovers, and the high Doppler shifts. Nonetheless, some of the above-mentioned contributions do allude to LEO satellites but do not accurately reflect their unique propagation properties, such as the time-variant received SNR, Doppler frequency shifts, and delay. The corresponding influence of these phenomena on the RIS design and on the performance achieved has not been investigated. Hence the employment of RIS-aided solutions in practical LEO systems requires substantial further research. Furthermore, security threats exist not only in satellite-terrestrial systems but also in satellite-satellite networks. 	
		
		Considering the high mobility and limited access time of LEO satellites, the working time of RISs deployed on the Earth or UAVs is also limited. Similar to RISs deployed on UAVs, it can also be considered to install RISs on LEO satellites, which undoubtedly increases the weight of the satellite and launch costs. Additionally, the hash space environment, such as intense temperature changes and cosmic rays, has to be considered. In a nutshell, the security of RIS-aided LEO satellites requires further exploration.
	
	\begin{table*}[ht]
		\tiny
		%	\scriptsize
		\centering
		\renewcommand\arraystretch{1.3}
		%		\begin{tabular}{|l|l|l|l|l|l|}
			\caption{The evolution of interference coordination}
			\begin{tabular}{|m{0.5cm}<{\raggedright}|m{0.6cm}<{\raggedleft}|m{2.4cm}<{\raggedleft}|m{1.4cm}<{\raggedleft}|m{4.4cm}<{\raggedleft}|m{5.5cm}<{\raggedleft}|}
				\hline
				\hline
				Year                  & Ref.                                       & Target Problem          & Ways                             & Proposed algorithm/scheme                            & Results                                                          \\ \hline
				\multirow{2}{*}{2015} & \multirow{2}{*}{\cite{powercontrol2015}}   & CCI between LEO SCSs    & \multirow{2}{*}{Power control}   & Presents three different efficient                   & Strikes a clear trade-off between channel                        \\
				&                                            & and terrestrial systems &                                  & power control methods                                & state information and rates.                                     \\ \hline
				\multirow{2}{*}{2016} & \multirow{2}{*}{\cite{powercontrol2016}}   & CCI between LEO SCSs    & \multirow{2}{*}{Power control}   & Investigates optimization approaches  to solve        & Formulates a multi-objective optimization                        \\
				&                                            & and terrestrial systems &                                  & the power and rate allocation problems               & problem and provides a  Pareto-optimal solution                  \\ \hline
				\multirow{2}{*}{2017} & \multirow{2}{*}{\cite{ModulationGEOLEO}}   & CCI between LEO         & Modulation                       & Presents a method combining modulation          & Improves the throughput of LEO SCSs compared                     \\
				&                                            & SCSs and GEO SCSs       & coding                           & and coding  based on power control                   & with traditional power control method                            \\ \hline
				\multirow{2}{*}{2018} & \multirow{2}{*}{\cite{LEOGEOCC}}           & CCI between LEO         & \multirow{2}{*}{Beam drifting}   & Presents an optimal method by tilting                & Guarantees the signal level of LEO satellite                     \\
				&                                            & SCSs and GEO SCSs       &                                  & the direction of PA of LEO satellite                 & with a simple method                                             \\ \hline
				\multirow{2}{*}{2018} & \multirow{2}{*}{\cite{Ruiding}}            & CCI between LEO         & \multirow{2}{*}{Beam drifting}   & Proposes an exclusive angle                          & Reduces the CCI level sacrificing the                            \\
				&                                            & SCSs and GEO SCSs       &                                  & strategy for CCI mitigation                          & coverage of LEO satellites                                       \\ \hline
				\multirow{2}{*}{2019} & \multirow{2}{*}{\cite{Broadband}}          & CCI between LEO         & \multirow{2}{*}{Beam drifting}   & Turns off the current beam and expands   & Ensures the throughput of LEO SCSs                               \\
				&                                            & SCSs and GEO SCSs       &                                  & its adjacent beam to take place          & as well as CCI mitigation                                        \\ \hline
				\multirow{2}{*}{2019} & \multirow{2}{*}{\cite{powercontrol1}}      & CCI between LEO         & \multirow{2}{*}{Power control}   & Proposes an adaptive beam power                      & Maximizes the throughput of LEO SCSs under the                   \\
				&                                            & SCSs and GEO SCSs       &                                  & control method based on optimization                 & premise of that the signal quality of GEO SCSs.                 \\ \hline
				\multirow{2}{*}{2019} & \multirow{2}{*}{\cite{ProgressivePitch}}   & CCI between LEO         & \multirow{2}{*}{Beam drifting}   & Adjusts the angle of the spot beams or even          & Reduces the CCI level with the limited                           \\
				&                                            & SCSs and GEO SCSs       &                                  & turns off some spot beams of LEO satellites          & throughput of LEO SCSs                                           \\ \hline
				\multirow{2}{*}{2019} & \multirow{2}{*}{\cite{CRTVT2019}}          & CCI between LEO         & \multirow{2}{*}{Cognitive radio} & proposes detailed spectrum strategies                & Adjusts the transmit power of LEO SCSs according                 \\
				&                                            & SCSs and GEO SCSs       &                                  & to detect the presence of the GEO SCSs               & to the signal power level of GEO SCSs                            \\ \hline
				\multirow{2}{*}{2020} & \multirow{2}{*}{\cite{CRDefination4}}      & CCI between LEO SCSs    & \multirow{2}{*}{Cognitive radio} & Integrates a distributed cooperative sensing         & Strikes a trade-off between the average throughput               \\
				&                                            & and terrestrial systems &                                  & network with satellite terrestrial network & and the average energy consumption                               \\ \hline
				
				\multirow{2}{*}{2020} & \multirow{2}{*}{\cite{CRPCAccess}}         & CCI between LEO         & Cognitive radio                  & Proposes an optimal method by combining              & Maximizes the throughput of LEO SCSs with power allocation  \\
				&                                            & SCSs and GEO SCSs       & power control                    & spectrum sensing and power allocation                & after optimizing the sensing time and the sensing interval             \\ \hline
				
				\multirow{2}{*}{2020} & \multirow{2}{*}{\cite{DLTVT2020}}          & CCI between LEO         & \multirow{2}{*}{Deep learning}   & Proposes a DL aided                       & Adjusts the operating frequency of LEO SCSs to avoid the CCI     \\
				&                                            & SCSs and GEO SCSs       &                                  & spectrum prediction method                           & by  diging the historical spectrum data of the GEO SCSs          \\ \hline
				\multirow{2}{*}{2021} & \multirow{2}{*}{\cite{BHPC}}               & CCI between LEO         & Beam hopping                     & Proposes a joint beam hopping                        & Maximizes the throughput of LEO SCSs  under the                  \\
				&                                            & SCSs and GEO SCSs       & power control                    & and power control scheme                             & premise of ensuring the signal quality of GEO SCSs               \\ \hline
				\multirow{2}{*}{2021} & \multirow{2}{*}{\cite{PowercontrolMinJia}} & CCI between LEO         & \multirow{2}{*}{Power control}   & Jointly optimizes the transmit power                 & Maximizes the throughput of LEO SCSs  under the                  \\
				&                                            & SCSs and GEO SCSs       &                                  & of LEO and GEO satellite beams                       & premise of ensuring the signal quality of GEO SCSs               \\ \hline
				\multirow{2}{*}{2021} & \multirow{2}{*}{\cite{CRTWC1}}             & CCI between LEO         & \multirow{2}{*}{Cognitive radio} & Proposes a low-complexity cognitive                  & Enhances  the throughput of LEO SCSs  under the                  \\
				&                                            & SCSs and GEO SCSs       &                                  & radio technique for CCI mitigation                   & premise of ensuring the signal quality of GEO SCSs               \\ \hline
			\end{tabular}
		\end{table*}

		\subsubsection{Interference Coordination} Interference coordination is a promising technique for mitigating the CCI between systems caused by the spectrum crunch. It typically mitigates interference by power control, beam drifting, cognitive radio techniques, etc., while improving spectral efficiency and meeting the ever-increasing capacity demands. Its evolution is chronologically arranged in Table \uppercase\expandafter{\romannumeral9}.
		
		The ITU specifies that GEO SCSs have priority over LEO SCSs with regard to frequency usage. Consequently, accurate power control is required in LEO SCSs to satisfy the interference constraints imposed by GEO SCSs. However, the power control also directly affects the throughput of LEO SCSs \cite{powercontrol2015, powercontrol2016}. As a remedy, the authors of \cite{powercontrol1} modeled this power control problem as an optimization problem aiming to maximize the sum rate of the LEO SCSs. Then, the popular fractional programming technique was employed to transform this nonconvex problem into a tractable form. By contrast, the authors of \cite{PowercontrolMinJia} conceived a joint multi-beam power control algorithm for optimizing the transmit power of LEO and GEO satellite beams. On the premise of ensuring the signal quality of GEO SCSs. This algorithm judiciously reduced the transmission power of GEO beams, thereby maximizing the throughput of LEO SCSs.
		
		Some schemes rely on so-called beam drifting in LEO SCSs, which force the LEO satellite users into the adjacent beam even before interference actually occurs \cite{Ruiding, Broadband, LEOGEOCC, ProgressivePitch}. The authors of~\cite{Ruiding} conceived a sophisticated strategy for reducing the downlink interference inflicted by LEO satellites on GEO satellite users. The authors of~\cite{LEOGEOCC} mitigated the interference between LEO and GEO satellites by appropriately tilting the transmission direction of the PA-based antennas of LEO satellites by solving a nonlinear programming problem used for finding the optimal direction. OneWeb adopted the method of \cite{ProgressivePitch} for LEO SCSs to avoid the risk of interference with GEO SCSs operating at the same frequency. Specifically, when an interference event occurs, some beams are briefly turned off as they cross the equator. Subsequently, when the LEO SCSs exit the GEO SCSs exclusion zone, the specific beams which were turned off are turned back on again. In the context of a hybrid-beam coverage scheme\footnote{There is a wide beam providing coverage for the whole service area and several spot beams for tracking users in each LEO satellite. The gain of a spot beam is designed to be much higher than that of a wide beam. Hence the spot beam is provided for supporting data transmission, while the wide beam is fixed and is suitable for control signals.}, the authors of \cite{Broadband} also proposed a so-called coverage-extension method for beam drifting, which relies on expanding the wide beam to cover the serving areas of adjacent satellites. When the coverage area of an LEO satellite is overlapped with that of the adjacent satellites, one of them can be turned off to avoid potential interference.
		
		Given the ever-increasing deployment density of LEO mega-constellations, a spectrum crunch is imminent. Cognitive radio \cite{InterferenceCR,CRHanzo,CRref1} techniques are capable of mitigating this problem. In cognitive radio networks, PUs have a higher priority or legacy rights on the usage of a specific spectrum. SUs, which have a lower priority, should not cause interference with PUs. Hence SUs must have cognitive radio capabilities for adapting their communications channel access to the dynamic environments in which they operate. Explicitly, cognitive radio devices can sense, detect, and monitor the surrounding opportunities, including spectrum, time, geographical space, code, as well as angle \cite{CRDefination1} and reconfigure the operating characteristics to best match those opportunities.
		
		Cognitive radios are capable of making autonomous real-time decisions for mitigating the spectrum scarcity problem in SAGINs. The authors of \cite{CRTVT2019} proposed a spectrum sensing scheme for LEO SCSs capable of mitigating the inter-system interference between GEO and LEO SCSs. Upon identifying the specific power level utilized by the GEO SCSs after differentiating the GEO signal from the interfering LEO signal and noise, the authors of \cite{CRTWC1} conceived a cognitive radio technique for improving the throughput of LEO SCSs, while guaranteeing that the signal quality of GEO SCSs can be satisfied. By applying sophisticated relaxation and approximation schemes, they significantly reduced the complexity of the related optimization problem. The authors of \cite{CRDefination4} proposed a cognitive satellite-terrestrial network relying on a distributed cooperative spectrum sensing technique by striking a trade-off between the average throughput and the average energy consumption under specific interference constraints.
		
		Additionally, the authors of \cite{CRPCAccess} conceived a two-stage spectrum-sharing framework by combining the advantages of cognitive radio and power control techniques. This framework jointly optimizes the spectrum sensing time and the LEO SCSs transmit power with the objective of enhancing spectral efficiency and seamless coexistence. The authors of \cite{BHPC} proposed a joint beam hopping and power control scheme for maximizing the throughput of LEO SCSs, while preserving the signal quality of GEO SCSs. A DL-aided spectrum prediction method was proposed in \cite{DLTVT2020} for mitigating the inter-system interference. A sophisticated combination of a convolutional neural network and a carefully dimensioned bespoke memory was harnessed for data mining from the historical spectrum usage of the GEO SCSs. This technique was used for predicting future spectral occupancy. Furthermore, an adaptive modulation and coding method was adopted in \cite{ModulationGEOLEO} for interference mitigation. Specifically, this method adopted the angle between LEO and GEO satellites for controlling the specific choice of modulation and coding scheme, with the objective of improving the spectral efficiency of LEO SCS, while limiting the interference inflicted upon the GEO SCSs to the maximum tolerable limit.
		
		{\it Lessons Learned:}
			Considering the high mobility and limited over-the-horizon time of LEO satellites, the active operating duration of RISs deployed on the Earth or UAVs is also limited. Similar to RISs deployed on UAVs, it may also be feasible to install RISs on LEO satellites, which undoubtedly increases the weight of the satellite and launch costs. Additionally, the hash space environment exhibiting intense temperature changes and cosmic rays has to be considered. In a nutshell, the security of RIS-aided LEO satellites requires further exploration.
			
			In fact, the establishment of standards at regional and global levels facilitates the efficient and economical use of the spectrum and the development of radio services. Hence the ITU provides a regional framework that allows sovereign nations to submit and discuss their spectrum requirements in different regions (Regions 1, 2, and 3). There are regional and sometimes country-specific differences in
			the way that spectrum band plans and radio system techniques are deployed. Because of the fixed orbital position of GEO satellites, GEO satellites can clearly be dealt with on a regional or country-by-country basis. Additionally, interference issues are generally related to fixed entities and are reasonably easy to manage.
			
			By contrast, LEO satellites fly over many regions and countries, requiring them to comply with many different regulatory regimes to allow them to provide services to users. Furthermore, they have to implement sophisticated interference coordination solutions relying on power control, CR, and beam drifting for mitigating CCI imposed on terrestrial communication systems and GEO SCSs, as required by ITU regulations.
			
			However, with the continued proliferation of LEO  mega-constellations, such as SpaceX, OneWeb, and Lightspeed, the CCI between these different constellations jostling for room in LEOs also has to be addressed, and even the interference between different orbital layers within the same constellation has to be given cognizance.
			
			AI is eminently suitable for analyzing the frequency usage in various regions by relying on spectrum sensing and combining the results of satellite flight trajectory prediction to assist interference coordination, making it a topic worth investigating.

		\begin{figure*}[]
			\centering
			
			\includegraphics[width=2.2\columnwidth]{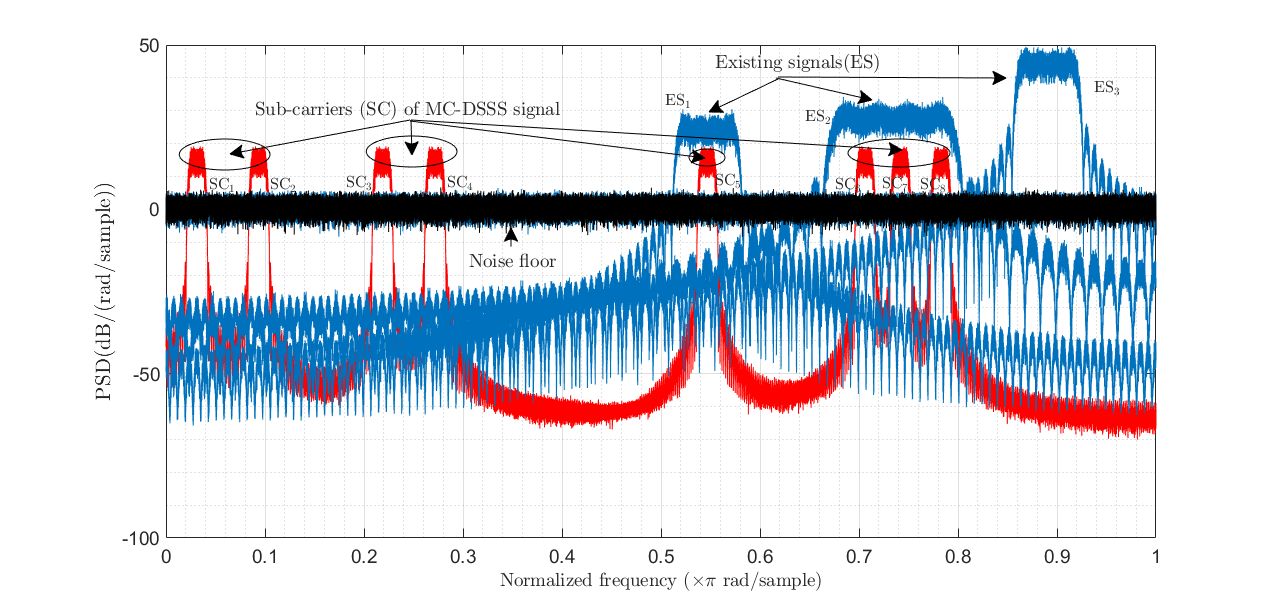}
			
			\caption{The trade-off between confidentiality and integrity in MC-DSSS systems.}
			\label{MCDSSS}
		\end{figure*}
		
		\subsubsection{SS Techniques}
		SS techniques have been routinely adopted as one of the secure techniques in military communications for more than 70 years \cite{SS1}, where the transmitted signal is spread to a much wide bandwidth than the information bandwidth. The common SS techniques include DSSS, Frequency Hopping Spread Spectrum (FHSS), and Multi-Carrier Direct Sequence Spread Spectrum (MC-DSSS). Unless the eavesdropper steals the random Frequency Hopping (FH) pattern or spreading code, it fails to detect the confidential information \cite{hanzo2003single}.
		
		Again, DSSS has been widely used in satellite communications \cite{DSSS1}. DSSS technique can prevent eavesdropping, thus guaranteeing confidentiality. Typically, the PSD of DSSS signal is low, and the received signal may be submerged in noise when arriving at the receiver, making it difficult for adversaries to eavesdrop. On the other hand, the DSSS technique is also immune to jamming to a certain extent. Whenever jamming contaminates the legitimate signal, the receiver correlator spreads the jamming to the entire bandwidth after despreading because the jamming and the local pseudo-noise code are uncorrelated. By contrast, the legitimate signal is despread back to its original narrower bandwidth. The Signal to Noise Ratio (SNR) of the baseband data increases after despreading by a factor of the Processing Gain (PG). By contrast, the PSD of jamming remains low in the baseband. Hence, the anti-jamming ability also depends on the PG. However, the payload rate is given by the ratio of the bandwidth and the spreading factor, which explicitly indicates the traffic rate versus anti-jamming capabilities trade-off in LEO SCSs. More specifically, when the jamming is strong, the DSSS sequence length should be increased to improve the anti-jamming capability controlled by its PG, hence leading to throughput reduction and $vice$ $versa$.
		
		Furthermore, FHSS constitutes another popular anti-jamming technique. In contrast to DSSS, the FHSS transceiver continuously jumps from one sub-carrier frequency to another during transmission according to the SS code. Hence, the FHSS signal bandwidth may be composed of discontinuous frequency bands, and it is often combined with cognitive radio techniques to avoid jamming at locations subject to severe jamming whilst relying on adaptive frequency hopping.
		
		Hopping across multiple frequencies within a single symbol leads to the concept of Fast Frequency Hopping Spread Spectrum (FFHSS). More explicitly, the dwell time of each hop is shorter than the symbol duration, and multiple frequency hops are completed within a single symbol duration, leading to strong anti-jamming capability. FFHSS may rely on low-complexity non-coherent dehopping and demodulation methods, but this results in a substantial loss of SNR \cite{noncorherentFHSS1,noncorherentFHSS2}. By contrast, the coherent reception of FFHSS exhibits better performance \cite{corherentFHSS}, at a substantially increased complexity.
			
			Compared to DSSS, the MC-DSSS receiver employs spectrum sensing to monitor and analyze the current operating frequency in real-time, identifying the available frequency bands (left side of Fig.~\ref{MCDSSS}) as well as occupied frequency bands (right side of Fig.~\ref{MCDSSS}), and flexibly devising sub-carrier allocation schemes. The MC-DSSS receiver can choose to avoid the existing signals to improve integrity by allocating each sub-carrier to available frequency bands. However, these signals are also easily eavesdropped by attackers, resulting in decreased confidentiality. By contrast, these sub-carriers could also be actively hidden in some of the existing signals to improve confidentiality. Specifically, the operating frequency of each sub-carrier can be set to the same as the existing signal. However, legitimate signals are also contaminated by existing signals, which undoubtedly degrades integrity. Fig.~\ref{MCDSSS} shows the MC-DSSS waveform with 8 sub-carriers, where four sub-carriers are allocated to the available frequency band and the other four sub-carriers are actively hidden in the existing signals. The SNR of each sub-carrier is $E_{\mathrm{s}}/N_{\mathrm{0}} = 10$ dB, and the Interference-to-Signal Ratio (ISR) of three existing signals ($\mathrm{ES}_{1}$,$\mathrm{ES}_{2}$,$\mathrm{ES}_{3}$ in Fig.~\ref{MCDSSS}) with each subcarrier interference signal is $18$, $38.5$, and $25$ dB, respectively.
			
\begin{figure}[ht]
			\centering
			
			\includegraphics[width=1.0\columnwidth]{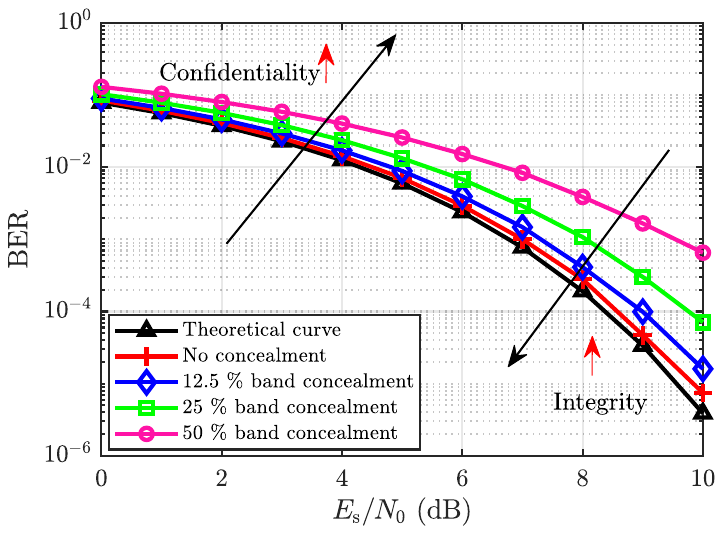}
			
			\caption{The variation of BER with band concealed in existing signals.}
			\label{MCDSSS1}
		\end{figure}

			To further illustrate this trade-off, we simulated the Bit Error Rate (BER) of MC-DSSS shown in Fig.~\ref{MCDSSS}. Taking the simulation conditions of the 8 sub-carriers in Fig.~\ref{MCDSSS}  as an example, we tested the BER with 0 to 4 sub-carriers actively hidden in the existing signals, and the BER is plotted in Fig.~\ref{MCDSSS1}. As shown in Fig.~\ref{MCDSSS1}, the BER degrades as the fraction of the total frequency band concealed in the existing signals from 12.5~\% to 50~\%.

		\subsubsection{Non-SS Jamming Suppression Techniques} When the jamming power exceeds the maximum tolerance level of the SS receiver, the SS system has to employ dedicated jamming suppression algorithms, such as temporal domain adaptive filtering \cite{LMS2} and transform domain adaptive filtering \cite{TransformSNR}.
		
		Temporal domain adaptive filtering algorithms are suitable for narrowband jamming suppression. The Least Mean Square (LMS) \cite{LMS1, LMS2} algorithm is a popular design option due to its low complexity. The basic idea behind the LMS algorithm is to mimic a causal Wiener filter by updating the filter weights until the least mean square of the error signal is approached. It is a stochastic gradient descent method, which means that the filter weights are only adapted based on the error at the current symbol instant. For a standard LMS algorithm, the convergence speed is determined by the step size parameter ($\mu$), which may be gradually reduced upon approaching convergence to the minimum.
		
\begin{figure}[ht]
			\centering
			
			\includegraphics[width=1\columnwidth]{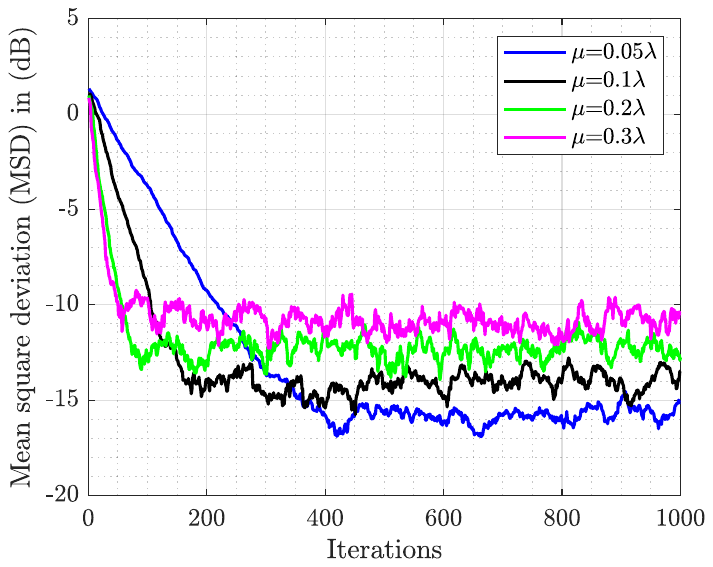}
			
			\caption{The convergence performance of weights. The number of taps is 4. $\mu $ is given as $0< \mu <\frac{2}{\lambda}$, where $\lambda$ is the greatest eigenvalue of the autocorrelation matrix $R=E\left\{X(n)X^{H} (n)\right\} $.}
			\label{LMS111}
		\end{figure}

		On the one hand, the higher the value of $\mu$, the faster the weights converge. Hence, we can promptly track and mitigate the fluctuating jamming. On the other hand, the higher $\mu$, the higher the variance of the weights will be, which affects the performance of jamming mitigation. Therefore, the realization of the LMS algorithm requires a trade-off, as seen in Fig.~\ref{LMS111}.
		
		By contrast, transform domain adaptive filtering is capable of promptly tracking the fluctuation of narrowband jamming without an iterative process \cite{Transformlowrank,LMSRL}. Transform domain adaptive filtering processes the received signal in the frequency domain. Briefly, it identifies the jamming and carries out the band-pass filtering before transforming the signal back to the temporal domain.
		
		\begin{figure*}[ht]
			\centering
			\subfigure[The scenario of single satellite]{
				\label{SingleRe1}
				\includegraphics[width=0.45\textwidth]{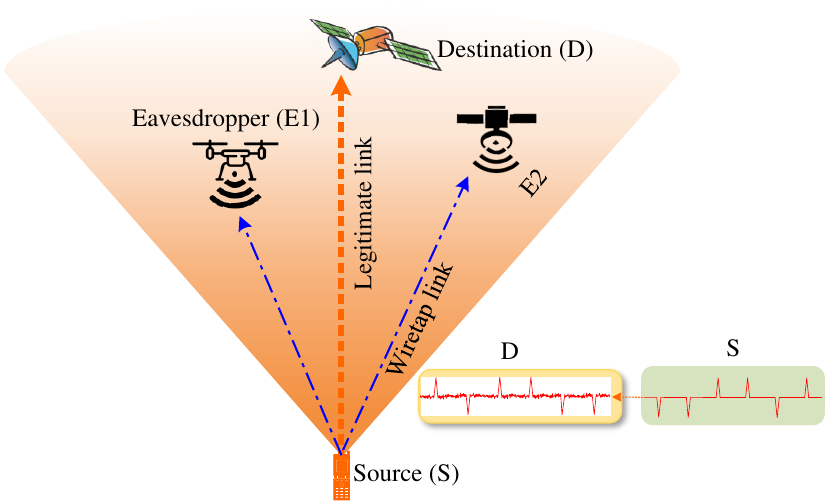}}	
			\subfigure[The scenario of multiple satellite cooperation]{
				\label{MulSatRe}
				\includegraphics[width=0.5\textwidth]{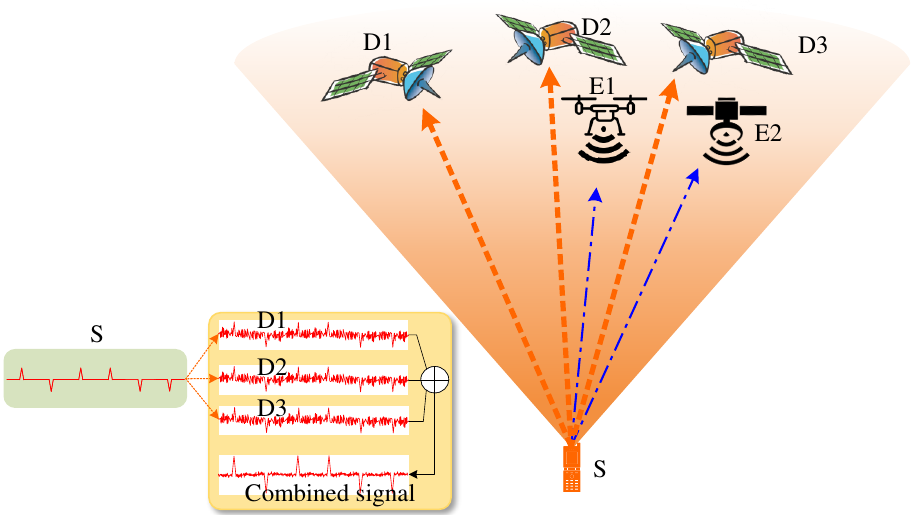}}	
			\caption{The concept of satellite cooperation.}	
			\label{MulSatRe11}
		\end{figure*}
		
		\subsubsection{Satellite Cooperation} 
			
			With the explosive proliferation of connected terminals and the emergence of diverse applications, a single satellite can hardly satisfy the confidentiality and integrity requirements simultaneously, especially for LEO satellites having limited onboard resources. 
			
			Again, the open nature of wireless propagation makes legitimate transmissions vulnerable to eavesdropping, as seen in Fig.~\ref{SingleRe1}. The source transmits its signal to D, while E is capable of overhearing the legitimate transmissions if it is located in the coverage area of S.

			Specifically, to improve the confidentiality of S, a common strategy is to reduce its transmit power. However, the system's integrity will also be reduced simultaneously. Conversely, by increasing the transmit power to improve the legitimate link's integrity, the probability of eavesdropping will also be inevitably increased \cite{YulongTradeoff}. This indicates a clear trade-off between confidentiality and integrity, which has been a long-term challenge for satellite communication researchers. 	
			
			With the continuous deployment of LEO mega-constellations, such as Starlink \cite{StarlinkTAES2023}, the number of satellites in space is increasing. Most locations on Earth's surface can be simultaneously covered by multiple satellites in these LEO mega-constellations, where a terminal can be covered by multiple satellites at the same time, as seen in Fig.~\ref{MulSatRe}. By beneficially combining the signals received from each satellite, the integrity of the combined signal can be maintained while allowing the terminal to transmit at lower power, thereby successfully tackling the previous trade-off between confidentiality and integrity. However, before combining, the delay, Doppler frequency shift, and phase offset of the signal received from each satellite have to be compensated. 
			
			Typically, each satellite can exploit the location information of the terminal to predict and compensate for the delay and Doppler frequency shift. However, due to the differences in satellite oscillators and startup times, estimation is the only way to deal with the phase offsets among these satellites. A modified SUMPLE \cite{Sumple} algorithm based on correlating the received signals was proposed in \cite{yue2022collaborative} to estimate and compensate the phase offsets, which is capable of coherently combining the DSSS signals received at different satellites. The authors of 	\cite{yue2022collaborative} also mentioned some promising future research directions, such as the exploration of cooperative detection and localization. 
		
		There are a number of LEO satellite cooperation research works that are based on the compensation of Doppler frequency shift, delay, and phase offset. In particular, the signals transmitted by several visible satellites to the target terminal can be combined through a specific combining scheme for reducing the jamming signal power \cite{weerackody2021satellite}, for enhancing the sensing accuracy \cite{anttonen2021space}, and for improving the overall performance \cite{yue2022collaborative}.  Specifically, the authors of \cite{weerackody2021satellite} proposed to combine all user-satellite links, where the transmit power of the user terminal targeted at each satellite was jointly optimized for maximizing the total data rate. The experimental results of \cite{weerackody2021satellite} demonstrated that the deleterious effects of jamming can be mitigated when at least 3 satellites are available. This demonstrated the benefits of cooperation diversity against jamming. 
		
		Furthermore, due to the uneven distribution of users over the world, some LEO satellites do not have sufficient resources to serve the users in their coverage, while some other satellites may have unused resources. As the number of satellites has proliferated, the concept of satellite collaboration has attracted research attention in the interest of resource-sharing.
		
		In general, cooperation among satellites is supported by information transmission among satellites through ISLs. For instance, in \cite{hao2021key}, the CSI estimated at each satellite is shared with others in the same constellation, which has the advantage of increasing the equivalent aperture of the satellite antenna. This sophisticated measure is capable of reducing the correlation between the legitimate channel and the wiretap channel, which has a positive impact on the PLS. In this context, the LEO satellites investigated in \cite{wang2020routing} served as a trusted relay to cooperatively realize QKD transmission between intercontinental ground stations. It was demonstrated that satellites deployed in different orbits were beneficial while placing more satellites in the same orbit did not introduce substantial security performance gains. The benefits of relaying satellites were also investigated in \cite{cui2022latency}, where an LEO satellite forwarded its task to another collaborative satellite or ground gateway via a GEO satellite to achieve load balancing among LEO satellites. The task offloading, communication, and computing resource allocation were jointly optimized for minimizing the task processing latency by a bespoke deep reinforcement learning solution in support of confidential delay-sensitive services.

		Additionally, the routing mechanism of satellite networks composed of a multitude of satellites also needs cooperation among satellites and the construction of ISLs. The cooperation mechanism of satellites should be carefully designed to ensure secure routing while taking both their dynamic topologies, constrained resources, and large coverage area into consideration. In the existing literature, routing is mainly secured by cryptography, and by some trust mechanism. More explicitly, typically encryption is utilized to ensure that the routing information transmitted among nodes achieves the required degree of confidentiality. Specifically, in \cite{yin2010qos}, hop-by-hop encryption is proposed for securing the multicast routing within a three-layer satellite network composed of both LEO, highly elliptic orbit (HEO), and GEO satellites. Briefly, each intermediate satellite node encrypts the routing packets with the aid of its own private key, which can be decrypted through its public key. In this case, the transmitted packets are safeguarded from malicious nodes, since the latter cannot access the routing information without the right key.
		
		From a trust mechanism perspective, the trust concerning a specific routing path is determined by the degree of trust attributed to the network nodes, which is typically inferred from their previous behavior. In general, the node having a higher degree of trust has a higher probability of being selected, while a node associated with a low degree of trust might be removed from the network. In fact, different satellite networks calculate the degree of trust based on various standards. For instance, Yu {\itshape et al.} \cite{yu2012trust} monitor both the packet forwarding rate and that of abnormal behaviors to evaluate the degree of trust in a cluster-based multi-layer satellite network. They also demonstrated the robustness of their technique against DoS and disruption attacks. The so-called signature mechanism, which is one of the cryptography algorithms, is utilized in the routing process to prevent packets from being tampered with or imitated. By contrast, the degree of trust in a micro-nano satellite node was evaluated by another metric, namely by observing the attacking behavior, and the residual energy \cite{ding2020secure}. To improve the robustness of the trust mechanism, the estimated degree of trust having a value lower than a certain threshold can be re-evaluated by a different node relying on the same evaluation standards. The degree of trust evaluated in a centralized manner in \cite{cai2020research} is based on the average estimated by the neighbors of each node according to the previous behaviors. Although this method requires more computations, it has a higher attack identification capability. Additionally, a load-balancing strategy is adopted in \cite{cai2020research} when performing path selection, which aims for preventing high-trust nodes from becoming congested, while others remain idle. However, both the cryptography-based and trust-based routing reviewed above relies on single-path routing, hence suffering from a high path failure risk. As a remedy, multi-path routing is proposed in \cite{zeng2021failure} for reducing the risk by transmitting redundant data copies along multiple paths at the same time, which outperforms its single-path counterpart by avoiding re-routing.
		
		In a nutshell, satellite cooperation generally outperforms stand-alone satellite operations, but there are numerous open problems awaiting further exploration.
		
		{\it Lessons Learned:} As it transpires from the literature, the important prerequisites of LEO satellite cooperation are the compensation of Doppler frequency shift, delay, and phase offset, which requires the synchronization of multiple satellites. However, the dimensionality of these tasks grows exponentially with the number of satellites. To make things worse, the high mobility of LEO satellites lead to an excessive range of Doppler frequency shift. Therefore, both the range and dimensionality of these tasks constitute severe challenges for the synchronization of multiple satellites. However, based on the existing solutions, the satellite exploits the known terminal location to predict and compensate for both the Doppler frequency shift and the delay, thus avoiding the explicit direct estimation of these two parameters. However, the user terminal location constitutes an important aspect of user privacy. Hence the synchronization of multiple satellites is worth exploring.

		%		Although a large number of LEO satellites provide convenience for security attacks, they also provide new solutions for improving security and reliability. On the one hand, as long as the Doppler frequency shift, delay, and phase offsets in different replicas of the uplink signals received by each satellite from the same terminal are estimated and compensated, coherent combining of multiple satellites can be achieved. The combined signal with improved SNR allows terminals to transmit at a lower power level, which deals with the trade-off between confidentiality and integrity in the previous standalone satellite.
		%			
		%		Multiple LEO satellites can integrate with blockchain technique with the merits of decentralized storage and processing for enhancing data security. Specifically, blockchain can distribute data across multiple LEO satellites, thus avoiding single-point failures. Even if some satellites are attacked, others can still maintain data security, ensuring the integrity of the data. But a large amount of personal privacy may be exposed to external attackers during this process. Hence there should exist a trade-off between data security and user privacy protection.}
	
	%低轨卫星多星协作充分发挥作用的前提是多普勒频偏，延时以及相位的估计和补偿，即多星同步。
	
	%事实上，这些参数的估计维度随着卫星数量的增加呈指数增长。然而，现有工作均是卫星利用已知的终端位置对多普勒频偏和延时进行预测和补偿，从而规避了这两种参数的估计。在此基础上对载波相位进行估计和补偿。事实上，终端位置属于用户隐私，在某些特殊场景，比如军事应用场景。在这种条件下，卫星无法获得终端位置先验信息也就无法对多普勒频偏和延时进行预测和补偿。事实上，在卫星位置已知条件下，终端到达每颗星的延时和多普勒频偏的纽带是终端位置，因此只要估计出终端位置，也就能得到多普勒频偏和延时的估计，降低了多普勒频偏和延时的估计维度，这是一个值得研究的话题。
	
	\subsubsection{Artificial Intelligence}
	Although the above-mentioned passive security enhancement solutions are indeed capable of enhancing the security of LEO SCSs, they often lead to challenging non-convex optimization problems under strict performance constraints, and to optimization objectives involving coupled variables. Hence only a near-optimal performance can be obtained at the cost of high computational complexity. Many of the existing strategies are difficult to implement in practice, especially when considering that LEO satellites have limited energy resources and computational capability. In the face of uncertainty, AI-aided active security enhancement solutions may be harnessed for prediction and detection. They can also be invoked as an efficient passive security solution for solving complex optimization problems.
	
	One of the representative application fields of AI is found in solving the beam-hopping problems of LEO SCSs, with the objective of minimizing the impact of harmful interference. Xu {\itshape et al.} investigated deep reinforcement learning aided dynamic beam hopping in multiple-beam satellite systems \cite{zhang2019dynamic,hu2022dynamic,hu2020dynamic,hu2020multi}. Specifically, in order to cope with the randomly fluctuating traffic demands and time-variant wireless channel conditions, deep reinforcement learning was combined with simulated annealing in \cite{hu2022dynamic}. As a further development, multi-objective deep reinforcement learning techniques were developed in \cite{hu2020dynamic,zhang2019dynamic}. Furthermore, a multi-agent deep reinforcement learning method was proposed in \cite{hu2020multi}, where each beam acted as an agent but operated cooperatively. The objectives and constraints of the beam-hopping optimization problem formulated are typically related to the SINR, which is directly influenced by the co-channel interference encountered. In order to mitigate the CCI between adjacent beams, the intelligent method developed in \cite{hu2020multi} estimated the received signal strengths in the overlapping areas and arranged the spatial relationship of the beams for ensuring that the adjacent beams do not adopt the same frequency resources.
	
	In \cite{cao2022robust}, interference-related sum-rate maximization problems have been formulated with the objective of optimizing the passive beamforming vector at the RIS, which were solved by graph attention networks belonging to the family of unsupervised offline learning techniques. They were shown to be capable of capturing the dynamic RIS-assisted LEO SCSs network topology at a low online complexity. Moreover, the authors of \cite{huangdeep} adopted a DL framework to find the optimal RIS coefficients in a secrecy capacity maximization problem, which achieved desired long-term goals at a high convergence efficiency and sample efficiency.
	
	On the other hand, intelligent passive security enhancement solutions can be utilized to prevent LEO SCSs from jamming attacks. As the wireless networks become smarter, so do the jamming attacks. For example, jamming actions may have the capability of learning and reasoning. The SCS associated with cyclic visibility and fixed orbit is exposed to these intelligent jamming attacks, potentially leading to congestion. The traditional anti-jamming methods, such as DSSS, FHSS, multi-beam antennas, and self-adaptive routing, cannot reliably handle these smart jamming attacks. A suite of intelligent anti-jamming designs has been conceived for SCSs by Han {\itshape et al.}. Specifically, in \cite{han2020spatial}, they formulated a hierarchical anti-jamming Stackelberg game to demonstrate the interactions between smart jammers and satellite users. They also proposed a two-stage anti-jamming scheme, where the first stage uses deep reinforcement learning to reduce the routing decision space. By contrast, the second stage relies on  Q-learning to promptly accomplish anti-jamming routing. In the follow-up work \cite{han2020dynamic}, the authors developed a distributed dynamic anti-jamming for a satellite-assisted military IoT network, which cut energy consumption without substantially eroding the performance. The jamming attacks were analyzed and they were counteracted by deep reinforcement learning-based anti-jamming policies. However, since their method relied on analyzing the confrontational interaction between jammers and legitimate users, the anti-jamming performance was inevitably influenced by the accuracy of detecting the existence of jamming attacks. More recently, the authors of \cite{yan2022cross} proposed a cross-layer anti-jamming method involving both the link layer and the network layer, which performs better than a single-layer anti-jamming technique. The link layer handled the particular channel jamming by harnessing sophisticated Q-learning, while the network layer tackled the inter-satellite link jamming by finding new routing paths with the aid of a deep Q network algorithm. However, this algorithm only fits the jamming problems associated with a low-dimensional scenario, bearing with the limited processing capability, and energy cost, which leaves numerous open problems for further investigation.
	
	{\it Lessons Learned:} AI has made substantial progress in safeguarding LEO SCSs in the face of both active and passive information domain perspectives. Both the specific features and evolutionary trends of the traffic can be accurately predicted by AI tools, which helps avoid future congestion and high CCI. Moreover, AI can act as a near-optimal solver for complex non-convex optimization problems encountered by hybrid satellite communication systems in beam-hopping solutions, RISs, anti-jamming techniques, and so on. However, the success of AI applications in LEO SCSs hinges on having a sufficiently large amount of training data, which might be inaccessible in practical communication scenarios. 	

	\subsection{Reliability Enhancement Solutions}
	
	Reliability solutions, including Space Situational Awareness (SSA), debris removal, and radiation resistance, are capable of supporting the stable operation of LEO satellites. Therein SSA employs lots of ground-based facilities or space-borne facilities and novel algorithms to detect and track debris. Moreover, debris removal is an effective means of cleaning up existing debris, thereby substantially reducing the risk of collision. In addition, radiation resistance can not only detect the occurrence of SEUs but also correct it.
	
	\subsubsection{Space Situational Awareness}
	
	The number of Resident Space Objects (RSO), including satellites, spacecraft, and space debris, orbiting the Earth has dramatically increased, hence posing a severe risk to space-based activities. To this end, governments, armed forces, and space agencies have set up SSA programs for collision warnings, and debris removal \cite{MLDebris}. SSA is based on accurate knowledge of the space environment, allowing the detection and tracking of the location of RSO at any time \cite{THzISAR}. Table \uppercase\expandafter{\romannumeral10} summarizes existing studies on SSA in terms of debris detection and tracking.
	
	\begin{table*}[]
		%	\tiny
		\scriptsize
		\renewcommand\arraystretch{1.3}
		\begin{center}
			%		\begin{tabular}{|m{0.5cm}<{\raggedright}|m{0.5cm}<{\raggedright}|m{1.5cm}<{\raggedright}|m{5cm}<{\raggedright}|m{6cm}<{\raggedleft}|}
				\caption{The evolution of SSA}
				\begin{tabular}{|l|r|r|r|r|}
					\hline
					\hline
					Year & Ref.                       & Target                     & Proposed algorithm/scheme                                & Results                                              \\ \hline
					2011 & \cite{ trajectorytracking} & \multirow{2}{*}{Tracking}  & A multiple satellite cooperation method to               & Adjusts the satellite's orbit to maintain camera     \\
					&                            &                            & obtain the 3D debris information                         & concentrating on target debris during tracking       \\ \hline
					2017 & \cite{EKFsdebris}          & \multirow{2}{*}{Tracking}  & The CDEKF as a variable                                  & Exploits discretization and linearization of         \\
					&                            &                            & discretization resolution                                & CDEKF to improve the tracking performance            \\ \hline
					2017 & \cite{formation}           & \multirow{2}{*}{Detection} & A satellite formation control algorithm to               & Calculates and adjusts actions of satellites to      \\
					&                            &                            & detect and track debris cooperatively                    & focus each the sensor carried on common debris       \\ \hline
					2017 & \cite{Bernoulli1}          & \multirow{2}{*}{Tracking}  & A consensus LMB filtering for                            & Solves the problems of debris tracking               \\
					&                            &                            & distributed debris tracking                              & and its dataa incest                                 \\ \hline
					2018 & \cite{THZradar}            & \multirow{2}{*}{Detection} & \multirow{2}{*}{A Space-borne THz Radar at 340 GHz}      & Provides high-resolution 3D imaging of               \\
					&                            &                            &                                                          & spinning space debris                                \\ \hline
					2018 & \cite{Radardebris}         & \multirow{2}{*}{Detection} & \multirow{2}{*}{A Space-borne MmWave Radar at 94 GHz}    & Employs COTS components and GaN solid-state          \\
					&                            &                            &                                                          & technology to demonstrate a space-borne radar        \\ \hline
					2019 & \cite{Lie}                 & \multirow{2}{*}{Tracking}  & A novel Lie-group based                                  & Derives an iterated EKF on Lie groups to             \\
					&                            &                            & parameterization method                                  & track a cluster of debris                            \\ \hline
					2019 & \cite{DeepDetection}       & Detection                  & A deep convolutional neural network based                & Improves the detection performance by deep           \\
					&                            & Tracking                   & space debris saliency detection method                   & convolutional neural network                         \\ \hline
					2020 & \cite{DetectionAcc}        & \multirow{2}{*}{Detection} & A feature learning of candidate regions                  & Removes hot pixels, ficker noise, and nonuniform     \\
					&                            &                            & method for space debris in optical image                 & background for improving detection performance       \\ \hline
					2020 & \cite{MLDebris}            & \multirow{2}{*}{Tracking}  & A ML-based approach for improving                        & Achieves at least 50$\%$ accuracy                   \\
					&                            &                            & orbit prediction in LEOs                                 & improvement of debris tracking                       \\ \hline
					2021 & \cite{Spaceborneradar1}    & Detection                  & \multirow{2}{*}{A Space-borne Ka-band Radar at 35.5 GHz} & Employs Filter Banks to combat Doppler shift for     \\
					&                            & Tracking                   &                                                          & improving the capabilities of detection and tracking \\ \hline
				\end{tabular}
			\end{center}
		\end{table*}
		
		SSA programs exploit a whole suite of sensors, including ground-based radar, optical telescopes, and space-based radar, for inferring the orbital features of objects with the inspiration of their classification and recognition \cite{RadarAESM}. In this context, ground-based radars and optical telescopes are eminently suitable for the observation of RSO. For example, the Tracking and Imaging Radar (TIRA) of \cite{TIRARL} was demonstrated to be capable of debris detection in LEO. The Herstmonceux telescope\cite{mehrholz2002detecting} having a small aperture also operated in good weather conditions. However, both ground-based radars and optical telescopes have their pros and cons. Ground-based radars can operate all the time free from weather conditions, but one of their problems is related to their high costs due to their high transmitter power. Additionally, the ground-based radar can not accurately observe debris in LEO scenarios with a diameter smaller than 10 cm \cite{THZradar} because of their large aperture. By contrast, optical telescopes have high sensitivity for observation, but their observation time is limited by weather conditions\cite{RadarAESM}.
		
		As a remedy, the idea of exploiting either Space-borne Radar (SBR) or Space-borne Cameras (SBC)\cite{Spaceborneradar1} was conceived for debris detection and tracking. They are closer to the target, hence they require lower transmit power. Numerous scholars have subsequently proposed a variety of SBRs and SBCs \cite{trajectorytracking, Spaceborneradar1, formation, THZradar, Radardebris}. Specifically, the coordination of multiple satellites carrying cameras was adopted in \cite{trajectorytracking, formation} for space debris detection and tracking. The estimated 3D position of the debris may be determined from two 2D images of cameras aboard the satellites flying in a formation \cite{trajectorytracking}. A two-stage asymptotically stable nonlinear robust tracking controller was adopted in the formation reported in\cite{trajectorytracking} for maintaining the target debris within the cameras' fields of view. A network of distributed space-borne optical sensors was shown to be able to detect and track debris in\cite{formation}. Torres {\itshape et al.} \cite{Radardebris} presented further technological developments for a space-based radar prototype operating at 94 GHz for detecting centimeter-sized debris. Maffei {\itshape et al.}  proposed a novel SBR payload architecture relying on the Ka-band. Moreover, a filter bank associated with a group of Doppler frequency shifts was also designed in \cite{ Spaceborneradar1} for improving the Doppler tolerance. Bayesian inference was adopted for precisely tracking the trajectory of a piece of debris for several hundreds of milliseconds. Yang {\itshape et al.} \cite{THZradar} designed a solid-state THz SBR operating at 340 GHz by relying on the so-called inverse synthetic aperture technique and obtained a high-resolution 3D image of spinning debris.

		Both Kalman filtering\cite{KalmanRL} algorithms and Bernoulli filtering\cite{BernoulliRL} algorithms have been used for debris tracking. Specifically, Dhondea {\itshape et al.} \cite{EKFsdebris} discussed the Continuous-discrete Extended Kalman Filtering (CDEKF) technique of debris tracking. For tracking a cluster of debris sufficiently close to each other, Labsir {\itshape et al.} \cite{Lie} formulated the problem as a filtering problem constructed over Lie groups\cite{LieGroupRL} and derived an iterated extended Kalman filtering for the tracking of debris. As a further advance, a consensus-based Labeled Multi-Bernoulli (LMB) filtering method was adopted in \cite{Bernoulli1} for estimating the state of debris. Wei {\itshape et al.} \cite{Random2} proposed a multi-sensor-based space debris tracking algorithm relying on $\delta$ generalized LMB filtering. This algorithm was also used for identifying unknown debris by involving a measurement-based `birth' model.	
		
		{\it Lessons Learned:} The capabilities of space-borne solutions relying on THz radars should be further improved. As detailed in \cite{THZRL}, the space-borne systems are potentially capable of safeguarding the information carried among LEO satellites by THz-based ISLs. As a benefit, they can also detect space debris for protecting the satellite with the aid of THz-based radar \cite{THZradar}. However, the maximum attainable transmission power of space-borne THz equipment severely limits both the communication distance \cite{140G5G12KM} and the radar detection distance, which calls for the conception of large-scale space-borne antenna arrays and high-power devices as part of future research.

		\subsubsection{Debris Removal}
		
		In practice, the LEO orbits are the most densely contaminated by space debris among all orbits. Therefore LEO satellites are at the most significant risk of being hit by debris. The accurate debris detection, tracking, and removal planning supported by the AI, sensors, and filtering algorithms introduced above have laid the foundations for our ensuing discussions on debris removal. Anecdotally, researchers in Japan are even experimenting with wooden spacecraft to minimize the amount of space debris \cite{woodensatellite}. At the time of writing, many institutes are contributing to the clean-up of space debris by harnessing the following techniques.
		
		\par\textbf{Nets and Harpoons:} The most famous initiative is that of European research institutions employing dedicated spacecraft to snare debris by firing harpoons and nets at them\cite{DUDZIAK2015509}. These space fishing nets are thousands of meters in diameter and are made of extremely fine wires that are woven together and strong enough to withstand the impact of space debris. The mesh is launched aboard a satellite to be deployed into space, and then it travels along Earth's orbit to sweep up space debris as it passes. Due to the gravitation of the Earth, it finally falls into the atmosphere and burns up. On September 16th, 2018, the RemoveDEBRIS satellite captured a nearby target probe that the vehicle had released a few seconds earlier, which verified the feasibility of this method \cite{SurreyRemoveDEBRIS}.
		
		Another alternative is to use space harpoons for `hunting' satellites. Specifically, such hunting satellites employ a lidar-based guidance system to locate space debris, and a pneumatic device is designed to control the harpoon while catching moving targets. The hunting satellites could also carry tiny sub-satellites that would push the debris into the atmosphere to burn it up.
		
		\par\textbf{Laser `Scavengers':} A new way to deal with space debris has been proposed by Australian scientists based on adopting firing lasers from the Earth to break up space debris \cite{Research}. There are two main ways of using lasers to clean up space debris. For tiny debris, high-power laser light can be used to melt and vaporize it. Larger pieces of debris can be hit at a point, generating a backlash like a rocket jet. Thus, its course changes accordingly, and then it will drop into the Earth's atmosphere and burn up.
		
		\par\textbf{Robotic Arms:} Japan's Aerospace Exploration Agency has also developed a robotic `cleaner' that can use a robotic arm to firmly grasp large pieces of space debris, e.g., dead satellites, and collect them for hurling into the atmosphere to burn them up. The robot, which weighs about 140 kg, has a robotic arm equipped with powerful magnets that can be used for slowing down space debris orbiting the Earth. However, the characteristics of most space debris are not precisely known beforehand, which results in measurement errors concerning the relative motion between the robotic arm and space debris. This makes capturing space debris complicated \cite{2015Residual}.
		
		\par\textbf{Giant Balloons:} It is generally possible for a satellite to fire up its engines at the end of its life and head towards the Earth to burn up in the atmosphere, which would require extra fuel and eventually increase the cost of launch. The new cheaper solution is to carry a folding balloon from launch filled with helium or other gases. Once the satellite exhausted its lifespan, it could blow helium bubbles to increase its drag through the atmosphere \cite{Giant2010}. It takes only a year for a 37-meter-diameter balloon to drag a 1200 kg satellite out of its initial 830 km orbit and to crash it into the Earth's atmosphere to burn it up.
		
		\par\textbf{`Suicide' Satellites:} The aforementioned methods of removing space debris, like using nets, harpoons, robotic arms, or lasers, are costly. Scientists in the UK developed a low-cost device called Cubic Sail to clean up space debris \cite{The2016}. CubeSail is a `suicide' micro-satellite, weighing just 3 kg, that can be launched into space. Once locked on to its target, it would deploy its kite-like solar sail, attach itself to space debris and slow its flight. Eventually, they will perish.
		
		Table \uppercase\expandafter{\romannumeral11} compares the advantages and disadvantages of these debris removal techniques. However, these solutions are currently in the design or experimental phase, and more engineering efforts are required to put these ideas into practice.
		
		\begin{table*}[ht]
			\centering
			\caption{A table comparison of debris removal techniques}
			\renewcommand\arraystretch{1.5}
			\begin{tabular}{|m{2.6cm}<{\centering}|m{5.4cm}<{\raggedright}|m{7cm}<{\raggedleft }|}
				\hline
				\hline
				Project                              & Advantages                                                             & Disadvantages                                                                         \\ \hline
				\multirow{2}{*}{Nets and harpoons}   & Able to handle irregular and spinning debris compared to a robotic arm   & Nets is not able to be reused                                                                \\ \cline{2-3}
				& Nets prevent further debris generation                                 & Smashing large space debris by harpoons  may generate further debris                  \\ \hline
				\multirow{3}{*}{Laser `scavengers'}   & Effective for small space debris                                       & May burn up the debris causing extra debris                                           \\ \cline{2-3}
				& Able to dexterously handle tumbling debris                                 & Large amount of beam energy, because it is hard to generate a small beam at a long distance \\ \cline{2-3}
				& Able to be reused                                                          & Sophisticated target detection and acquisition system                                 \\ \hline
				\multirow{2}{*}{Robotic arms}      & Able to grasp space debris firmly                                          & Sophisticated control                                                                 \\ \cline{2-3}
				& Able to be reused                                                          & Easily penetrated by debris, especially sharp debris                                  \\ \hline
				\multirow{2}{*}{Giant Balloons}      & Effective large space debris such as failing or inoperative spacecraft & Easily penetrated by debris, especially sharp debris                                  \\ \cline{2-3}
				& Preventing further debris generation                                   & Slow response because of balloon inflation                                            \\ \hline
				\multirow{2}{*}{`Suicide' Satellites} & Preventing further debris generation                                   & Not able to be reused                                                                     \\ \cline{2-3}
				& Low cost                                                               & Suitable for larger debris                                                            \\ \hline
			\end{tabular}
		\end{table*}

		\subsubsection{Artificial Intelligence} The AI family, especially ML, and DL, also find wide-ranging applications in collision avoidance and debris identification as well as removal planning.

		\textbf{Collision Avoidance:} The rapidly escalating number of mega-constellations inevitably increases the risk of collision, especially in the LEO orbit, which has numerous objects traveling at high speed in an uncontrolled manner, including rocket body parts, dead satellites, shrapnels, and debris. In fact, collisions could generate additional orbiting debris that, in turn, produce further collisions and thereby trigger an avalanche-like debris growth chain reaction, which prompts space institutions and agencies to intensify their collision avoidance actions. The release of real-world datasets in the form of messages containing information about collision times and risks of near-miss events lays the foundation for the utilization of ML tools to avoid collisions \cite{acciarini2021kessler, uriot2022spacecraft}. Specifically, the authors of \cite{acciarini2021kessler} described an open-source Python package named Kessler to predict the evolution of conjunction events in a reliable manner by relying on Bayesian neural networks. A milestone in solving space collision challenges was achieved by the European Space Agency \cite{uriot2022spacecraft}, by organizing an ML competition based on a large curated dataset to inspire competing teams to find the best collision risk estimation model. The competition results demonstrated the difficulties in finding a generic training set and highlighted the benefits of ML techniques in this research field.

		\textbf{Debris Identification and Removal Planning:} Although the amount of debris can be reduced by adopting effective collision avoidance strategies, the LEO orbits are still contaminated by space debris that comes from explosions, impacting other space objects or launch activities. As Wyler, the founder of OneWeb, said: ``My epitaph should say `Connect the World' instead of `Making Orbital Garbage'." To exploit the space debris and effectively exploit the LEO for future exploration, we must make concerted, collaborative efforts to both prevent the generation of future debris and eliminate existing space debris. The above reliability enhancement solutions, such as robotic arms, are capable of pushing the failing or inoperative spacecraft into Earth's atmosphere and burning them down. This is an effective means of mitigating the generation of space debris. However, given the dynamically time-varying factors in the space environment, the practical operational feasibility of the above reliability enhancement solutions should be carefully verified before any action in the face of the associated uncertainties. This requires substantial online or offline computing capability to identify targets, as well as to plan and track their trajectories before capturing moving targets \cite{zhang2023review}. 
		
\begin{figure*}[]
			\centering
			
			\includegraphics[width=2.0\columnwidth]{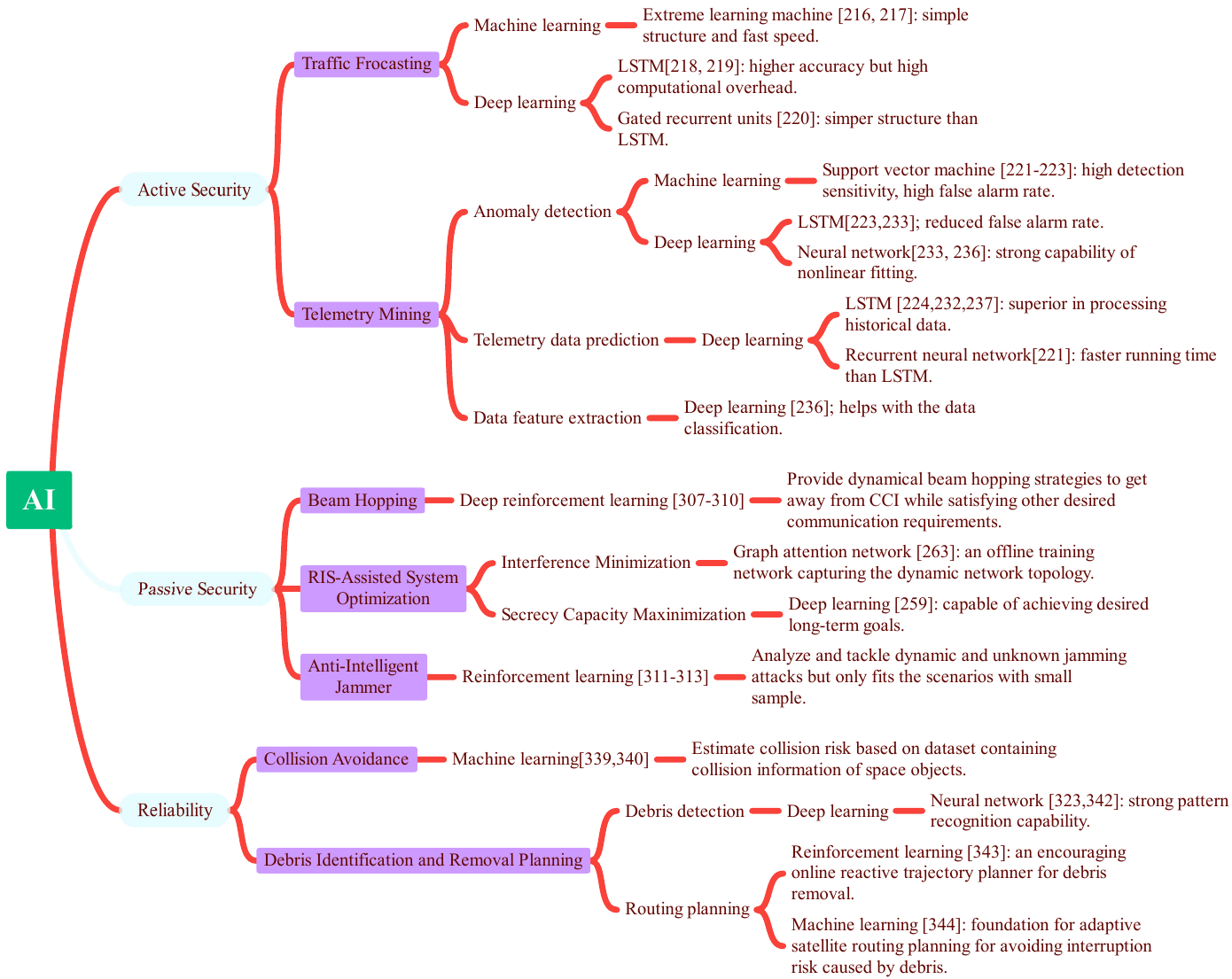}
			
			\caption{Literature review for AI-based enhancement solutions.}
			\label{AIactive}
		\end{figure*}

		In this context, the first step, namely space debris detection, is associated with a considerable challenge, since debris appears as a blob without visual features. Moreover, the reflectivity of debris is weak both due to its mobility and owing to the noise in the cosmic space due to the cluttered starry background. This leads to extremely low SNR. By exploiting the strong pattern recognition capability of DL, the authors of \cite{DeepDetection,GridLearn} constructed neural networks to detect space debris. The input of the convolutional neural network proposed in \cite{DeepDetection} was a local contrast map derived from the space-based surveillance video. The spatiotemporal saliency information captured from a local contrast map enhanced the robustness when facing time-varying noisy background. To increase the detection speed, the authors of \cite{GridLearn} split the space image captured into small tiles of the same size, where a binary label was assigned to each tile to show whether there is space debris located in it. Once the debris is identified, the remaining task is to decide how to remove it. Since reinforcement learning relies on reward collection, it fits the objective of the active multi-debris removal mission planning problem of LEO SCSs \cite{yang2019application}. The experiments relying on the Iridium 33 system confirmed that reinforcement learning constitutes a beneficial online reactive planner. However, not all the debris can be `de-orbited' in time to avoid causing interruption to the inter-satellite laser links. The authors of \cite{ma2021adaptive} thus discussed several common laser link interruption scenarios, followed by an interruption risk perception model relying on a powerful ML tool, which lays the foundation for developing adaptive routing strategies. In summary, the AI family, especially the ML and DL techniques have diverse wide applications both in active and passive security provision and in reliability enhancement solutions. The existing literature on AI-based enhancement solutions is summarized at a glance in Fig. \ref{AIactive}.

		{\it Lessons Learned:} Reducing the probability of collision risk caused by space debris requires each country and research institution to bear corresponding responsibilities and establish an international cooperation mechanism as well as a shared resource platform to jointly research and carry out space debris clean-up actions. This is the only way of effectively reducing the impact of space debris for ensuring the smooth and sustainable development of future space activities. First of all, each country and institution should reach a consensus and adopt measures for preventing the creation of additional space debris. 
			
			Firstly, during the satellite launch process, consider the reusability and recyclability of the spacecraft exemplified by SpaceX's Falcon series rockets \cite{zhang2021}. Once the lifespan of a satellite ends, we must consider the safe and effective removal of the satellite from its orbit. This removal process aims for ensuring that the satellite is completely burned up in the Earth's atmosphere, thus preventing any potential space debris from posing a future threat to other spacecraft.
			
			To deal with the existing space debris, all countries should introduce effective debris removal measures, such as robotic arms, giant balloons, nets as well as harpoons, to collect large space debris and pull it back into the atmosphere for direct burning.
			
			During a satellite's operation, it is necessary to detect and track space debris by SSA. Ground-based radar and optical telescopes are commonly used for detecting and tracking space debris. In the event of a collision risk, the satellite can be controlled by the ground segment to perform emergency avoidance maneuvers. Meanwhile, the satellite itself should be equipped with SBRs or SBCs, which can also help reduce collision risks. Furthermore, the exploitation of AI tools for learning and training on data generated by these SBRs and SBCs results in the improvement of detection and tracking.

		\subsubsection{Radiation Resistance}
\begin{figure}[h]
			\centering
			
			\includegraphics[width=0.7\columnwidth]{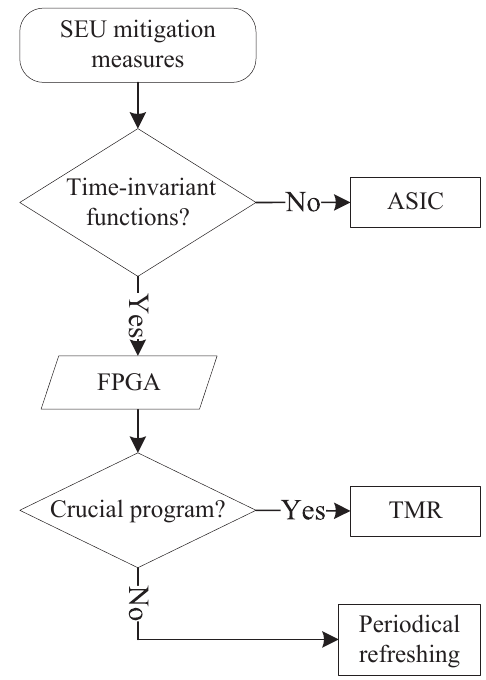}
			
			\caption{A flow chart of radiation resistance measures.}
			\label{SEUmitigationpic}
		\end{figure}

		Radiation resistance is an engineering problem involving advanced chip technology and different forms of redundancy for ensuring the reliable operation of the space-borne payload in harsh space environments. The formulation of radiation resistance measures usually obeys the process shown in Fig.~\ref{SEUmitigationpic}. The time-invariant functions should be implemented by ASICs, while the programs that have to be upgraded or iterated should be implemented using FPGAs because of their flexibility.
		
		For the program implemented in FPGAs, usually, TMR is adopted for preventing the impact of SEUs \cite{TMR}. Briefly, TMR is a fault-masking scheme based on feeding the outputs of three identical copies of the original program module to a majority voter. If the output of the three modules is the same, the system will be regarded to operate normally. If any faults occur in one of the modules, the other modules can mask the fault. Thus, TMR can efficiently prevent single faults from propagating to the output.
		
		However, there is a trade-off between resource consumption and integrity. The resource consumption of TMR is three times that of the original program module. Hence, designers usually apply the TMR philosophy only to the key part of the program, such as the control part.
		
		Another radiation resistance solution is periodical refreshing \cite{refreshing0}, which can correct errors by refreshing the program without interrupting its execution, as detailed in \cite{refresh}. However, the block Random Access Memory (RAM) used in FPGAs will be initialized during the periodical-refreshing operation when its real-time state is lost. Hence the block RAM should also adopt TMR for coping with the impact of SEUs \cite{refreshingRAM}. 
		
		{\it Lessons Learned}: Based on our detailed literature review, we can clearly identify the advantages and disadvantages of both TMR and periodical refreshing. TMR is capable of detecting the occurrence of SEUs and correcting their effect. Thus, there is no doubt that TMR mitigates the impact of SEUs, but at the cost of certain additional resource consumption. The resource consumption of TMR may be as high as three times that of the original program module \cite{MEOTMR}. Although periodical refreshing does not impose additional resource consumption, it cannot cover all types of FPGAs resources, such as block RAM. As a remedy, the combination of partial TMR and periodical refreshing may improve the FPGAs' reliability.

			\subsection{Trade-offs in Quality of Service (QoS) guarantees}
			
			As mentioned above, in consideration of service type, inherent characteristics of LEO SCSs, security and reliability issues as well as solutions, there exists substantial trade-offs in QoS guarantees.
			
			\begin{itemize}
				\item[$\bullet$] It is of vital importance to determine the specific choice of solutions to be employed by the different segments of LEO SCSs according to the specific trade-offs between the security improvement attained and its cost in terms of the overhead imposed. As detailed in \cite{PLSSIN, PLS6G, DroneRL}, data confidentiality can be maintained by traditional mathematics-based encryption schemes. However, LEO satellites are considered to have limited computation capabilities \cite{SINCM}, hence the encryption schemes relying on excessive computational complexity are unsuitable for them, but they are routinely used at ground segments for improving the security level.

				\item[$\bullet$] There is a trade-off between the integrity related to traffic rate and anti-jamming capabilities indicated by the integrity when using the DSSS technique. More specifically, given the communication bandwidth, the traffic rate is related to the spreading factor. When the jamming is strong, the spreading factor should be increased to improve the anti-jamming capability, hence leading to integrity reduction and $vice$ $versa$.
				
				There is also a trade-off between integrity and the confidentiality of MC-DSSS systems, as seen in Fig.~\ref{MCDSSS} and Fig.~\ref{MCDSSS1}. As reported in \cite{MengRL}, a fraction of the sub-carriers may be hidden in the existing signals, which improves the confidentiality of the transmitted signal, but potentially degrades the integrity indicated by BER, and \textit{vice versa}.
				
				\item[$\bullet$] Non-SS jamming suppression techniques, including transform domain adaptive filtering and temporal domain adaptive filtering, can be used for improved jamming mitigation. Between them, the LMS algorithm, which is a low-complexity design option for the temporal domain adaptive filtering technique, is eminently suitable for space-borne payloads. However, there is a trade-off concerning its iterative step size. The authors of \cite{LMSRL} provided evidence that a higher step size leads to faster convergence but also to a higher variance of the weights. Furthermore, the more rapid convergence of the weights results in low latency, hence promptly tracking and mitigating malicious jamming. However, the resultant higher variance of the weights may affect the performance of jamming mitigation, which undoubtedly degrades the integrity.
				
				\item[$\bullet$] TMR is capable of not only detecting the occurrence of SEUs but also correcting its effect. Thus, there is no doubt that TMR mitigates the impact of SEUs, but at the cost of a certain additional resource consumption \cite{refreshing0}. The resource consumption of TMR is three times that of the original program module. As a remedy, the combination of partial TMR and periodical refreshing may improve the FPGAs' reliability \cite{refresh, refreshingRAM}.
				
			\end{itemize}
		
		\section{The Road Ahead}
		Given the rapid developments of satellite technologies, Integrated Sensing and Communication (ISAC) has great potential in terms of mitigating some of the challenges of LEO SCSs. This section will address their new opportunities in stimulating future research. Computer Vision (CV)-aided communication may provide new perspectives for secure space communications since accurate target detection, identification, and tracking can be offered by exploiting the information extracted. Additionally, the development of efficient and low-cost satellite production lines expedited the Mega-constellation planning and commercialization, which however exacerbates their security challenges. This section will address the emerging new opportunities for stimulating future research.
		
		\subsection{ISAC-aided Secure Transmission}
		With the rapid proliferation of connected devices and satellites, the available frequency spectrum assigned for wireless communications tends to be increasingly congested, which motivates network designers to seek spectrum reuse opportunities for the better exploration of bands originally assigned to other technologies. In recent years, the similarities between communications and sensing, such as their hardware components, antenna architecture, and signal processing modules attracted scholars to study a technology combining the two functional modules, leading to the concept of ISAC. The philosophy is to allow the communication systems to access large portions of the spectrum available at radar frequencies \cite{liu2020joint, liu2022survey}.
		
		The advantages offered by ISAC for SCSs include alleviating the shortage of radio frequencies, reducing overall system costs, cutting energy consumption, and miniaturizing the devices. In particular, the strong directivity, low side-lobe, and anti-interference capability of radar are capable of enhancing the information security, transmission reliability, communication quality, and coverage of the LEO SCSs \cite{feng2020joint}. Additionally, the sensing part in ISAC can partially characterize the propagation environment, which potentially improves the CSI estimation accuracy and reduces the channel estimation overhead. Moreover, the sensed movement and location information of high-mobility objects facilitates LEO SCSs to improve their beam alignment strategies and routing protocols \cite{tan2021integrated}. 		
		
		On the other hand, the information obtained from communications, in turn, assists high-accuracy localization, real-time tracking, and high-precision imaging, as well as activity recognition \cite{tan2021integrated}, which has great potential in safeguarding the security of LEO SCSs. For instance, high-accuracy localization and tracking lay solid foundations for the debris removal operation. Additionally, the high-precision imaging relying on the support of AI allows LEO SCSs to be trained to dramatically reduce the risk of space collisions. The LEO SCSs are also expected to have high classification accuracy to detect anomalies and impending faults by continually sensing their surroundings.

		The application of ISAC brings along additional challenges. The bottlenecks faced by ISAC in other wireless networks also exist in LEO SCSs. For instance, it has to investigate unified performance metrics that examine the communication-sensing trade-off, clarify the fundamental limits of ISAC under practical considerations, and develop low-complexity, yet accurate ISAC signal processing algorithms \cite{ISACCOMST}. Moreover, the mobility and hardware limitations of LEO satellites introduce further challenges, including limited synchronization resolution, restricted channel coherence time, high-speed moving targets, and insufficient computational capability. These factors make it more challenging to employ ISAC in LEO SCSs efficiently and reliably. Additionally, the massive amounts of data generated by ISAC, which contain sensitive information, impose further challenges on securing LEO SCSs. However, there is a paucity of literature on the integration of ISAC into LEO SCSs. Only a brief conference paper [361] has considered the simultaneous communication and sensing capabilities, highlighting the inevitable propagation delay and significant Doppler shifts in LEO SCSs that necessitate sophisticated solutions. Therefore, substantial further investigations are required to explore how to leverage the unique advantages of ISACs to enhance the security of LEO SCSs, while addressing the accompanying challenges.

		\subsection{CV-aided Space Communication}	
		Recently, some researchers proposed CV-aided \cite{Tianyu, Tianyu1,Tianyu5,Tianyu6} communication schemes for mmWave or THz transmission systems in which LoS propagation is critically important. In contrast to ISAC, the core idea of CV-aided methods is to extract, recognize, and estimate useful information about the associated static system topology, including the terminals' positions, distances among themselves, and their number. It is also beneficial to keep track of their velocity, direction, and their number. The objective is to achieve new potential benefits in terms of improving wireless system design/optimization, such as resource scheduling and allocation, algorithm/protocol design, and so on.
		
		Therefore, a pair of salient features of CV-aided schemes, which are beneficial for the security of LEO SCSs, can be summarized as follow: 1) One-way sensing capability: By employing optical cameras/devices \cite{Tianyu2}, no RF signals have to be transmitted and received, hence resulting both in low detection probability by adversaries and in low resource consumption; 2) Hostile target/non-partner detection: Targets can be detected, identified, and distinguished via optical processing algorithms, and thus secure/covert information delivery schemes can be designed and implemented to avoid/hinder potential eavesdropping/perturbation \cite{Tianyu3}.
		
		Situational awareness in space has already been established by applying various optical and radar sensors, e.g., electro-optical/infrared systems and optical telescopes, they only survey, identify, and predict objects in orbit. However, at the time of writing, no literature exists on applying CV-aided methods to safeguard information transmissions in space. Given the unique high-dynamic and large-scale space scenarios of LEO SCSs, a number of challenges have to be tackled for realizing CV-aided space secure/covert communications in the field of ultra-high-speed target detection, 
		carrying out reliable on-broad data processing and robust as well as secure transmission.

		\subsection{Mega-Consetellations}
		In light of the impending mega-constellation launch proposals, the number of active satellites in orbit will soar to around 50,000 in ten years, leading to an unprecedented scale of LEO SCSs. The operators of mega-constellations thus suffer from a heavy computational burden, since they have to supervise and manage the operational status and diverse functions of hundreds or even thousands of satellites in real-time. A minor computational or command error might have severe consequences for this giant network. Furthermore, large LEO SCSs require numerous ground stations and gateways. The authors of \cite{Starlink} estimate that around 123 ground station locations and 3500 gateway antennas are necessary for the 4400-satellite version of Starlink to approach the throughput limits. Such a large-scale deployment will require highly automated and secure management systems.

		Additionally, the information exchange between the mega-constellations and ground stations relies on the inter-satellite network constructed by ISLs within the constellation. This is different from the traditional small-scale constellations, where the information exchange can be realized by the satellite-ground links or by the relaying assistance of the GEO satellites. Although the ISL strategy reduced the deployment costs of ground stations and relay satellites, the multi-hop inter-satellite networks are more vulnerable to malicious attacks due to their predominantly LoS propagation. The authors of \cite{zhan2022networked} also highlighted that the total transmission delay of the multi-hop inter-satellite network should be taken into account in the context of security problems. In fact, a malicious node might succeed in masquerading as one of the legitimate nodes in ISLs to deliberately extend the data forwarding delay. Emergencies may even aggravate disasters, such as collisions and large-scale destruction. The solution proposed in \cite{zhan2022networked} was based on the investigation of suitable routing algorithms by exploiting the knowledge of the degree of trust concerning each satellite, combined with other existing security technologies, including encryption, digital signatures, and so on. Nevertheless, the investigation of the security problems in mega-constellations is at an early stage, which urgently requires the researchers' attention to fill this gap.

		\subsection{LEO SCS Commercialization}
		Although there are still numerous open problems, the LEO SCS has reached a certain maturity. The vibrant LEO economy is attracting companies and investors. However, the crowded space and limited channel resources have resulted in intense competition among major companies. It will be hard to unify the quality of the satellites and their associated products with more and more partners entering the satellite communication industry.

		Moreover, the commercialization also assists the development of LEO satellite applications, such as the IoTs, smart cities, and intelligent manufacturing, which provides tech giants with more business opportunities for combining the space industry with hybrid network applications to boost their profit. However, the resultant trend imposes further aggravated security challenges, because the diverse nature of terminals, standards, and operational policies are more prone to attackers. For example, the ground segments of satellites in the Arctic are of strategic importance to the North Atlantic Treaty Organization, given their ability to collect intelligence from some leasehold commercial satellites. Hence the ground segment of these commercial satellites is increasingly, employed for both civilian and defense purposes, which makes them vulnerable to military targets \cite{Comerical2022}. In fact, the booming commercial applications, in turn, stimulate the production of LEO satellites. Their small size is a clear advantage from a financial perspective, but generally, this is achieved at the cost of a shortened life span. Therefore, companies also must have end-of-life plans before launching new satellites.
		
		Although the future of the LEO SCS market seems bright at the time of writing, the experience due to financial issues should not be forgotten. Many companies, such as LeoSat, and OneWeb, have to scale back or even cancel their intended constellations unless they secure additional investment. The COVID-19 pandemic has inflicted uncertainty and challenges upon LEO SCS commercialization.  On the other hand, their high cost makes the satellite-connectivity options expensive, which can only be afforded by a limited market segment, where terrestrial solutions are uneconomical. Investors might provide low-tariff space services at reduced profits to attract business at the beginning. Clearly, the LEO SCS market requires substantial upfront investment and cannot provide immediate positive cash flow, which thus increases the risk of financial challenges. Hence it is necessary to reduce costs, from materials to manufacturing, from the launch to the user equipment. Clearly, the cost reduction option should be carefully investigated to avoid low-quality products improving security problems. 
		
        \section{Design Guidelines}
			The design of an LEO SCS is complex because it must consider numerous potentially conflicting design factors. In this section, we provide tangible design guidelines for LEO SCSs from a security and reliability perspective, which is derived from our critical review of the literature and the lessons learned concerning the security and reliability requirements, issues, and their corresponding solutions. The iterative procedure of our design guidelines is as follows.

			\begin{figure*}[ht]
				\centering
				
				\includegraphics[width=2\columnwidth]{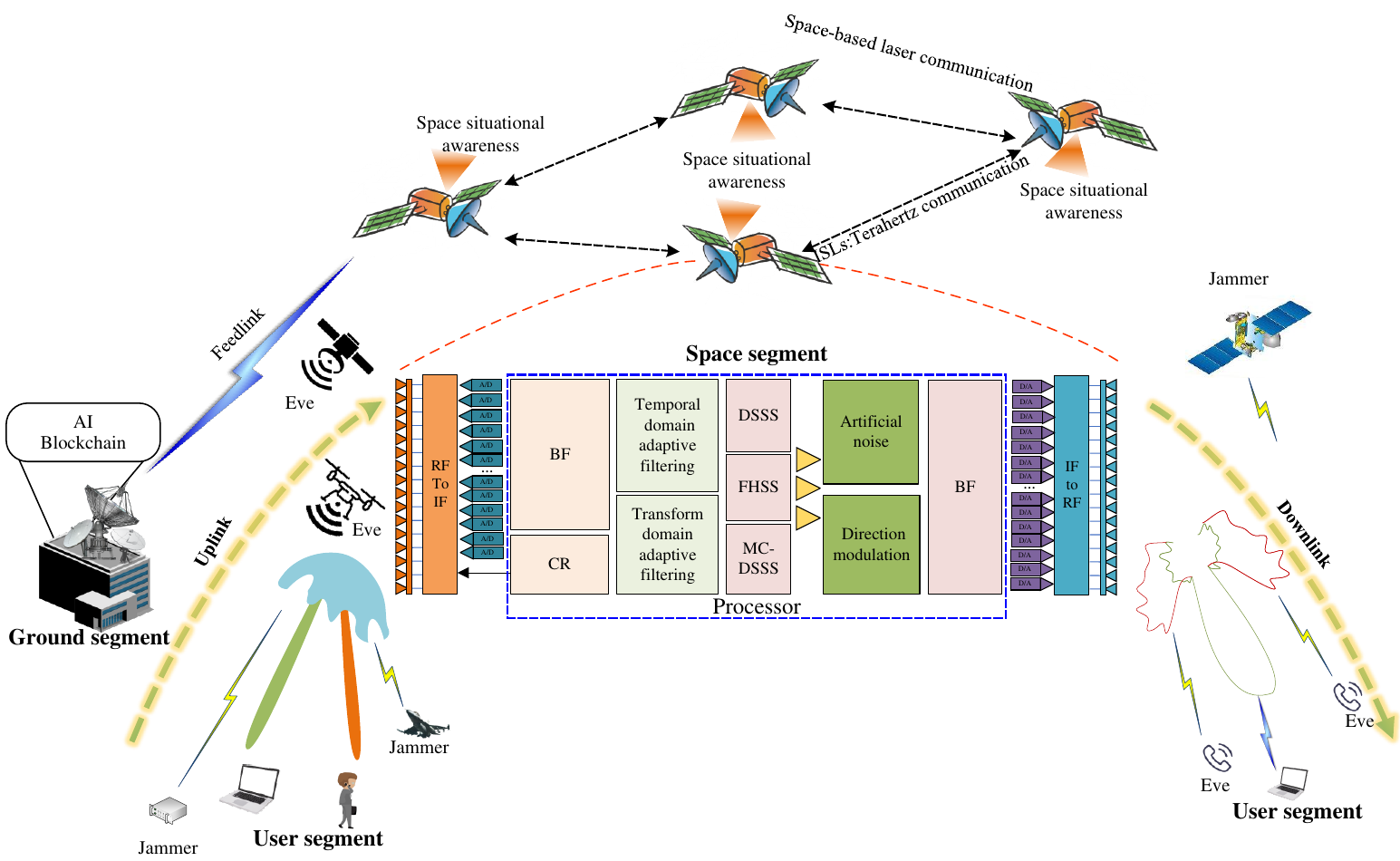}
				\caption{Design guidelines of secure and reliable LEO SCSs.}
				\label{Designguidelines1}
			\end{figure*}
			
			\subsection{Orbit Selection}
			The selection of satellite orbit is extremely complex. It is necessary to submit an application to the ITU as a prerequisite. From the perspective of reliability, the orbit selection has to consider two aspects: orbit altitude and orbit inclination. The selection of orbit altitude has to consider the distribution of already approved or deployed LEO satellites. The intensifying deployment of LEO mega-constellations has made low orbits more crowded, as shown in Fig.~\ref{Debrispic}. On the other hand, the orbit inclination affects the probability of SEUs. In low Earth orbits, the higher the orbit inclination, the higher the probability of SEUs. Therefore, the orbit inclination has to be considered from two perspectives: collision avoidance (the number of satellites deployed in existing orbit inclinations) and reducing the probability of SEUs.			
			
			\subsection{Frequency Selection}
			Similar to orbit selection, the frequency resources of LEO satellites also have to be requested from the ITU. From the perspective of security and reliability, frequency selection has to be based on both the business type and usage scenarios. For example, given the limited energy support of the IoT devices and narrow band low-speed communication of LEO satellites empowering the Internet of Remote Things as shown in Fig.~\ref{application}(a), the L-band with low propagation loss is a good choice, whereas the K-band frequency is the preferred choice for high-throughput backhaul services supported by LEO satellites, as seen in Fig.~\ref{application}(b).
			
			When selecting specific frequencies, the existing frequency allocation should also be considered to avoid inter-system interference, as shown in Fig.~\ref{Frequency}. Additionally, the expanding emerging frequency bands can alleviate the problem of inter-frequency interference caused by the spectrum crunch. This is also one of the important reasons why THz and laser communications are gradually replacing the original K-band for ISLs \cite{Iridium}.

			\subsection{Waveform Selection}
			
			Considering their inherent characteristics of concealment and anti-interference, SS waveforms, including DSSS, FHSS, and MC-DSSS, remain the most competitive. Among them, MC-DSSS has the feature of flexible sub-carrier allocation, which can be combined with CR to dynamically adjust its sub-carriers for mitigating the inter-frequency interference between systems.
			
			\subsection{Considerations before Design: On-board Processing is More Secure than Transponder}
			
			Attackers utilize satellite transmissions with no onboard processing to forward their own information, which is defined as transponder stealing in this paper, but the algorithms or programs running on the onboard processing system effectively prevent these illegal transmissions. Hence the onboard processing system avoids this eavesdropping behavior. In fact, the benefits of the onboard processing system in terms of improving security and reliability are not limited to this. A transparent forwarding system requires a large number of connected ground segments for achieving global coverage. The sheer number of ground segments is appealing to potential attackers, while the onboard processing system can achieve global coverage by relying on ISLs, and only a few ground segments are needed for stable operation. Additionally, in a transponder system, security and reliability enhancement solutions are concentrated on the ground segment, while onboard processing satellites can carry out extra security and reliability enhancement actions.

			According to the inherent characteristics and the serious security challenges of LEO SCSs, the pertinent security and reliability requirements were summarized in Sec. \uppercase\expandafter{\romannumeral3}, which are regarded as the most crucial requirements. Hence, the particular security specifications constitute the initial guiding policies for a designer. As an overriding principle, usually, complex encryption algorithms are harnessed in the ground segments of LEO SCSs as a benefit of their abundant resources, but they are typically unsuitable for the resource-limited satellites. They tend to require lightweight and low-power solutions.
			
			\subsection{LEO SCSs Secure and Reliable Design}	
			
			After selecting the satellite orbits, the frequencies, and the waveforms, the secure and reliable LEO SCSs design has to proceed by bearing the aspects in Fig.~\ref{Designguidelines1} in mind. 
			
			\subsubsection{Uplink}	
			
			Advanced security-oriented antennas are adopted for mitigating eavesdropping and jamming. Furthermore, the increasing number of LEO satellites is laying the foundation for satellite cooperation. By combining the signals received from each satellite, the integrity of the combined signal with improved SNR can be maintained while allowing the terminal to transmit at reduced power, thereby improving security. Additionally, non-SS jamming suppression techniques \cite{LMS2,TransformSNR}, including temporal domain adaptive filtering \cite{LMS1} and transform domain adaptive filtering \cite{Transformlowrank}, can be further used for improved jamming mitigation.
			
			\subsubsection{Downlink} 
			Advanced security-oriented antennas can also be adopted for mitigating the eavesdropping and jamming probability in the downlink. The artificial noise can be released in directions other than that of legitimate users to drown out potential eavesdroppers. Moreover, the amplitude and phase of the downlink signals can be adjusted with the aid of RISs to further hinder the malicious actions of eavesdroppers and jammers.
			
			\subsubsection{The Processor of Space Segment} 
			The increased probability of SEUs poses serious challenges to the reliable operation of programs or algorithms running onboard the satellites. Therefore, the processors have to be radiation-resistant, as intimated in Fig.~\ref{SEUmitigationpic}.
			
			\subsubsection{Ground Segment} 
			Again, the ground segment is the center of operation, management, and control for the entire LEO SCS. Any malfunction of the ground segment can potentially bring the system to a halt, hence it often becomes the preferred target of attackers. The ground segment having abundant power makes it possible to run more complex algorithms, such as encryption, machine learning, and blockchain, in the interest of improving security.
			
			These algorithms can effectively address challenges like message modification, node compromise, DoS attacks, etc.
			
			\subsubsection{User Segment} 
			Each user should update the patches at regular intervals for reducing the probability of successful attacks. Additionally, each user has to cooperate with the space segment, for example by adjusting the operating frequency or transmit power, for dealing with intra-system interference, such as MAI and CCI between beams. 
			
			\subsection{LEO SCSs Secure and Reliable Operation} 
			After completing the design of LEO SCSs, measures must also be taken to ensure their reliable and stable operation.
			
			\subsubsection{Collision Avoidance} 
			Again, the increasing number of spacecraft and space debris in low Earth orbit poses a serious challenge to the reliable operation of LEO satellites. To address this issue, the maintenance personnel of the ground segment has to closely monitor the operational status of spacecraft and provide advance warning of potential collisions. Additionally, each satellite should be equipped with SBRs or SBCs to detect sudden, erratic space debris movements, and take immediate action to adjust the satellite's altitude to avoid collisions in case of risk.
			
			\subsubsection{Interference Coordination} 
			According to ITU regulations \cite{ITU1}, LEO SCSs shall not impose unacceptable interference on GEO SCSs. Under the coordination of LEO satellites, users carry out precise power control of the uplink signal to reduce their impact on the GEO system. Additionally, LEO SCSs may perform beam drifting for forcing the LEO satellite users to use their other beams in the downlink before interference actually occurs.

		\begin{table*}[]
			
			\scriptsize
			\caption{List of acronyms}
			\renewcommand\arraystretch{1.3}
			\centering
			%	SEUs	
			\begin{tabular}{m{1.5cm}<{}m{6cm}<{}|m{1.5cm}<{}m{6cm}<{}}
				\hline
				\hline
				Acronyms & Definitions                                   & Acronyms & Definitions                                   \\
				6G    & Sixth-generation                              & MAI     & Multiple Access Interference                  \\
				ADS-B & Automatic Dependent Surveillance-Broadcast    & MC-DSSS & Multi-Carrier Direct Sequence Spread Spectrum \\
				AI    & Artificial Intelligence                       & MEO     & Medium Earth Orbit                            \\
				AN    & Artificial Noise                              & MIMO    & Multiple Input Multiple Output                \\
				ASIC  & Application Specific Integrated Circuit       & ML      & Machine Learning                              \\
				BER   & Bit Error Rate                                & MmWave  & Millimeter Wave                               \\
				BF    & Beamforming                                   & ms      & Milliseconds                                  \\
				CCI   & Co-channel Interference                       & NASA    & National Aeronautics and Space Administration \\
				CCSDS & Consultative Committee for Space Data Systems & NCC     & Network Control Center                        \\
				CDEKF & Continuous-discrete Extended Kalman Filtering & NTN     & Non-terrestrial Network                       \\
				CIR   & Carrier to Interference Ratio                 & OGS     & Optical Ground Station                        \\
				COTS  & Commercial Off The Shelf                      & PA      & Phased Array                                  \\
				CV    & Computer Vision                               & PG      & Processing Gain                               \\
				DDoS  & Distributed DoS                               & PLS     & Physical Layer Security                       \\
				DoS   & Denial of Service                             & PSD     & Power Spectral Density                        \\
				DSSS  & Direct Sequence Spread Spectrum               & PU      & Primary User                                  \\
				ESA   & European Space Agency                         & PUF     & Physical Unclonable Function                  \\
				FDA   & Frequency Diverse Array                       & QKD     & Quantum Key Distribution                      \\
				FH    & Frequency Hopping                             & QSDC    & Quantum Secure Direct Communications          \\
				FHSS  & Frequency Hopping Spread Spectrum             & RAM     & Random Access Memory                          \\
				FFHSS & Fast Frequency Hopping Spread Spectrum        & RIS     & Reconfigurable Intelligent Surface            \\
				FPGA  & Field Programmable Gate Array                 & RSO     & Resident Space Objects                        \\
				GEO   & Geostationary Earth Orbit                     & SAGIN   & Space-air-ground Integrated Network           \\
				GHz   & Gigahertz                                     & SBC     & Space-borne Camera                            \\
				GPS   & Global Position System                        & SBR     & Space-borne Radar                             \\
				HTS   & High Throughput Satellites                    & SIN     & Space Information Network                     \\
				IDS   & Intrusion Detection Systems                   & SCS     & Satellite Communication System                \\
				IoRT  & Internet of Remote Things                     & SDR     & Software Defined Radio                        \\
				IoT   & Internet of Things                            & SEU     & Single Event Upset                            \\
				IoV   & Internet of Vehicles                          & SNR     & Signal to Noise Ratio                         \\
				ISAC  & Integrated Sensing and Communication          & SS      & Spread Spectrum                               \\
				ISL   & Inter-satellite Link                          & SSA     & Space Situational Awareness                   \\
				ISRO  & Indian Space Research Organisation            & SU      & Secondary User                                \\
				ITU   & International Telecommunications Union        & TIRA    & Tracking and Imaging Radar                    \\
				kg    & kilogram                                      & THz     & Terahertz                                     \\
				km    & kilometer                                     & TMR     & Triple Module Redundancy                      \\
				LEO   & Low Earth Orbit                               & TPC     & Transmit Precoding                            \\
				LFDA  & Linear Frequency Diverse Array                & TT$\&$C & Telemetry, Tracking, and Command              \\
				LMB   & Labeled Multi-Bernoulli                       & UAV     & Unmanned Aerial Vehicle                       \\
				LMS   & Least Mean Square                             & WLAN    & Wireless Local Area Networks    \\   \hline          
			\end{tabular}
		\end{table*}

		\section{Summary}
		LEO SCSs have attracted increasing attention as a benefit of their seamless global coverage with low latency. However, there are many open issues in the course of exploiting the full potential of LEO SCSs, including their security issues. Due to inherent characteristics such as special location, high mobility, and so on, LEO SCSs suffer severe security challenges. Not only security attacks, such as eavesdropping and DoS but also reliability risks, such as collisions and SEUs, affected the safe operation of LEO SCSs.
		
		In this paper, we classified the issues encountered by LEO SCSs, summarized their characteristics, and discussed their lessons learned. To deal with these issues, we then introduced and summarized some corresponding solutions, which can be divided into security and reliability enhancement solutions. Moreover, we also provided numerous trade-offs and lessons. Based on this, we highlighted ISAC-aided secure transmission, CV-aided space communication, mega-constellation security problems, and commercialization issues for future research. Finally, we presented high-level design guidelines for secure LEO SCSs.

		\bibliographystyle{IEEEtran}
		\bibliography{ref0530}
		
		\vspace*{-0mm}
		\begin{IEEEbiography}[{\includegraphics[width=1in,height=1.25in,clip,keepaspectratio]{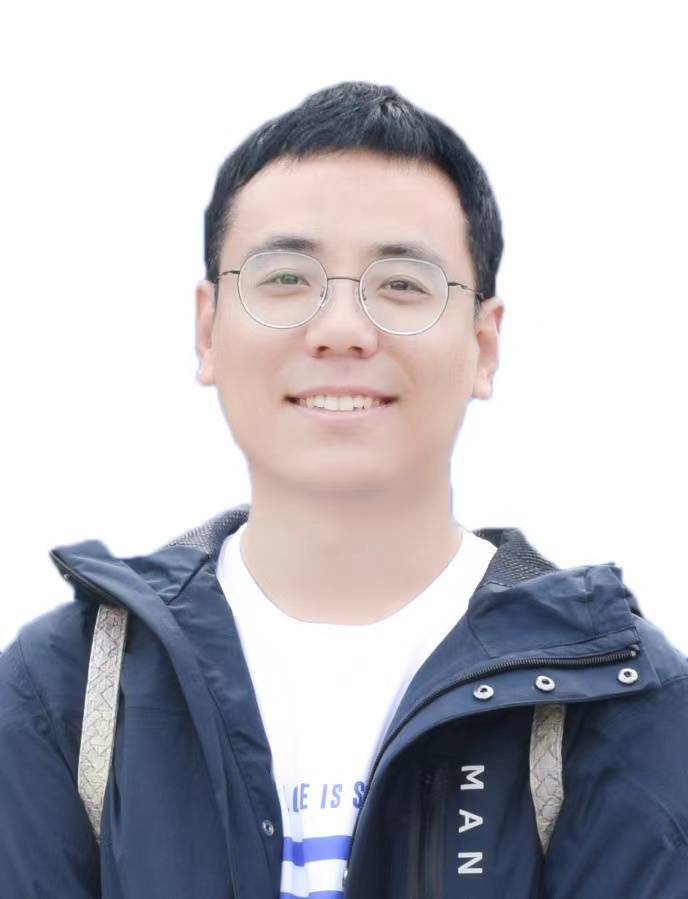}}]
			{Pingyue Yue} (Student Member) received his B.S. degrees from the Zhengzhou University , Zhengzhou, China, in 2016. He is currently a Ph.D. student in the School of Information and Electronics, Beijing Institute of Technology. His research interests include satellite communication, physical-layer security, and interference suppression.
		\end{IEEEbiography}
		
		\vspace*{-0mm}
		\begin{IEEEbiography}[{\includegraphics[width=1in,height=1.25in,clip,keepaspectratio]{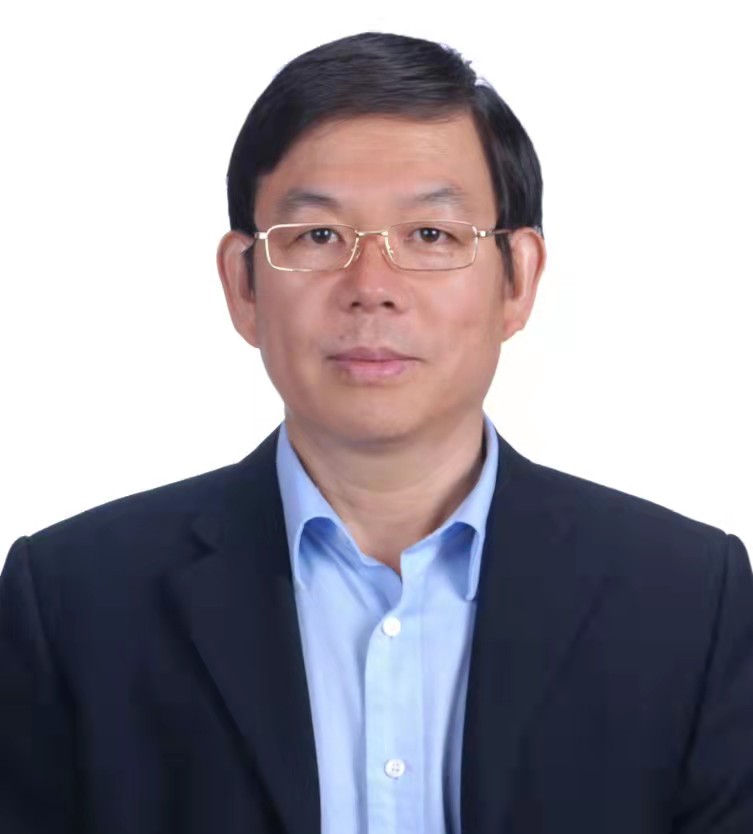}}]
		{Jianping An} (Senior Member, IEEE) received his Ph.D. degree from Beijing Institute of Technology, China, in 1996. He joined the School of Information and Electronics, Beijing Institute of Technology in
		1995, where he is now a full professor. He is currently the Dean of the School of Cyberspace Science and Technology, Beijing Institute of Technology. His research interests include digital signal processing,
		wireless communications, and satellite networks. He has received two national awards for technological
		inventions and science and technology progress.
		\end{IEEEbiography}	
		
		\vspace*{-0mm}
	\begin{IEEEbiography}[{\includegraphics[width=1in,height=1.25in,clip,keepaspectratio]{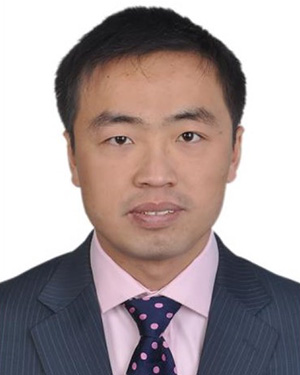}}]
	{Jiankang Zhang} (Senior Member, IEEE) is a Senior
	Lecturer of Computer Science with Bournemouth
	University. Prior to joining in Bournemouth
	University, he was a Senior Research Fellow with
	the University of Southampton, U.K. He serves as
	an Associate Editor for IEEE ACCESS.
	\end{IEEEbiography}		
			
		\vspace*{-0mm}
		\begin{IEEEbiography}[{\includegraphics[width=1in,height=1.25in,clip,keepaspectratio]{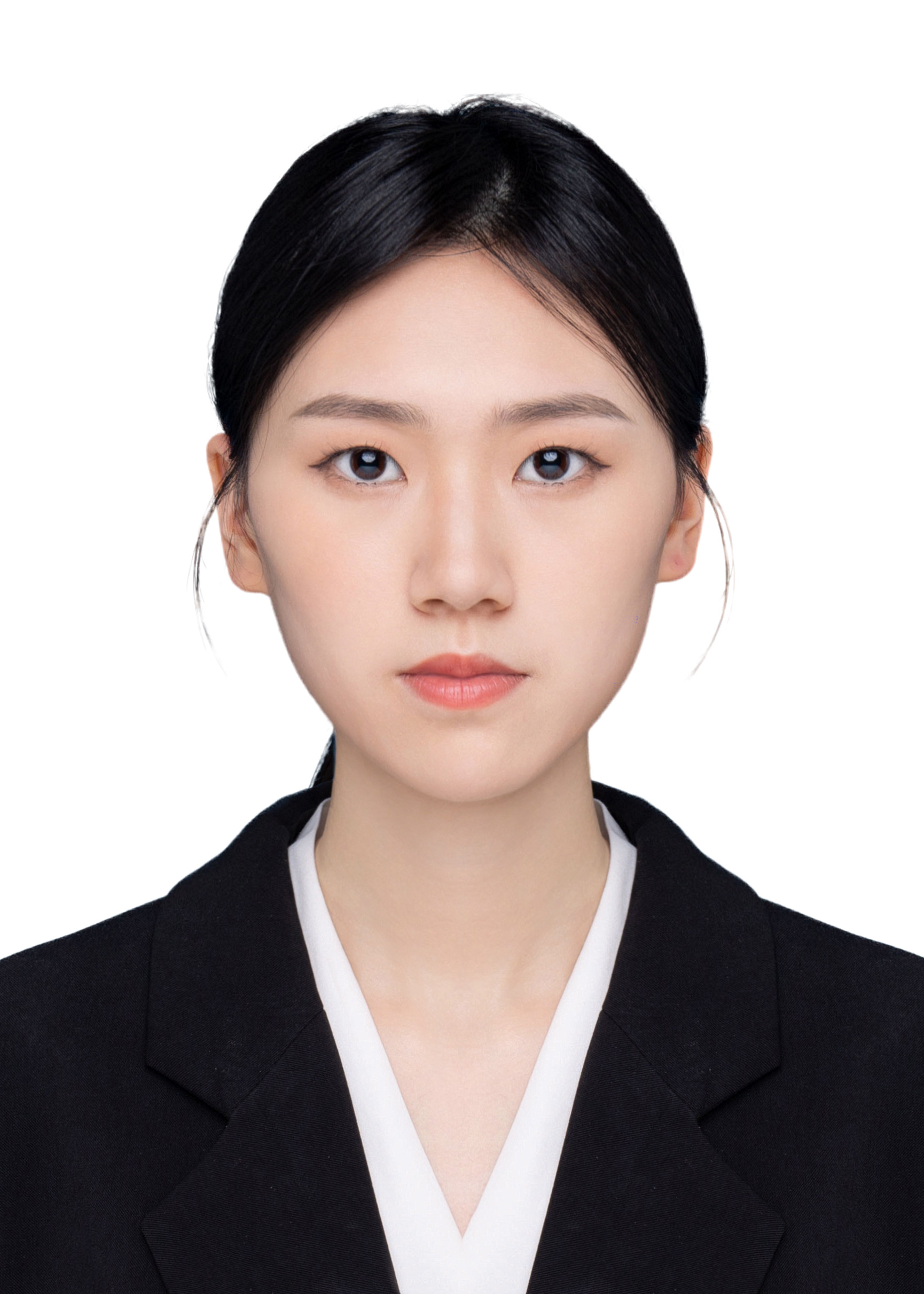}}]
			{Jia Ye} (Member, IEEE) was born in Chongqing, China. She received the Ph.D. degree in electrical and computer engineering
			from the King Abdullah University of Science and Technology (KAUST), Saudi Arabia, in 2022. Since January 2023, she has been with the School of Electrical Engineering, Chongqing University, Chongqing, China, as an Associate Professor. Her main 
			research interest includes the performance analysis and modeling of wireless information and energy transfer systems.
		\end{IEEEbiography}
		
		\vspace*{-0mm}
		\begin{IEEEbiography}[{\includegraphics[width=1in,height=1.25in,clip,keepaspectratio]{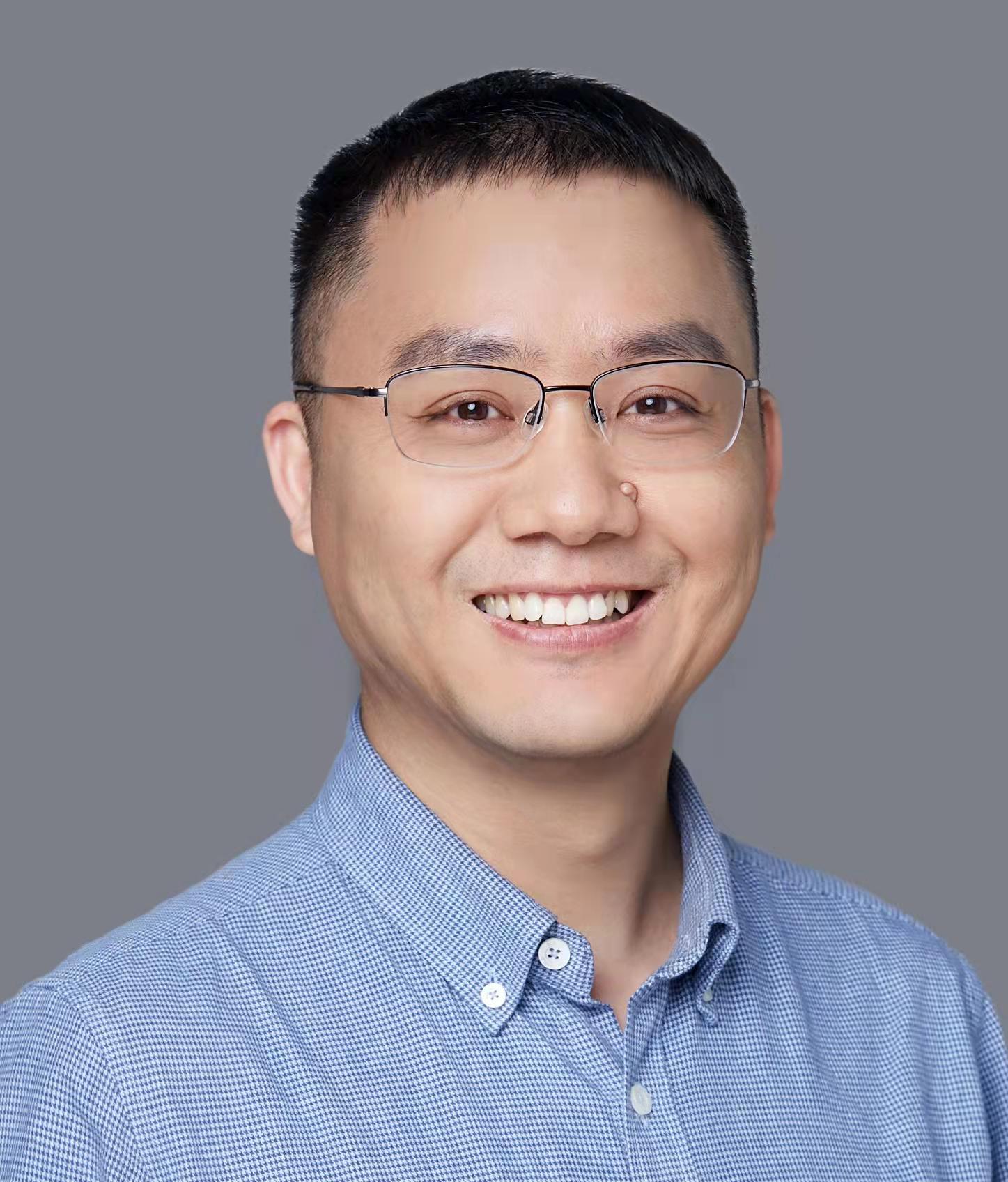}}]
			{Gaofeng Pan} (Senior Member, IEEE) received his B.Sc in Communication Engineering from Zhengzhou University, Zhengzhou, China, in 2005, and the Ph.D. degree in Communication and Information Systems from Southwest Jiaotong University, Chengdu, China, in 2011. He is currently with the School of Cyberspace Science and Technology, Beijing Institute of Technology, China, as a Professor. He is also serving as an Editor for several jounals, e.g., IEEE TRANSACTIONS ON GREEN COMMUNICATIONS AND NETWORKING, PHYSICAL COMMUNICATION, etc. His research interest spans special topics in communications theory, signal processing, and protocol design. 
		\end{IEEEbiography}

		\vspace*{-0mm}
		\begin{IEEEbiography}[{\includegraphics[width=1in,height=1.25in,clip,keepaspectratio]{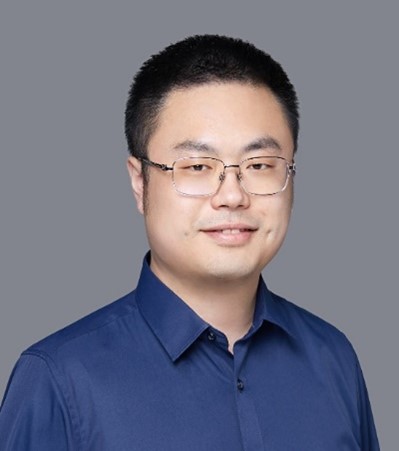}}]
			{Shuai Wang} (Member, IEEE) received the Ph.D. degree in communications systems from the Beijing Institute of Technology (BIT), China, in 2012. Upon his graduation, he joined the Faculty of the School of Information and Electronics, BIT. In 2021, he transferred to the new-founded School of Cyberspace Science and Technology, where he has been appointed as the Chair Professor of the Department of Information Security and Countermeasures. He has contributed more than 40 peer-reviewed articles, mainly in leading IEEE journals or conferences and holds more than 60 patents. His research interests include satellite communications, anti-interference communications, and datalink technologies for space platforms. He was a co-recipient of the Second Class National Technical Invention Award of China in 2019. He has served as an Editor for IEEE WIRELESS COMMUNICATIONS LETTERS. He is serving as an Editor for China Communications.
		\end{IEEEbiography}

		\vspace*{-0mm}
	\begin{IEEEbiography}[{\includegraphics[width=1in,height=1.25in,clip,keepaspectratio]{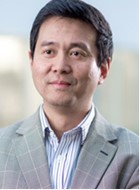}}]
		{Pei Xiao}(Senior Member, IEEE) is a professor of Wireless Communications at the Institute for Communication Systems (ICS), home
	of 5GIC and 6GIC at the University of Surrey. He received the PhD degree from Chalmers
	University of Technology, Gothenburg, Sweden
	in 2004. He is currently the technical manager of 5GIC/6GIC, leading the research team
	in the new physical layer work area, and coordinating/supervising research activities across all
	the work areas (https://www.surrey.ac.uk/institutecommunication-systems/5g-6g-innovation-centre). Prior to this, he worked at
	Newcastle University and Queen’s University Belfast. He also held positions
	at Nokia Networks in Finland. He has published extensively in the fields of
	communication theory, RF and antenna design, signal processing for wireless
	communications, and is an inventor on over 15 recent 5GIC patents addressing
	bottleneck problems in 5G systems.	
\end{IEEEbiography}	
		
		\vspace*{-0mm}
	\begin{IEEEbiography}[{\includegraphics[width=1in,height=1.25in,clip,keepaspectratio]{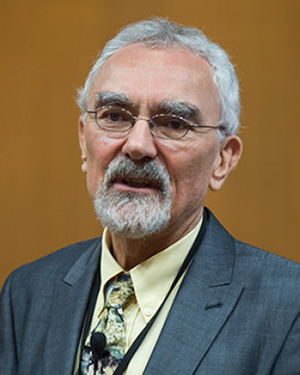}}]
	{Lajos Hanzo}  (Life Fellow, IEEE) received the
	master’s and Doctoral degrees from the Technical
	University (TU) of Budapest in 1976 and 1983,
	respectively, the Doctor of Sciences (D.Sc.) degree
	from the University of Southampton in 2004, and the
	first Honorary Doctoral degree from TU of Budapest
	in 2009 and the second Honorary Doctoral degree
	from the University of Edinburgh in 2015. He is the
	recipient of the 2022 Eric Sumner Field Award. He
	is a Foreign Member of the Hungarian Academy of
	Sciences and a former Editor-in-Chief of the IEEE
	Press. He has served several terms as Governor of both IEEE ComSoc and
	of VTS. He has published over 2000 contributions at IEEE Xplore, 19 Wiley-
	IEEE Press books and has helped the fast-track career of 123 Ph.D. students.
	He is also a Fellow of the Royal Academy of Engineering, IET, and of
	EURASIP.
	\end{IEEEbiography}

	\end{document}